%% file: idr2_pspec.tex
\documentclass[twocolumn]{aastex63}

\usepackage{mathtools,hyperref}
\usepackage{amsmath}
\usepackage{graphicx}
\usepackage[figuresright]{rotating}
\usepackage{natbib}
\usepackage[all]{hypcap}								% Workaround for hyperref
\usepackage{fancyhdr}
\usepackage{bm}
\usepackage{booktabs}
\usepackage{array}
\def\q{\mathbf{q}}
\def\p{\mathbf{p}}
\def\x{\mathbf{x}}
\def\R{\mathbf{R}}
\def\W{\mathbf{W}}
\def\F{\mathbf{F}}
\def\E{\mathbf{E}}
\def\H{\mathbf{H}}
\def\G{\mathbf{G}}
\def\C{\mathbf{C}}
\def\bS{\mathbf{S}}
\def\N{\mathbf{N}}
\def\Cinv{\mathbf{C}^{-1}}
\def\I{\mathbf{I}}
\def\Q{\mathbf{Q}}
\def\M{\mathbf{M}}
\def\Cov{{\rm Cov}}

\def\kpara{k_{\parallel}}
\def\kperp{k_{\perp}}

\def\bd{\mathbf{d}}
\def\bSigma{\mathbf{\Sigma}}

\newcolumntype{L}{>{$}l<{$}}
\newcolumntype{C}{>{$}c<{$}}
\newcolumntype{R}{>{$}r<{$}}

\shorttitle{HERA Phase I Upper Limits on the 21\,cm EoR Power Spectrum}
\shortauthors{The HERA Collaboration}

\begin{document}
\title{First Results from HERA Phase I: Upper Limits on the Epoch of Reionization 21 cm Power Spectrum} 

\collaboration{100}{The HERA Collaboration:}

\input{author-list}

\begin{abstract}
We report upper-limits on the Epoch of Reionization (EoR) 21\,cm power spectrum at redshifts 7.9 and 10.4 with 18 nights of data ($\sim36$ hours of integration) from Phase I of the Hydrogen Epoch of Reionization Array (HERA).
The Phase I data show evidence for systematics that can be largely suppressed with systematic models down to a dynamic range of $\sim10^9$ with respect to the peak foreground power.
This yields a 95\% confidence upper limit on the 21\,cm power spectrum of $\Delta^2_{21} \le (30.76)^2\ {\rm mK}^2$ at $k=0.192\ h\ {\rm Mpc}^{-1}$ at $z=7.9$, and also $\Delta^2_{21} \le (95.74)^2\ {\rm mK}^2$ at $k=0.256\ h\ {\rm Mpc}^{-1}$ at $z=10.4$.
At $z=7.9$, these limits are the most sensitive to-date by over an order of magnitude.
While we find evidence for residual systematics at low line-of-sight Fourier $k_\parallel$ modes, at high $k_\parallel$ modes we find our data to be largely consistent with thermal noise, an indicator that the system could benefit from deeper integrations.
The observed systematics could be due to radio frequency interference, cable sub-reflections, or residual instrumental cross-coupling, and warrant further study.
This analysis emphasizes algorithms that have minimal inherent signal loss, although we do perform a careful accounting in a companion paper of the small forms of loss or bias associated with the pipeline.
Overall, these results are a promising first step in the development of a tuned, instrument-specific analysis pipeline for HERA, particularly as Phase II construction is completed en route to reaching the full sensitivity of the experiment.
\end{abstract}

%%%%%%%%%%%%%%%%%%%%%%%%%%%%%%
%%%%%%%%%%%% Introduction %%%%%%%%%%%%
%%%%%%%%%%%%%%%%%%%%%%%%%%%%%%

\section{Introduction}
\label{sec:intro}	

The 21\,cm line of neutral hydrogen has emerged in the past few decades as a theoretically powerful probe of cosmology and astrophysics by tracing the growth of structure across cosmic time \citep{Hogan1979, Madau1997}.
At high redshift ($z > 6$), the 21\,cm line is a tomographic probe of the intergalactic medium (IGM) and allows us to directly trace its ionization, density, and temperature state.
This is particularly important for understanding the Cosmic Dawn and the subsequent Epoch of Reionization (EoR), when radiative feedback from the first generation of stars and galaxies heated and ionized the IGM over the course of baryonic structure growth from $40 < z < 6$.
Understanding these two milestones will give us a much better understanding of the formation of the first luminous sources, their environments, and the growth of large-scale structure.

Our current understanding of the timing of the EoR is largely based on measuring the absorption and scattering of the IGM against background sources. 
These can be either (i) astrophysical sources, such as galaxies and quasars whose Lyman series transmission is sensitive to the ionization state of the IGM; or (ii) the Cosmic Microwave Background (CMB), whose fluctuations are sensitive to the integrated column density of free electrons. 
Recent analyses of both (i) and (ii) point towards a EoR that is evolving rapidly between $5.5 < z < 7$.
In particular, IGM damping wing absorption in QSO spectra (e.g. \citealt{MH04, Bolton11, Greig17b, davies18}) and the rapid decline of Lyman alpha emitting galaxies (e.g. \citealt{Stark10, Schenker12, Jensen13, Caruana14, Pentericci14, Mesinger15, mason18, Mason19}), suggest that the EoR is ongoing at $z \sim 7$, albeit with significant systematic uncertainties. 
The Lyman-$\alpha$ forest at $z \sim 6$ suggests that reionization has mostly completed by then \citep{McGreer2015}.  However, the sizable sightline-to-sightline scatter in the forest transmission \citep{Becker15, Bosman18} seems to require the final, overlap stages of the EoR to extend down to $z\sim5.5$ \citep{Kulkarni19, keating20, Choudhury20, Qin21}.
This is broadly consistent with the CMB constraints on the electron scattering optical depth \citep{Planck18} that supports a relatively late EoR with a midpoint around $z\sim7$ (e.g. \citealt{Mitra15, Douspis15,  Greig17, Millea18, Qin20}), and it is also consistent with the high-$z$ galaxy luminosity functions, given reasonable assumptions about their properties \citep{Robertson2015, Price16, Gorce18, Hazra19, Park19, Qin20b}.
However, much remains to be understood about the EoR, such as when it began, the properties of the astrophysical sources that drove it, and what impact it had on future generations of star-forming galaxies.
Meanwhile, considerably less is understood about Cosmic Dawn, including when the first Population III stars formed, what their physical and spectral properties were, what their impact was in heating and enriching the IGM, and how this ultimately enabled the emergence of Population II stars and modern galaxies as we understand them.

Low-frequency radio experiments are aiming to directly measure the high redshift 21\,cm signal and in doing so place constraints on the timing and duration of the EoR to understand the underlying physical processes driving the heating and reionization of the IGM \citep[for reviews, see][]{Ciardi2005, Furlanetto2006c, Morales2010, Pritchard2012, Mesinger2016b, Liu2020}.
Prior and ongoing interferometric experiments include the Donald C. Backer Precision Array for Probing the Epoch of Reionization \citep[PAPER;][]{Parsons2010, Cheng2018, Kolopanis2019}, the Murchison Widefield Array \citep[MWA;][]{Tingay2013, Dillon2014, Dillon2015b, Ewall-Wice2016b, Beardsley2016, Barry2019b, Li2019, Trott2020}, the Low Frequency Array \citep[LOFAR;][]{vanHaarlem2013, Patil2017, Gehlot2019, Mertens2020}, the Giant Metre Wave Radio Telescope \citep[GMRT;][]{Paciga2013}, and the Long Wavelength Array \citep[LWA;][]{Eastwood2019}, which have put increasingly stringent constraints on the power spectrum over the past decade.
At the same time, total-power measurements of the sky-averaged signal (or global signal) have similarly put upper limits on the monopole component of the 21\,cm signal \citep{Bernardi2016, Singh2017, Monsalve2017}, with a tentative first-detection of the 21\,cm global signal at $z\approx 17$ from the Experiment to Detect the Global Eor Signature \citep[EDGES;][]{Bowman2018}, 
%although complementary analyses contend that
with a robust discussion of whether
this signature may be due to instrumental systematics \citep{Bradley2019, Hills2018, Singh2019, Sims2020, Mahesh2021}.

The 21\,cm power spectrum at the EoR and Cosmic Dawn contains a wealth of statistical information that can be used as both a cosmological and astrophysical probe \citep{Mao2008, Patil2014, Pober2014, Liu2016b, Greig2016, Ewall-Wice2016a, Kern2017}.
While radio interferometric experiments have indeed made substantial progress over the past decade, a first power spectrum detection has yet to be made.
As second-generation 21\,cm experiments are designed and built, such as the Hydrogen Epoch of Reionization Array \citep[HERA;][]{DeBoer2017} and the Square Kilometre Array \citep[SKA;][]{Koopmans2015}, the understanding and control of instrumental systematics will be the crucial factor in enabling their ultimate success in making a first robust detection of the 21\,cm power spectrum.

HERA is an interferometric array of fixed, zenith ponting dishes located in the Karoo desert, South Africa.
The dishes are 14 meters in diameter and packed hexagonally into a nearly continuously covered core 300m across. 
% The array configuration is optimized to support 
% averaging of redundant interferometric baselines for improved sensitivity\citep{Parsons2012b}.  The redundant
% configuration is broken into three offset ``shards'' which mitigates the worst impacts on uv coverage\citep{Dillon2015b}. 
The array is being built in a series of phases with simultaneous construction and observing.
Phase I used the feeds and correlator from the PAPER experiment \citep{Thyagarajan2016, Ewall-Wice2016c, Patra2018, Fagnoni2021}, while Phase II will use a new feed as well as analog and digital systems \citep{Fagnoni2020}.
%HERA Phase II is currently still under construction.
The data reported here come from the second internal data release (IDR2) taken from the first Phase I observing season, which commenced with roughly 50 antennas.

A series of recent papers describe the Phase I analysis pipeline in detail, which includes redundant calibration \citep{Dillon2020}, absolute calibration \citep{Kern2020b}, systematic modeling \citep{Kern2019, Kern2020a}, power spectrum estimation and error propagation \citep{Tan2021}, and pipeline validation \citep{Aguirre2021}.
Complementary studies on the data set discussed in this work also include foreground modeling \citep{Ghosh2020}, imaging \citep{Carilli2019}, power spectrum analysis of the bispectrum phase \citep{Thyagarajan2020}, antenna primary beam characterization \citep{Nunhokee2017}, and electromagnetic modeling of the front-end signal chain \citep{Fagnoni2021}.
Here, we give an overview of the full analysis pipeline, discuss the criteria used for data selection, present the power spectrum limits, and describe statistical tests used to characterize the performance of the system and our analysis techniques.

In this paper, we report the first upper limits on the 21\,cm power spectrum from HERA Phase I at redshifts 7.9 and 10.4 with 18 nights of observations.
These observations were made with only a fraction of the array built ($\sim50$ out of 350 antennas), as it was under active construction at the time.
Our analysis shows that the data are largely consistent with the expected thermal noise level at large line-of-sight Fourier $k$ modes, acheiving a dynamic range with respect to the peak foreground emission of $10^9$ in power.
However, at low Fourier $k$ modes we see evidence for residual, low-level systematics that begin to exceed the thermal noise.
Nonetheless, the limits presented here are to-date the most sensitive at $z\sim8$ by over an order of magnitude.
A companion astrophysical interpretation paper shows that, under standard galaxy astrophysics, these limits disfavour cold reionization scenarios, where the IGM temperature is not substantially heated above its adiabatically cooled limit (HERA Collaboration in prep.).

The paper is organized as follows.
In \S\ref{sec:obs} we give a summary of the instrument and the observations analyzed in this work.
In \S\ref{sec:reduction} we discuss the data reduction pipeline.
In \S\ref{sec:pspec} we discuss our power spectrum estimation pipeline and present our integrated power spectrum limits.
In \S\ref{sec:validation} we discuss a suite of statistical tests used to assess the quality and stability of the final data products.
Finally, in \S\ref{sec:summary} we summarize our results.

%%%%%%%%%%%%%%%%%%%%%%%%%%%%%%%
%%%%%%%%%%%% Observations %%%%%%
%%%%%%%%%%%%%%%%%%%%%%%%%%%%%%%

\section{Observations}
\label{sec:obs}

Here we discuss the observational parameters used for this analysis, as well as the state of the HERA instrument when these observations were made.
As noted, the data discussed in this work was taken with the Phase I instrument, which was a hybrid HERA/PAPER system.
Phase I re-purposed the radio frequency (RF) signal chains and correlator from PAPER and attached them to new HERA dishes.
The antenna consists of a 14-meter dish with a cross-dipole feed at its focal point measuring two linear polarizations \citep{DeBoer2017}.
At 150\,MHz the beam has a full width at half maximum (FWHM) of $\sim$10$^\circ$ \cite{Fagnoni2021}.
Not all PAPER front ends could be salvaged and as a stopgap new signal chain components---feed baluns and post-amplifiers---were manufactured to be backward compatible with the 75\,$\Omega$ cables carried over from PAPER.

\begin{figure*}
\centering
\includegraphics[width=\linewidth]{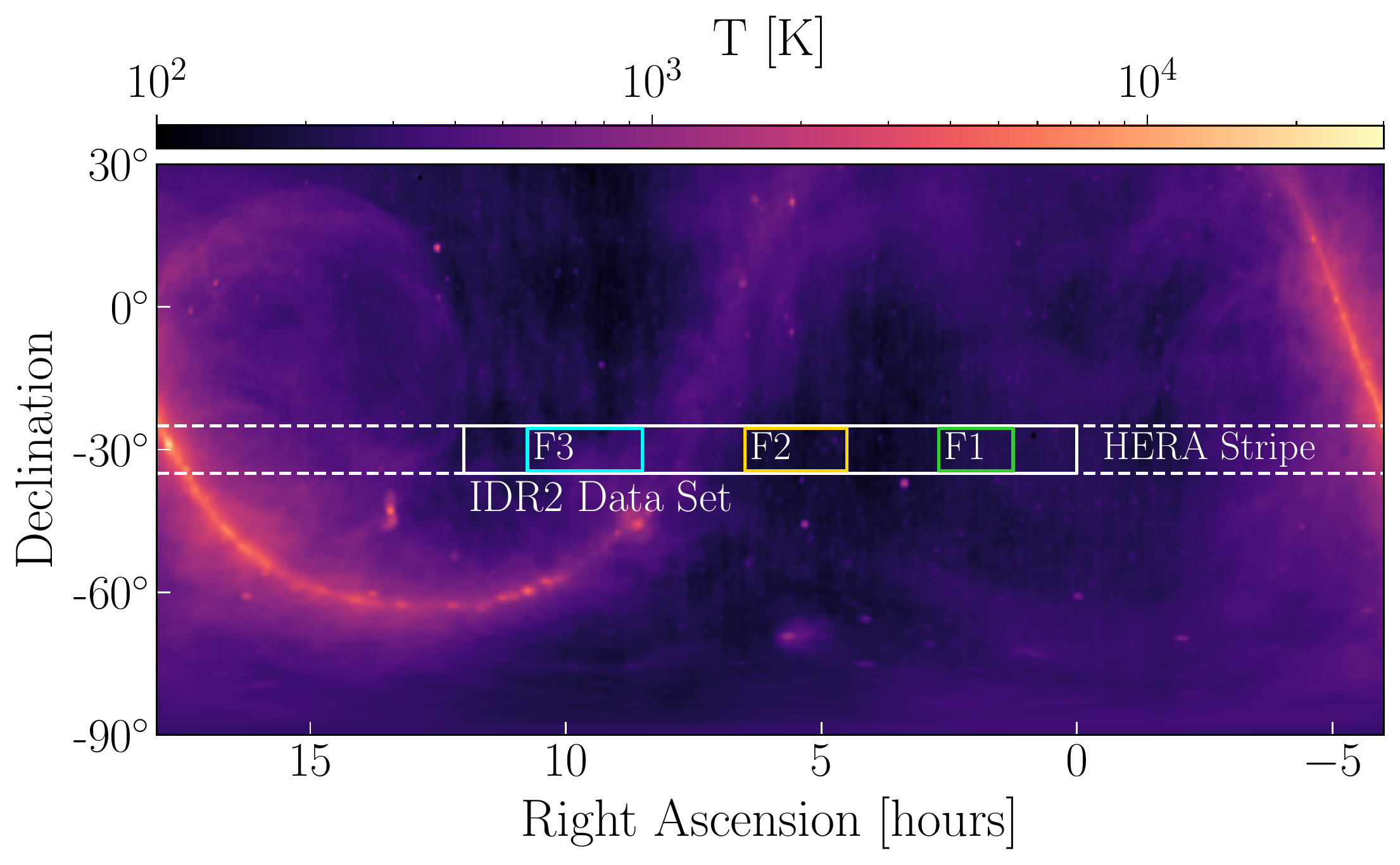}
\label{fig:hera_stripe}
\vspace{-5mm}
\caption{The Global Sky Model at 150 MHz \citep{Oliveira2008} showing the dominant diffuse foregrounds from the galaxy. Being a non-tracking, zenith pointed array, HERA's field of view is centered at $\delta=-30.7^\circ$ and in theory spans a full $360^\circ$ range in right ascension.
With a primary beam full width at half maximum of $\sim$10$^\circ$ (at 150 MHz), HERA's sensitivity is primarily focused on a narrow strip of sky. The data reported here spans a range of right ascensions from 0 to 12 hours (IDR2), although only three specific fields (colored boxes) largely devoid of bright foregrounds are used for power spectrum estimation.
%We also plot the observational weight across the IDR2 data set--a unitless relative scale of total sensitivity--which shows reduced sensitivity at the LST boundaries as well as during the transit of Fornax A at $\sim3.3$ hours LST.
}
\end{figure*}

\begin{figure}
\centering
\includegraphics[width=0.45\textwidth]{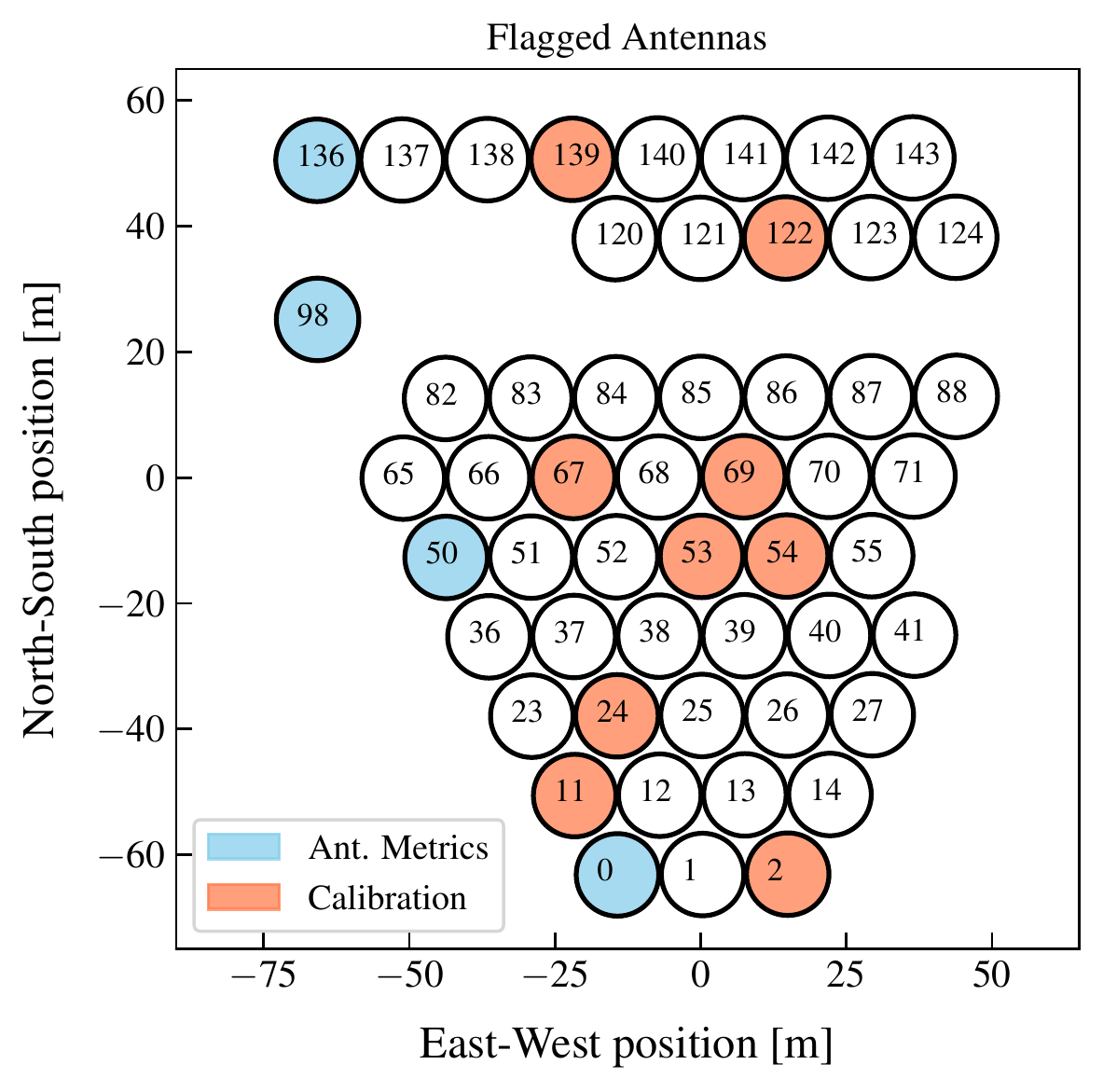}
\label{fig:array_layout}
\caption{The HERA Phase I array layout. Antennas flagged by the antenna metrics stage are shown in blue, while those flagged during calibration are shown in red, leaving a total of 39 unflagged antennas. For details on this process, see \citet{Dillon2020}.}
\end{figure}

\citet{Kern2020a} characterized the performance of the new and old signal chains for the Phase I system, finding spectral structure across a range of delays that cover the EoR window, which they attribute partially to cable reflections.
They also present methods for modeling and suppressing these systematics in the data.
\citet{Fagnoni2021} also performed electromagnetic simulations of the Phase I signal chains, finding a broad range of structure induced by both instrument cross-coupling, cable reflections, as well as dish reflections.

The HERA Phase I correlator, re-used from the PAPER experiment, employs an FX architecture, which was housed in a radio frequency interference (RFI)-shielded container in the field.
In the F-engine, analog-to-digital (ADC) units on a ROACH2 board \citep{Parsons2009} digitize each antenna's linear polarization voltage stream, which are then fed to a field programmable gate array (FGPA) that Fourier transforms the signal into voltage spectra with a 100 MHz bandwidth from 100--200\,MHz.
This is done across 1024 channels, leading to a spectral resolution of 97.66 kHz.
The spectra are then sent to a Graphics Processing Unit (GPU)-based X engine that correlates all $N(N-1)/2$ cross antenna-polarization pairs and all $N$ auto antenna-polarization pairs, where $N$ is the number of feeds.
The correlator then integrates the data for 10.7 seconds before writing them to disk.
Finally, the data are chunked into 10-minute files and sent to the on-site Karoo Array Processing Center for storage, and eventually transported to the National Radio Astronomy Observatory (NRAO) Array Operations Center in New Mexico, USA via internet connection.
Other observational parameters are summarized in \autoref{tab:obs}.

The HERA Phase I observing season ran from October 2017 to April 2018 while the array was under active construction.
Observations were taken throughout the South African summer at night when the galactic center and Sun are below the horizon, while active construction and commissioning proceeded during the day.
A major focus of development was building an online data quality assessment pipeline that reported on radio interference, calibration quality, and other metrics. 
These metrics, some of which we discuss in the next section, were used to select a subset of data taken during a relatively stable epoch in the observing season when construction activities were at a minimum and data quality was consistently high.
This resulted in an 18-day dataset that is the basis of this work, referred to as the second internal data release (IDR2).
These observations spanned December 10th - 28th, 2017 (Julian Dates 2458098--2458116).
For each of these days, observations were made for 12 hours per night, of which roughly 10 hours are used for science data when the Sun was below the horizon.

Together, these observations span a local sidereal time (LST) of 0 to 12 hours (\autoref{fig:hera_stripe}), which overlaps with a radio cold patch at 4 hours LST, as well as the galactic plane at 8 hours LST for HERA's latitude of -30.7$^\circ$ South.
In total, this work considers an 18-night data set (for a total of 180 hours), however, as discussed in \autoref{sec:reduction}, after additional rounds of data cuts we are left with three fields, each with $\sim36$ hours of integration for power spectrum analysis.
The area of the sky observed in the IDR2 dataset is highlighted in \autoref{fig:hera_stripe}, which shows the diffuse foreground emission from the galaxy and also identifies the three main fields used for power spectrum analysis (colored boxes).
The edges of the data set are not included due to reduced sensitivity of the nightly-averaged data products caused by LST drift of the observations from night-to-night.
%We also plot the observational weight of the fully integrated data, which is a relative, unitless scale that accounts for the flagging enacted on the data prior to binning.
%This shows reduced sensitivity at the boundaries of our lST range, which is simply due to the evolving LST range of each night in the dataset, as well as reduced sensitivty at $\sim3.3$ hours LST when Fornax A transits the beam, which causes our flagging routines to overflag the data (flagging is discussed in \autoref{sec:reduction}).

At the time of observations, the array had 52 fully deployed antennas arranged in a close-packed fashion, which constitues only a fraction of the full 350 that will eventually be built, and corresponds to a corner of the full array (\autoref{fig:array_layout}).
However, as we will discuss next, only 39 of these antennas were deemed science quality and were used for data analysis.

%\begin{figure}
%\centering
%\includegraphics[width=0.45\textwidth]{imgs/hera_antenna.jpg}
%\label{fig:hera_ant}
%\caption{A single HERA Phase I antenna in the field. This specific antenna is an outrigger. Though no outriggers are used in this analysis we use it here to clearly depict the dish. \dcj{Might want to consider adding a picture of the array too.} \jsd{Agreed with Danny... I'd consider just showing the array, but pick a picture where the antennas in front are very clear.}}
%\end{figure}

\begin{table}
\centering
\caption{HERA Phase I array and correlator parameters.}
\vspace{-3mm}
\begin{tabular}{l c c} 
 \hline
 \hline
 {\bfseries Array Parameters} \\
 Array coordinates (Lat, Lon) & -30.7$^\circ$S, 21.4$^\circ$E \\
 Total antenna number & 52 \\
 Unflagged antenna number & 39 \\
 Min. baseline length & 14.6 & meters \\
 Max. baseline length & 140 & meters \\
 Dish diameter & 14 & meters \\
 \hline
 {\bfseries Correlator Parameters} \\
 Min. frequency & 100 & MHz  \\ 
 Max. frequency & 200 & MHz  \\
 Number of channels & 1024  \\
 Channel width & 97.66 & kHz \\
 Integration time & 10.7 & sec  \\
 Nighttime recording duration & 12 & hours  \\ [1ex] 
 Total data duration & 18 & nights \\
 \hline
 \hline
\end{tabular}
\label{tab:obs}
\end{table}

%%%%%%%%%%%%%%%%%%%%%%%%%%%%%%%%
%%%%%%%% Data Reduction %%%%%%%%
%%%%%%%%%%%%%%%%%%%%%%%%%%%%%%%%

\section{Data Reduction}
\label{sec:reduction}

\begin{figure*}
\centering
\includegraphics[scale=0.4]{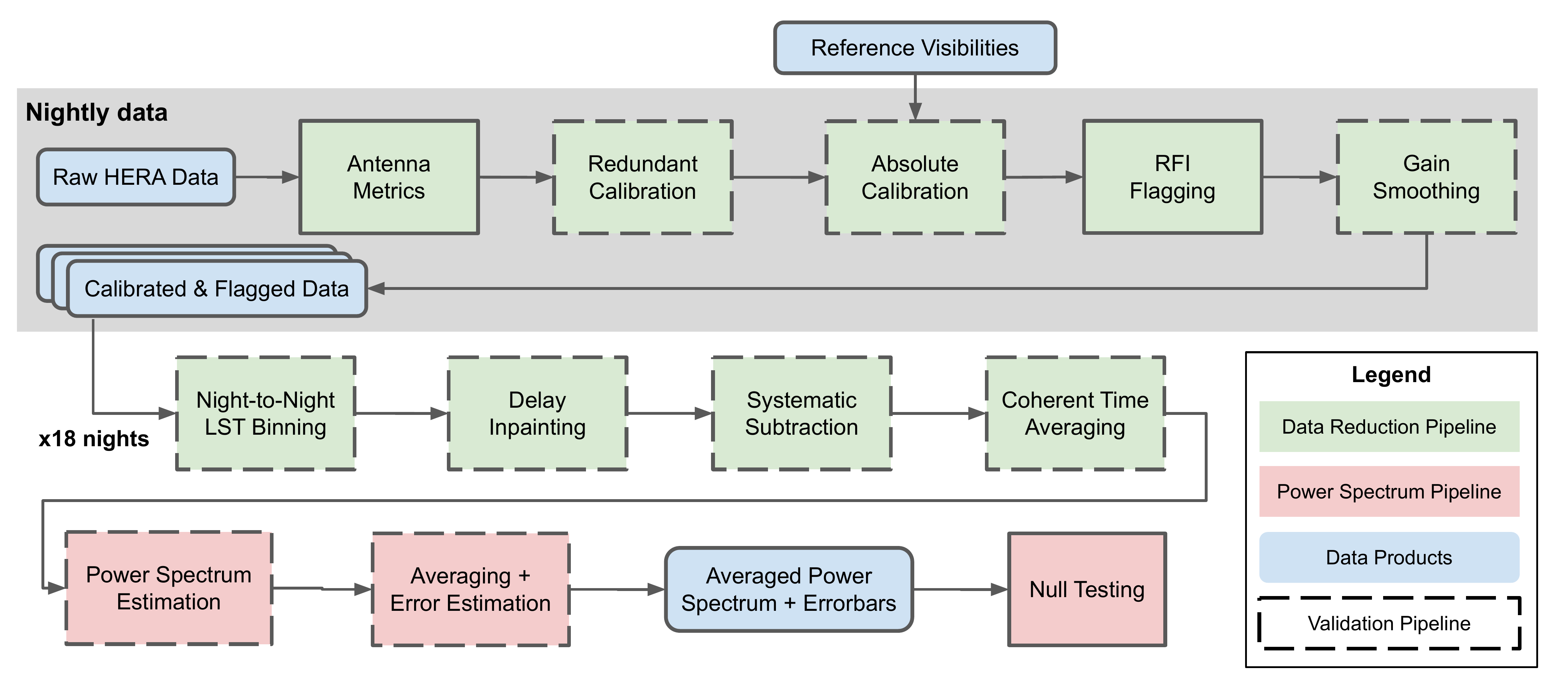}
\label{fig:data_flow}
\vspace{-3mm}
\caption{A diagram of the data reduction and power spectrum estimation pipelines, starting with raw HERA data and ending with the averaged power spectrum and their associated null tests.
The blue boxes represent data products while the green and red boxes represent steps in the data reduction and power spectrum pipelines, respectively.
Boxes with dashed borders represent elements tested by the validation pipeline \citep{Aguirre2021}.
Noise simulations are generated after the data reduction pipeline and fed through the power spectrum pipeline for diagnostic purposes when evaluating null tests.
}
\end{figure*}

Our analysis pipeline is summarized in \autoref{fig:data_flow}, which highlights the data reduction pipeline (green), the power spectrum pipeline (red), the data products that are input to and output by the pipeline (blue), and the steps that are tested in the companion validation analysis \citep[dashed;][]{Aguirre2021}.
Here we will focus only detailing the data reduction pipeline, which we do in order of their appearance in the pipeline.
The primary role of the reduction pipeline is to identify and flag faulty data, solve for the complex gain solutions imparted by the instrument and invert them, average the visibilities coherently across nights, and treat the data for known systematics.
Note that, unlike other works, we do not perform any explicit foreground subtraction or filtering in this analysis, although future work may benefit from such techniques \citep[e.g.][]{Ewall-Wice2021}.

Note that all of the important analysis packages can be found in the publicly accessible \texttt{HERA-Team}\footnote{\url{https://github.com/hera-team}} software repository, including the \texttt{hera\_cal}, \texttt{hera\_pspec},  \texttt{hera\_qm}, and \texttt{hera\_opm} packages used extensively in this analysis.

\subsection{Antenna Metrics}
\label{sec:ant_metrics}

Raw data from the correlator are first sent through an antenna metrics stage where faulty antennas are identified and flagged.
This is an important pre-calibration step to prevent obviously malfunctioning antennas from adversely affecting the calibration solutions.
Metrics are calculated on every 10-minute file for each of the dual-polarization feeds of every antenna, which gives us a sense for how antennas and their linear polarization feeds are behaving throughout any given night.

While a few different metrics were trialed, one metric was found to be the most useful and was the primary metric used to flag faulty antennas.
This metric was the mean visibility amplitude for each antenna, which we define for a given antenna $i$ as
\begin{align}
M_i = \frac{\sum_{j\neq i,\nu,t}|V_{ij}(\nu, t)|}{N-1},
\end{align}
where $V_{ij}(\nu, t)$ is the visibility between antennas $i$ and $j$ at frequency $\nu$ and time $t$, $N$ is the number of antennas in the array, and the sum is taken over all antennas $j\neq i$, times and frequencies in the observation.
For notational brevity, we will drop the explicit frequency and time dependence of $V_{ij}$.
This metric measures when an approximate estimate of the antenna gain is low compared to most other antennas, which can occur, for example, when the signal chain is not connected to the feed or a gain stage has lost power.
It was also used to help determine when an antenna was cross-polarized, or when its East and West polarization signal chains were accidently interchanged in the cable connections downstream.

Because the metric is computed before calibration, and thus uses raw data, we cannot make an absolute cut based on the expected noise level of the visibilities in Jansky.
Instead, we estimate the standard deviation of the metric across antennas and compute a $Z$-score, which is defined as the deviation of any given point from its sample mean in units of the sample standard deviation.
To account for outliers, we use the more robust, modified $Z$-score, defined as
\begin{align}
Z_i^{\rm mod} &= \frac{x_i - {\rm med}(x)}{\sigma^{\rm mad}} \label{eq:modz}\\
\sigma^{\rm mad} &= 1.482\times{\rm med}|x - {\rm med}(x)|, 
\end{align}
where $x$ is the mean visibility amplitude metric defined above.\footnote{Using the modified $Z$-score still relies on the majority of the array being healthy in order to detect outlier antennas. When this is not true, for example during correlator errors or partial power outage, it is very obvious from visual inspection of the data products.}
The median absolute deviation ($\sigma^{\rm mad}$) is a robust estimator of the standard deviation but requires a correction factor of 1.482 to reproduce the standard deviation in the case of white Gaussian noise \citep{NIST_Stats}.
%\autoref{fig:ant_metrics} shows an example of these metrics for all antennas for one 10-minute observation from night 2458101.

It is generally clear from these metrics which antennas are poorly behaving, nonetheless, we flag antennas that deviate from the sample mean beyond a 5$\sigma$ level.
Antennas that were repeateadly flagged by the antenna metrics stage for nights throughout the data set are flagged outright for all nights.
If either of an antenna's polarizations is thus flagged, we flag the entire antenna (i.e. we flag both linear polarizations).

The faulty antennas caught the antenna metrics stage are really only the very worst offenders, shown in blue in \autoref{fig:array_layout}.
Other sub-nominal antennas are flagged subsequently in the process of calibration, which we describe next.

\subsection{Redundant-Baseline Calibration}
\label{sec:redcal}

One of the foremost challenges for 21\,cm cosmology is the task of precision instrumental gain calibration.
In general, the measured visibility $V_{ij}^\text{obs}$ between antennas $i$ and $j$ is related to the true, uncorrupted visibility $V_{ij}^\text{true}$ via a product of each antennas gain, $g_i$,
\begin{align}
\label{eq:di_cal}
V_{ij}^\text{obs} = g_i g_j^\ast V_{ij}^\text{true} + n_{ij},
\end{align}
where $n_{ij}$ is the thermal noise in the measurement \citep{Hamaker1996, Smirnov2011}. Note that we solve \autoref{eq:di_cal} for both linear polarizations independently (EE and NN), which are also implicitely a function of time and frequency.
This representation ignores intra-feed D-terms as well as direction dependent gain calibration, which we do not solve for in this analysis (a simulated primary beam model is used for sky-based calibration discussed in \autoref{sec:abscal}).

HERA's redundant array configuration opens the possibility of calibrating the relative gains of antennas using internal degrees of freedom among measurements of groups of redundant baselines \citep{Dillon2016}.
Redundant-baseline calibration uses the principle that every redundant baseline should measure the same visibility, they in practice do not due to antenna-based gain terms, which allows one to constrain the relative gain of each antenna \citep{Wieringa1992, Liu2010}.
Specifically, redundant-baseline calibration seeks to minimize a $\chi^2$ written as
\begin{align}
\label{eq:chisq}
\chi^2 = \sum\limits_{i<j}\frac{\left|V_{ij}^\text{obs} - g_ig_j^\ast V_{i-j}^\text{sol}\right|^2}{\sigma_{ij}^2},
\end{align}
where $V_{i-j}^\text{sol}$ is the visibility solution for the redundant baseline type with the same vector separation as $V_{ij}$.
The key distinction in redundant-baseline calibration highlighted here is that the visibility solutions are left as free parameters to be solved for in the process of calibration along with $g_i$ and $g_j$.
For highly redundant configurations like HERA, this leads to an overconstrained system of equations that can be solved to yield estimates of the gains in addition to the model visibilities. 

% commented out for now b/c these are details well-summarized by companion papers, and not functionally necessary to understand this paper 
%Redundant-baseline calibration does not suffer from some of the now well-studied effects in traditional sky based calibration such as incomplete point source catalogues and inaccurate diffuse emission templates that lead to erroneous spectral structure in calibration solutions \citep{Barry2016, Ewall-Wice2017}.
%Although sky-based spectral gain errors can be mitigated by enforcing smoothness of the gain solutions \citep{Yatawatta2015, Barry2016, Kern2020b}, they also limit the spectral modes of the true gains that can be calibrated-out accurately.
%However, deviations from the ideal scenario of baseline-redundancy can also cause spectral gain errors if not properly addressed \citep{Orosz2019}.
%Furthermore, redundant-baseline calibration only solves for the relative gain between antennas and needs a reference sky model to solve for three fundamental degeneracies per frequency and time that set the overall amplitude and phasing of the array \citep{Zheng2014, Dillon2018, Byrne2019, Li2018}.

Due to inherent degeneracies in the system of equations, redundant-baseline calibration cannot solve for the overall amplitude and directional phasing (the ``Tip-Tilt" phasor) of the antenna-based gains \citep{Liu2010, Zheng2014, Dillon2018, Li2018, Byrne2019}.
Consequently, these must be solved for independently and added to the redundant calibration gains, commonly referred to as absolute calibration.
%Because the two instrumental polarizations were calibrated independently, there are three such degeneracies per visibility polarization: the absolute amplitude, the East-West phase gradient, and the North-South phase gradient \citep{Dillon2018}.
%Because redundant-baseline calibration is insensitive to these parameters, the gains that come from our redundant-baseline calibration pipeline generally have some non-zero component in the degenerate space, set by the starting point of the raw data and the convergence of the solver.
%To simplify later absolute calibration (\autoref{sec:abscal}), we project out these degeneracies, which amounts to setting the absolute amplitude to one and the phase degeneracies to the initial degeneracies after solving for cable delays (i.e.\ \texttt{firstcal}, see \citealt{Dillon2020}).

A complete description of the redundant-baseline calibration of HERA appears in \citet{Dillon2020}, including a quantitative assessment of the redundancy of HERA via, e.g.\ the deviation of $\chi^2$ for the expected value in a perfectly redundant array. For more detail on the series of algorithms with which we minimize $\chi^2$ (\autoref{eq:chisq}), see \citet{Dillon2020}.
Furthermore, the subsequent process of solving for the degenerate modes (i.e. absolute calibration) is described in \citet{Kern2020b}.

\begin{figure}
\centering
\includegraphics[width=.48\textwidth]{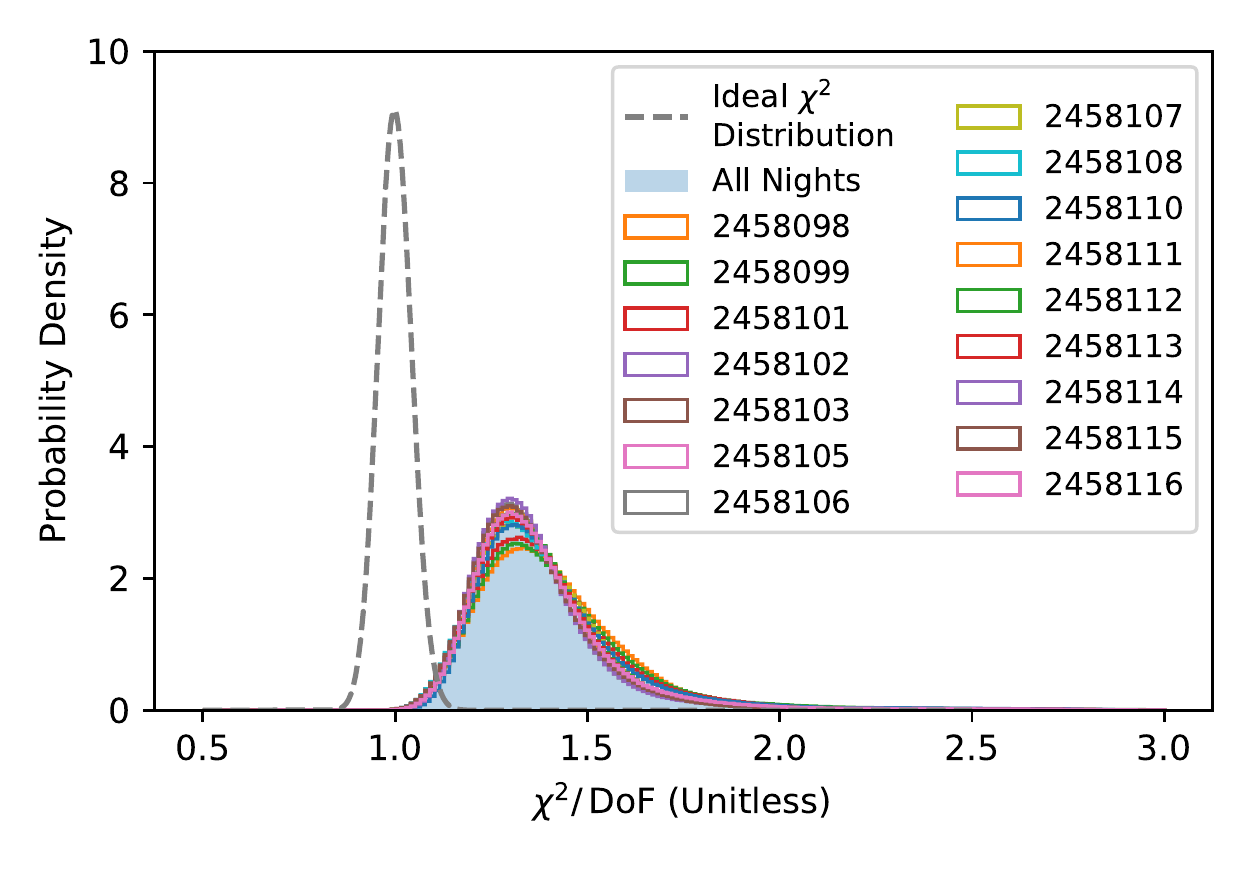}
\label{fig:redcal_chisq}
\vspace{-20pt}
\caption{Redundant calibration attempts to minimize $\chi^2$, defined in \autoref{eq:chisq}, using a model that assumes that redundant baselines observe the same true visibilities up to their antenna-dependent gains and the thermal noise. Were this true, we exect that $\chi^2$ normalized by the number of degrees of freedom (DoF) in the model (approximately 520 in this analysis, see \citet{Dillon2020} for a precise quantification) would have a mean of one and follow a $\chi^2$-distribution. Here we show the distribution of $\chi^2/$DoF over polarizations and unflagged (see \autoref{sec:xrfi}) times and channels for different nights in our analysis. While consistant from night-to-night, the overall mean of 1.389 is clearly inconsistent with one, indicating persistant  non-redundancy at levels roughly 20\% in excess of the thermal noise. \emph{Figure reproduced from \citet{Dillon2020} with permission.}} 
%\gb{This sentence seems contradictory, i.e. ``clearly inconsistent" with ``mild"... I don't mind too much, but I am left thinking if this is really not a problem...}} 
\end{figure}

In \autoref{fig:redcal_chisq} we show the results of redundant-baseline $\chi^2$-minimization for different nights in our analysis. 
Our results show clear evidence for non-redundancy in excess of the thermal noise at the $\sim$20\% level in the visibility. \citet{Dillon2020} attributes much of this excess to small deviations in the placement of dishes and to minor antenna-to-antenna variation of the primary beam. Some of it is also likely attributatble to baseline-dependent systematics \citep{Kern2020a} which break the assumption in \autoref{eq:di_cal}.
As discussed in \citet{Dillon2020}, the statistic is not consistent with one, meaning there are extra sources of variance in the data beyond noise that cannot be constrained by antenna-based  calibration. We assess the effect of non-redundancy on our final power spectrum limits in \autoref{sec:loss}.

\subsection{Absolute Calibration}
\label{sec:abscal}

To finish calibration, we need to solve for the degeneracies of redundant-baseline calibration by referencing to the sky.
One way to do this is to compare a model of the true visibilities to redundantly-calibrated visibilities and varying the degenerate parameters to make the two match.
In this work, we use a series of previously calibrated visibilities over the full LST range as our set of reference (or model) visibilities that we use to extract the degenerate components of the calibration gains.

We build our reference visibilities following the sky-based calibration methodology described in \citet{Kern2020b}, which uses the Common Astronomy Software Applications (CASA; \citealt{CASA}) software to construct visibility models at specific fields where we expect to be able to model the dominant contributions of the flux on intermediate and long baselines.
Specifically, we select three fields transiting our FoV at 2.0, 5.2, and 14.4 hours LST.
These fields are chosen to have minimal diffuse foregrounds within the FoV, and have a bright point source from the MWA GLEAM catalog \citep{Hurley-Walker2017} near the field's zenith pointing, which is used to confirm the absolute flux density scale of the calibrated data.

For each of the three fields we pick a different night in the Phase I observing season, selected such that the field transits roughly halfway through the nighttime observations.
We then select-out five minutes of observations centered at the field transit, average the data and perform calibration using standard CASA routines (e.g. \texttt{gaincal} and \texttt{bandpass}).
These gains are then transferred to every integration for that observing night, leaving us with three nights of calibrated night, each of which span a slightly different range of LSTs.
Drift in the gain amplitude throughout the nighttime observation is not corrected for, but is expected to drift at the $\sim3\%$ level \citep{Kern2020b}.

\begin{figure*}
\centering
\includegraphics[width=\linewidth]{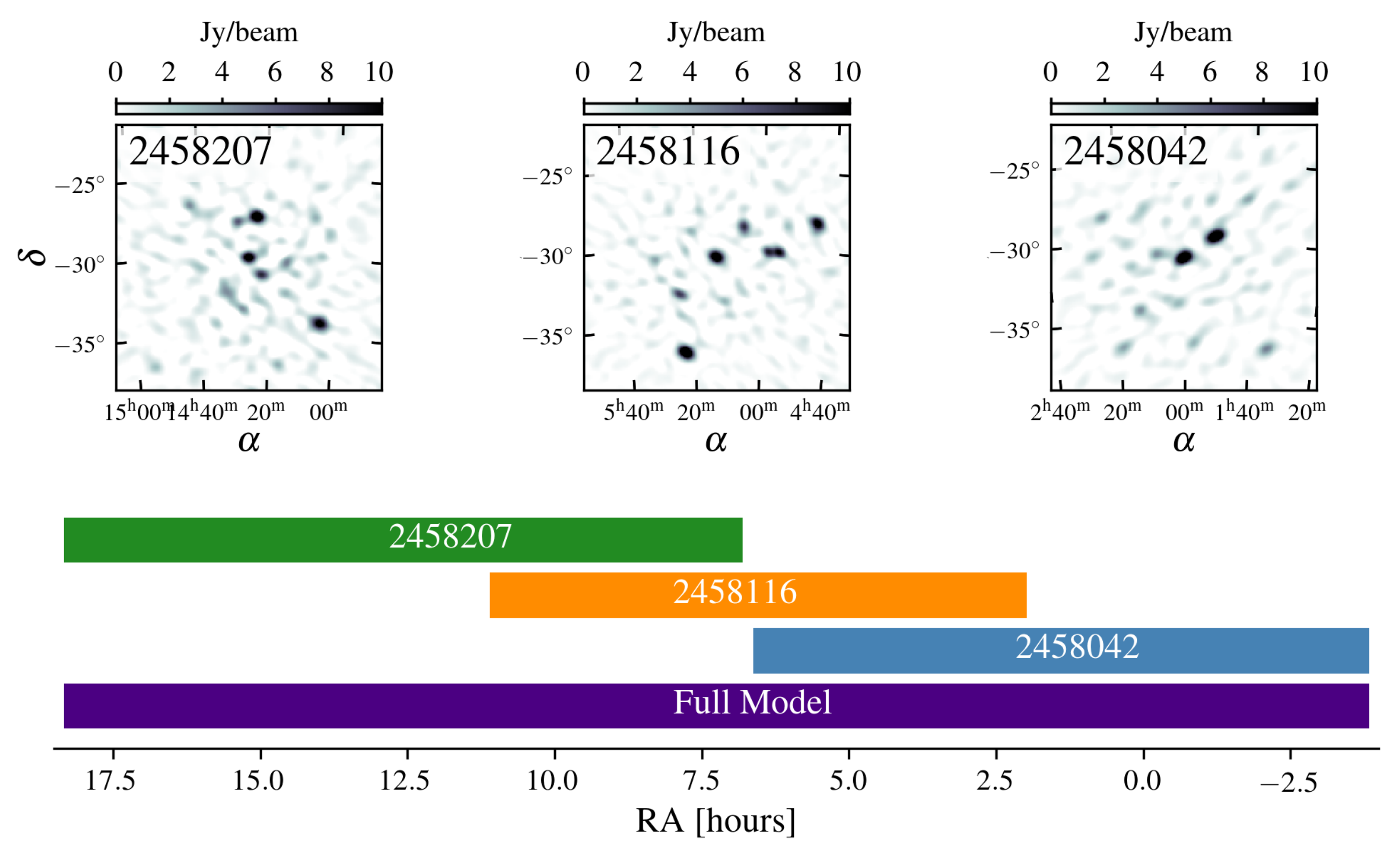}
\label{fig:abscal_model}
\vspace{-5mm}
\caption{Construction of the absolute calibration reference visibilities (purple) from a set of three nights throughout the observing season.
Each night is calibrated at a single field that transits near midway through the night. The calibration is transferred to all integrations from that observing night, and the visibilities are then LST binned onto a common grid to form the full model spanning nearly 22 hours in LST. Drift in the gain amplitude through each night is not corrected for, but is expected to drift at the $\sim$3\% level \citep{Kern2020b}.} 
\end{figure*}

Finally, these three sets of visibilities are averaged onto a common LST grid spanning nearly 20 hours of LST, which yields our full set of reference (or model) visibilities that we use for absolute calibration.
After averaing, the visibilities are passed through a Fourier low-pass filter across frequency, which filters out all signals with delays beyond the baseline horizon delay plus an additional 50 nanoseconds.
This suppresses noise in high-delay structure that is not associated with foregrounds, and also fills-in flagged frequency channels with a best-guess of the foreground structure using a delay-domain deconvolution \citep[similar to the inpainting algorithm described later;][]{Parsons2009, Kern2020b}.

\autoref{fig:abscal_model} shows CLEANed, multi-frequency synthesis images (across $135--165$ MHz) of the three fields we use for constructing the reference visibilities, demonstrating the point source distribution in each field.
It also shows the Julian Date (JD) of the night used to calibrated each field as well as range in LSTs of each of the nights (green, orange, and blue bars).
Stacking these visibilities onto a common LST grid gives us our final ``Full Model'' (purple bar) that we use in performing absolute calibration of the nightly data in the Phase I data set considered here.
It should be noted that the gains solved for are inherently direction-independent gains, with no cross-feed D terms.

\subsection{Radio Frequency Interference Flagging}
\label{sec:xrfi}

The Karoo Astronomy Reserve is an radio frequency interference (RFI) quiet zone, and as such much of the frequency band between 120 to 180 MHz is relatively clean of interference.
Nonetheless, we still see narrowband RFI due to satellite and terrestrial transmitters, as well as rare occurances of wideband RFI above 180 MHz.
Our RFI detection algorithm relies on the detection of local outliers, using the distribution of surrounding data points in time and frequency to distiguish RFI from thermal noise fluctuations.
Our pipeline uses a number of data products to identify not just interference but also general problems with the array, such as correlator malfunctions.
As its input it takes raw HERA data, the gain solutions from redundant-baseline calibration and the redundant visibility solutions, its $\chi^2$ distributions, the absolute calibration gains, and their $\chi^2$ distributions.

For each data product (except the raw visibilities) a corresponding modified Z-score metric is computed, $Z^{\rm mod}$, which is formed by median-subtracting the data product with a median filter and then normalizing it by an empirical estimate of its noise.
%detrending the data product with a median filter and then normalizing by an estimate of the per time and frequency noise using the local median absolute deviation.
%Both the filter and the noise estimate are performed using a 17 integration by 17 channel region around the data point of interest.
Performing the median filter across frequency and time helps to identify occasional sources of wideband RFI from broadcast televsion that can be as wide as 8 MHz \citep{Wilensky2020Memo}.
%Wideband outliers are identified in an analogous way using the median and median absolute deviation of the redundant-baseline calibration $\chi^2$ across time and frequency. This helps identify occasional RFI from broadcast television which can be as wide as 6\,MHz \citep{Wilensky2020Memo}.
These metrics are averaged in quadrature across all baselines, antennas, and polarizations to increase the sensitivity to low-level contamination, resulting in a single metric as a function of time and frequency for each data product.
%We calculate one additional metric by combining all individual metrics together in quadrature.

Next we flag these metric waterfalls with an initial threshold of $Z^{\rm{mod}} \ge 5$, followed by a watershed algorithm where any time and frequency pixel adjacent to a flagged pixel is itself flagged if it exceeds a lower threshold of $Z^{\rm{mod}} \ge 2$ \citep{Kerrigan2018}.
We found that different types of contaminants will manifest more brightly in certain metrics, so each metric is flagged individually.

Having removed the brightest sources of RFI, we now apply the resultant flags and repeat the flagging procedure; however, this time we compute the standard $Z$-score metric using a mean filter rather than a median filter.
The mean filter is considerably faster, so at this stage we are also able to include the raw visibility data as an additional metric.
Just as before, we compute these metrics, average them in quadrature, and apply the same watershed flagging routine on the averaged metrics.
The result is a single waterfall of flags for all baselines.
This entire routine is performed on every 10-minute file over the course of a nightly observation.

Our last step is to examine each metric waterfall over the entire night and use use medians to collapse them to a single time series or spectum.
These are again flagged by computing a modified $Z$-score for each channel or integration, flagging any above 7 or any above 3 that are adjacent to a previously flagged channel or integration.
This step helps us find global outliers in time that may not be obvious from the analysis of a signal file and to find channels that are so frequently occupied by RFI that it safer to assume they are always RFI-contaminated.

\begin{figure*}
\centering
\includegraphics[width=\linewidth]{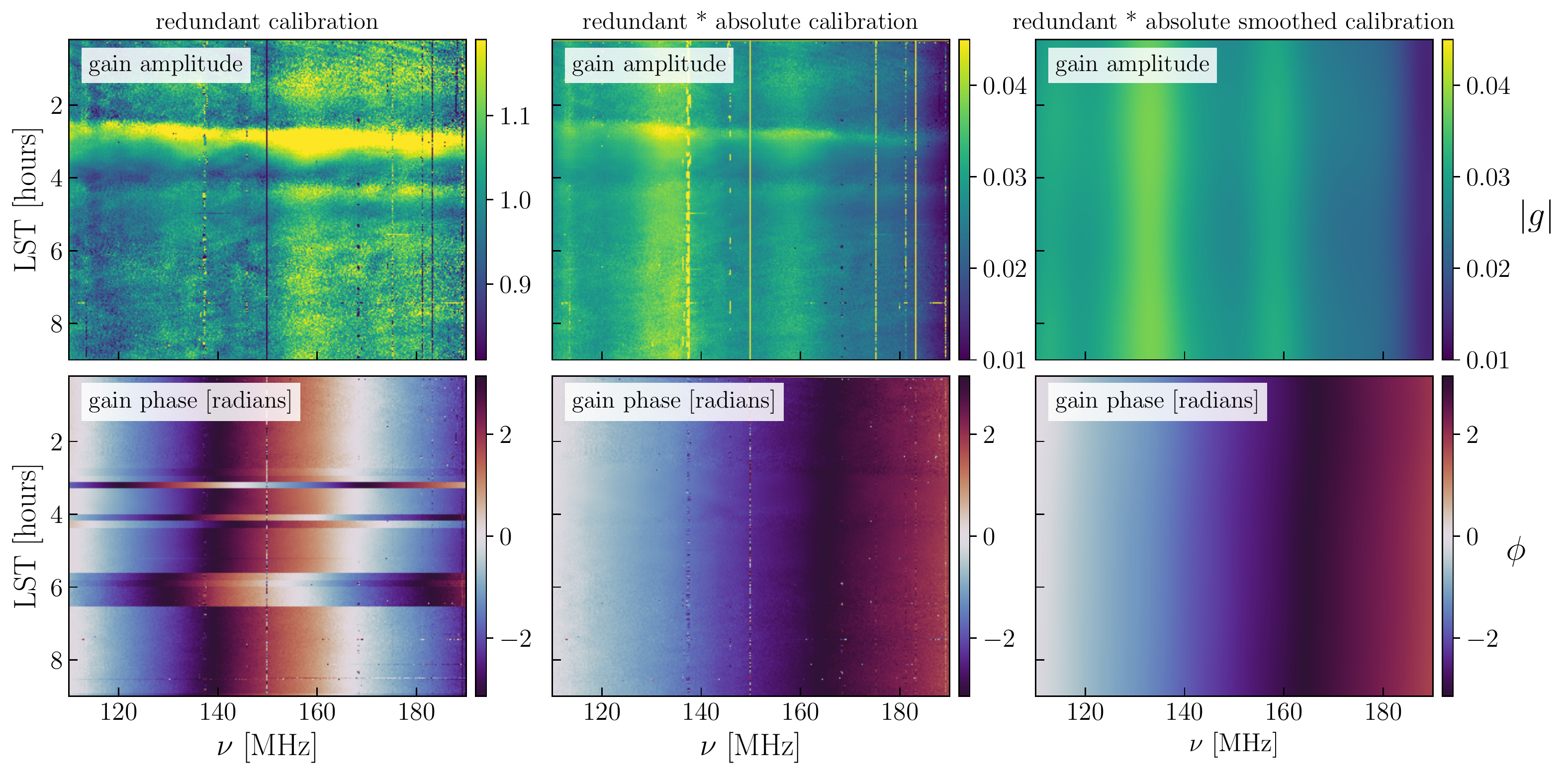}
\label{fig:gain_progression}
\vspace{-5mm}
\caption{The progression of a single antenna based gain through the calibration process of a single night, showing the gain amplitude (top panels) and phase (bottom panels).
Redundant calibration (left) solves for the relative gain but is susceptible to phase jumps at certain times and requires certain degeneracies, like the overall gain amplitude, to be filled in. Absolute calibration (center) solves both of these problems by pinning the degenerate components and removing the phase jumps.
Lastly, the final gains are smoothed across frequency and time (right) to limit excess structure \citep{Kern2020b, Dillon2020}---for example, amplitude fluctutations caused by the passage of Fornax A around 3.4 hours seen in the top left and middle panels. Note that the process of gain smoothing is applied to the center column, not to the first column where phase jumps are apparent.}
\end{figure*}

The strongest forms of indentified RFI are persistent, narrowband emitters, which can be clearly seen in the calibration gain solutions (e.g. \autoref{fig:gain_progression}).
However, the final flagging mask applied to the nightly data excises both narrow and wideband features (e.g. the left panel of \autoref{fig:lstbin_inpaint}).
The contaminating features in the data include FM radio ($\nu < 111$\,MHz), ORBCOMM satellite communication ($\nu=137$\,MHz), broadcast television ($\nu > 174$\,MHz), as well as intermittent correlator integration failures (wide band).

In addition to flagging the data for RFI and correlator failures, we also flag 50 channels on either edge of the frequency band where the bandpass filter falls off steeply, as well as all integrations where the Sun is at or above the horizon due to issues it creates in our calibration.
Lastly, after LST binning (described below), the averaged visibilities are manually inspected for low-level narrowband RFI that was missed by the nightly RFI excision. In this process we find and flag a handful of frequency channels (for all baselines and times), which amounts to $<1\%$ of the total data.

On average, we flag roughly 15\% of the band in our RFI flagging step. This estimate excludes our routine flagging of the band edges due to the rolloff of the bandpass filter ($\sim10$ MHz on either side of the band), as well as the complete flagging of certain times due to problems with correlator failures.
Through visual inspection, we see areas where our pipeline is likely overflagging the data.
This suggests that our flagging strategy is somewhat more aggresive and could be improved to reduce the false positive rate.
Future work will investigate how more precise RFI excision techniques, like the matched-filter SSINS package \citep{Wilensky2019}, can improve the overall RFI identification performance.

\subsection{Gain Smoothing}
\label{sec:gain_smoothing}

While the goal of gain calibration is to remove spectral structure introduced by the instrument, it is a double-edged sword and can also impart spurious spectral structure.
Many sources of uncertainty come into play when calibrating a real instrument, including baseline-to-baseline non-redundancies \citep{Orosz2019}, incomplete sky flux density models \citep{Byrne2019}, and various baseline-dependent instrumental systematics.
A number of tehcniques have been proposed to mitigate this effect, including the consensus optimization technique \citep{Yatawatta2015, Yatawatta2016} used in LOFAR power spectra (e.g.\ \citealt{Patil2017} and \citealt{Mertens2020}) and fitting low-order polynomials to averaged gains, a technique employed with the MWA (e.g.\ \citealt{Barry2019a, Barry2019b}).

\citet{Kern2020b} discussed some of these effects for HERA, and found their impact to be most severe for $|\tau| > 100$ ns.
They introduce a Fourier filtering procedure for low-pass filtering the gains after calibrating out a per-antenna delay, which mitigates the impact these spectrally-dependent gain errors have on the data.
This filter is a two-dimensional frequency and time iterative deconvolution algorithm that both acts as a low-pass filter and also fills-in missing calibration solutions due to RFI flagging.
It is conceptually similar to the Hogbom CLEAN algorithm \citep{Hogbom1974}.

\citet{Dillon2020} showed that per-integration redundant-baseline calibration exhibits time dependent structure that is LST-locked and argued that this variation was due to bright sources moving through direction dependent non-redundancy.
Examining the temporal power spectra of gain solutions after dividing out an LST-locked nighly average revealed no evidence for intrinsic gain fluctuations on timescales shorter than 6 hours.
Likewise, \citet{Kern2020b} show that the gains drift in amplitude with changes in the ambient temperature, which varies slowly over the course of the 10-hour nightly observation. Therefore, we also use the 2D low-pass filter to remove temporal structure in the gain solutions on all timescales shorter than 6 hours.
This substantially mitigates sky-dependent effects, since the filtering timescale is much longer than the beam-crossing time---approximately 40 minutes at 150\,MHz, given a $10^\circ$ FWHM \citep{Neben2016}.

\autoref{fig:gain_progression} shows the progression of the antenna gains from redundant-baseline calibration (left), absolute calibration (center), and gain smoothing (right).
Redundant-baseline calibration is prone to jumps in the degenerate subspace of calibration solutions, particularly in phase.
These are solved by absolute calibration, which fills-in the degenerate components, thus mitigating the phase discontinuities in time introduced by redundant calibration.
Next the fast and localized spectral and time-dependent features seen in the redundant and absolute gains are  filtered out by the smoothing step.
Gain filtering occurs after RFI identification, meaning we can input our flag masks into the Fourier filtering algorithm to deconvolve them out and fill in the missing pixels with a model of the smoothly varying components.\footnote{While we can infer gain solutions for times and frequencies that are flagged, we only do this for the purpose of finding a smooth gain model. Data that are flagged remain flagged.}
This is reflected in the sharp spikes in the gains in RFI-dominated pixels that are smoothed over after gain filtering (\autoref{fig:gain_progression}).
Gain smoothing is performed independently for each antenna and linear polarization.

We are left with a set of gains spanning all frequencies and times of our data on a nightly basis that has been filtered 
from the initial solution at each freuqency and time sample
to a restricted number of degrees of freedom for each antenna and linear polarization.
The total bandwidth of our data leads to a delay resolution of $\Delta\tau=10$ ns meaning we keep 21 spectral degrees of freedom, and the nightly observation duration is 10 hours, leading to 3 temporal degrees of freedom, for a total of $\sim63$ degrees of freedom in each antenna-polarization gain.\footnote{The gains are complex quantities and our filter extends from $-100<\tau<100$\,ns, giving us 21 independent Fourier modes: 10 on each side of the $\tau = 0$\,ns mode. Likewise, we keep the temporal modes correspoding to the $-1$st, 0th, and 1st Fourier modes.}

\subsection{LST Binning}
\label{sec:lstbin}

Having calibrated each night independently, we next coherently average them at fixed LST.
Due to sidereal drift night to night, each observing night has a slightly offset LST grid, which must be accounted before LST binning.
To do this, we establish a single LST grid spanning 0 to 24 hours at a cadence of 21.4 seconds, which is double the native correlator integration time and helps to increase the sample count in the subsequent outlier rejection routine.
For each of these LST bins, we take data from all nights that fall within the bounds of this bin and rephase their pointing centers to the center of the LST bin.
We then perform another round of outlier rejection to help flag missed RFI and other problems with the data that do not repeat night-to-night at the same LST (as the cosmological signal should).
We do this by looking at the distribution of visibilities to be binned together for every particular LST, frequency, and baseline combination and rejecting individual samples across the nights that have a modified $Z$-score (as defined in \autoref{eq:modz}) greater than $4\sigma$.
We set the threshold at $4\sigma$ such that the clipping has an effectively negligible impact on the Gaussian noise distribution, yet still rejects strong outliers that were for whatever reason not caught in our initially flagging round.
Clipping is performed on the real and imaginary component of the visibility separately, but if either are clipped in the process we flag the pixel entirely.
Indeed, in practice we find that the sigma clipping routine tends to catch clear broadband artifacts due to correlator failures that were not initially caught.
For the $\sim$18 nights of nightly data sampled at a cadence of 10.7 seconds, each LST bin contains on average 36 samples as input to this sigma clipping routine.
All data in the bin that were not originally flagged or flagged during the sigma clipping routine are then uniformly averaged onto a single LST grid.

\begin{figure}
\centering
\includegraphics[width=\linewidth]{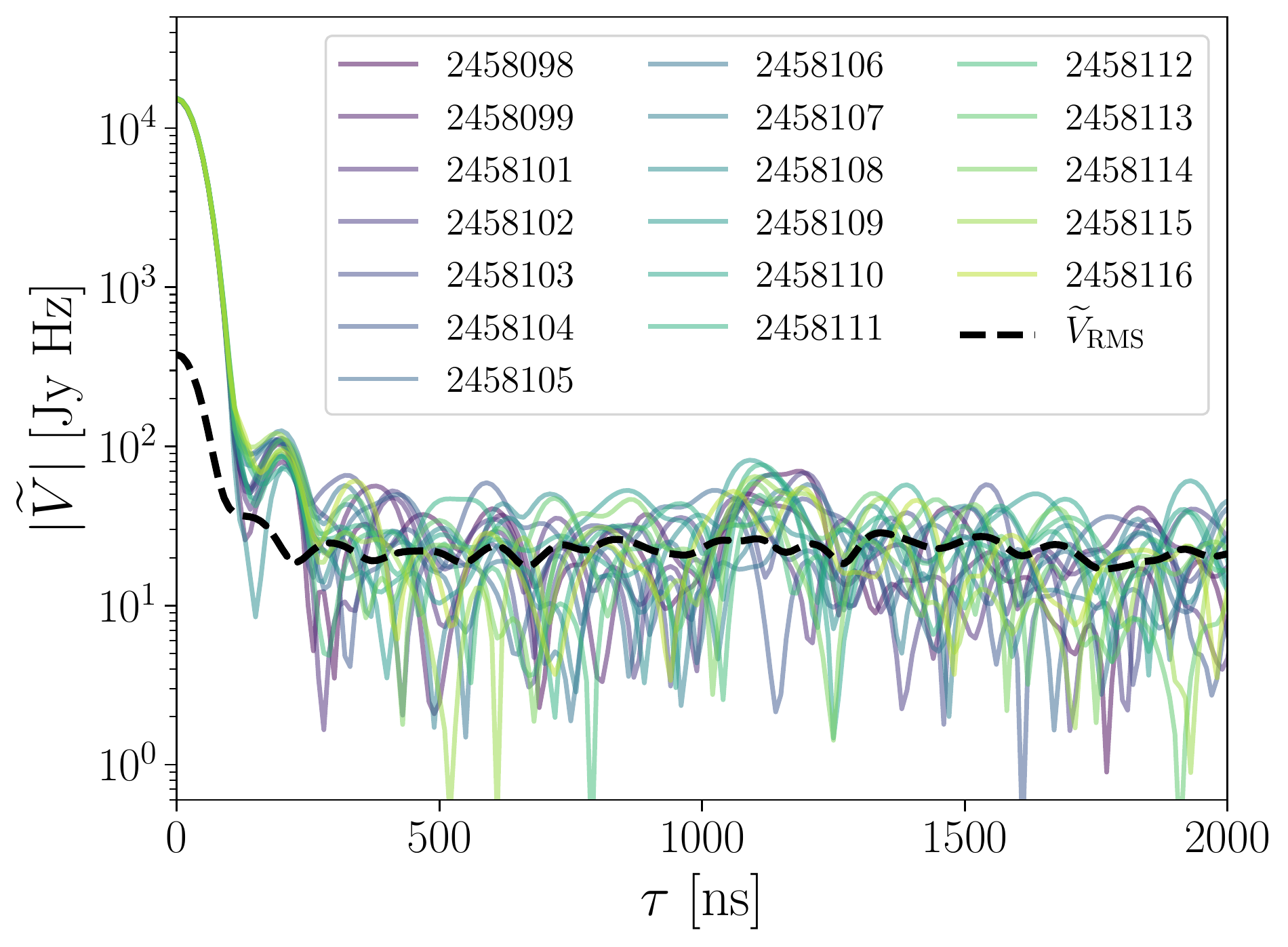}
\label{fig:lst_residuals}
\vspace{-5mm}
\caption{Delay transformed visibilities from each night after calibration but before LST binning, showing the consistency of the calibrated visibilities from night-to-night for a fixed baseline and LST. We also plot the RMS of the complex delay transformed visibilities across nights (black dashed), showing excess variability above the noise at low delays (at $\sim3\%$), indicative of night-to-night variation in calibration} errors at the $\sim$1.5\% level.
\end{figure}

A key requirement for LST binning is that each night is calibrated accurately, otherwise the averaged data will decohere.
We assess this by looking at the variability of the calibrated data across all nights in the data set.
\autoref{fig:lst_residuals} shows the delay-transformed visibilities across each night from a single baseline and at a fixed LST.
The dashed black line shows the root-mean-square (RMS) of the visibilties across nights.
At high delays we see the RMS is consistent with the apparent noise level of the nightly data; however, at low delays the RMS rises to a level that is $\sim$3\% of the peak foreground power. 
The most likely origin of this is the night-to-night variability of the gain calibration, which would put the gain errors on the $\sim$1.5\% level.
This is consistent with the fidelity of the Phase I calibration pipeline assessed in \citet{Aguirre2021}.

\begin{figure*}
\centering
\includegraphics[width=\linewidth]{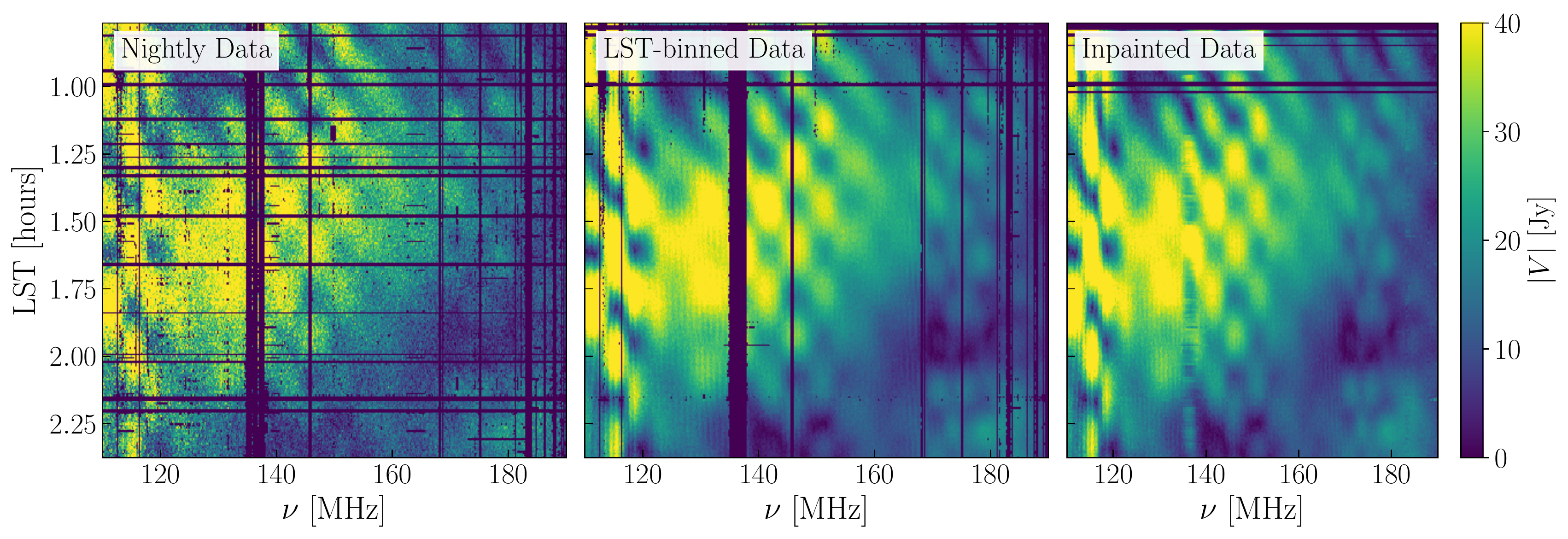}
\label{fig:lstbin_inpaint}
\vspace{-5mm}
\caption{A progression of a visiblity waterfall over a small time range through LST binning and flag inpainting, showing the reduction in flags (masked regions).
Note the marked reduction in flags after LST binning, due to different nights having independent flagging masks.
Data inpainting reconstructs a best-guess of the data in flagged regions, but demonstrates poor performance over wideband flags like the ORBCOMM feature at 138 MHz. Because it operates on each integration independently, integrations that were fully flagged remain flagged.} 
\end{figure*}

\subsection{Data Inpainting}
\label{sec:inpainting}

Flagging masks due to RFI and other instrumental problems are generally fairly complex, with a non-trivial frequency, time, and antenna dependence.
Flagging masks presents a challenge for power spectral analyses that make measurements in the Fourier domain as they induce strong sidelobes in non-periodic signals.
This is especially troublesome for 21\,cm measurements, which require high dynamic range signal separation against bright foregrounds in the Fourier domain.
A related effect is produced when time averaging masked visibility data with differing frequency flags as a function to time.
The resultant spectrum contains a complex weight structure as a function of frequency which can manifest as sidelobe structure of bright foregrounds \citep{Offringa2019}.

A common solution to the problem of masked or missing data is to inpaint the masked data with a best-guess of the signal that should have been measured in that bin.
A wide variety of algorithms have been developed to solve this kind of problem; the CLEAN algorithm \citep{Hogbom1974}, for example, is one way to do this, and is in fact how we addressed the problem of masked data in our gain smoothing procedure of \autoref{sec:gain_smoothing}.
In this work, we use a similar delay-based iterative deconvolution algorithm to fill-in the missing data in the LST-binned dataset \citep{Parsons2009, Ali2015, Kerrigan2018}.
The deconvolution is performed across nearly the full bandwidth, spanning 115 -- 185 MHz, and is performed on a per-integration basis down to the thermal noise of each visibility.
The algorithm finds model components in delay space across a window spanning $-2000 < \tau < 2000$, which is chosen to encapsulate all of the strong features seen in the data across all baselines.

\autoref{fig:lstbin_inpaint} shows an example of the LST-binned data products and the results of delay deconvolution-based inpainting.
Inpainting only affects the pixels where the data are masked, and because it is done only across frequency, integrations that are fully flagged remain fully flagged.
\citet{Aguirre2021} study the accuracy of the inpainting algorithm described here, showing that its performance is degraded for widechannel gaps, like the 138 MHz ORBCOMM features (this is furthermore seen in our spectral window null test \autoref{sec:spw_stability}).

\subsection{Systematic Modeling}
\label{sec:systematics}

\citet{Kern2020a} and \citet{Fagnoni2021} outline the major systematics seen in the HERA Phase I system, which are attributed to cable reflections in the 150-meter coaxial cable between the feed and the node and in the 20-meter cable between the node and the digitization stage, in addition to signal chain cross-coupling systematics (e.g. mutual coupling).
\citet{Fagnoni2021} use electromagnetic simulations of the HERA front end to show that mutual coupling (i.e. dish-to-dish communication) can appear as a decaying shoulder in the auto-correlation delay spectrum from $100-500$ ns.
They also predict the presence of a cable reflection at the termination of the coaxial cables, as well as a series of complicated sub-reflections in the 150-meter cable spanning $100-1200$ ns that can be unresolved by the native delay resolution of the data.
Inspection of the data by \citet{Kern2020a} affirm the presence of a shoulder in the auto-correlation delay spectrum, however, its exact origin is not quite clear.
\citet{Kern2020a} speculate that the shoulder could in fact be due to the cable sub reflections and not mutual coupling, in part because it can be effectively modeled and suppressed via a reflection calibration procedure, which is not to be expected from a highly direction-dependent systematic like mutual coupling.
Future work will seek to more clearly understand these systematics via improved electromagnetic beam modeling and visibility simulations, which explicitly include mutual coupling effects.

%Cable reflections can be modeled as a complex passband term $\epsilon$ which can be solved for each feed-polarization $i$ and antenna $q$
%\begin{align}
%\label{eq:cable_reflection}
%\epsilon_{i,q}  = A_{i,q}\exp\left[2\pi i\tau_{i,q}\nu + i\phi_{i,q}\right],
%\end{align}
%where $A$ is the amplitude, $\tau$ is the delay, and $\phi$ is the overall phase of the reflection.
%We solve for these reflection terms on both the 150-m and 20-m cable using the autocorrelations only.
%This has the advantage of not requiring a sky model \citep{Li2019, Kern2019}.
%These are then converted into into antenna-based gains as $g_{i,q} = (1+\hat{\epsilon}_{i,q})$, which are then applied to the LST-binned data in the usual manner.

To deal with the sub-reflections, we employ the same reflection fitting algorithm on all antenna auto-correlations from the LST binned data, iteratively solving for 25 individual reflection terms within a delay range of 150 -- 1500\,ns.
These parameters were chosen manually after visual inspection of the residual visibilities.
All reflection systematics are modeled at a 21.4 second cadence and then smoothed in time with a similar gain smoothing procedure employed in the first round of calibration.

\begin{figure*}
\centering
\includegraphics[width=\linewidth]{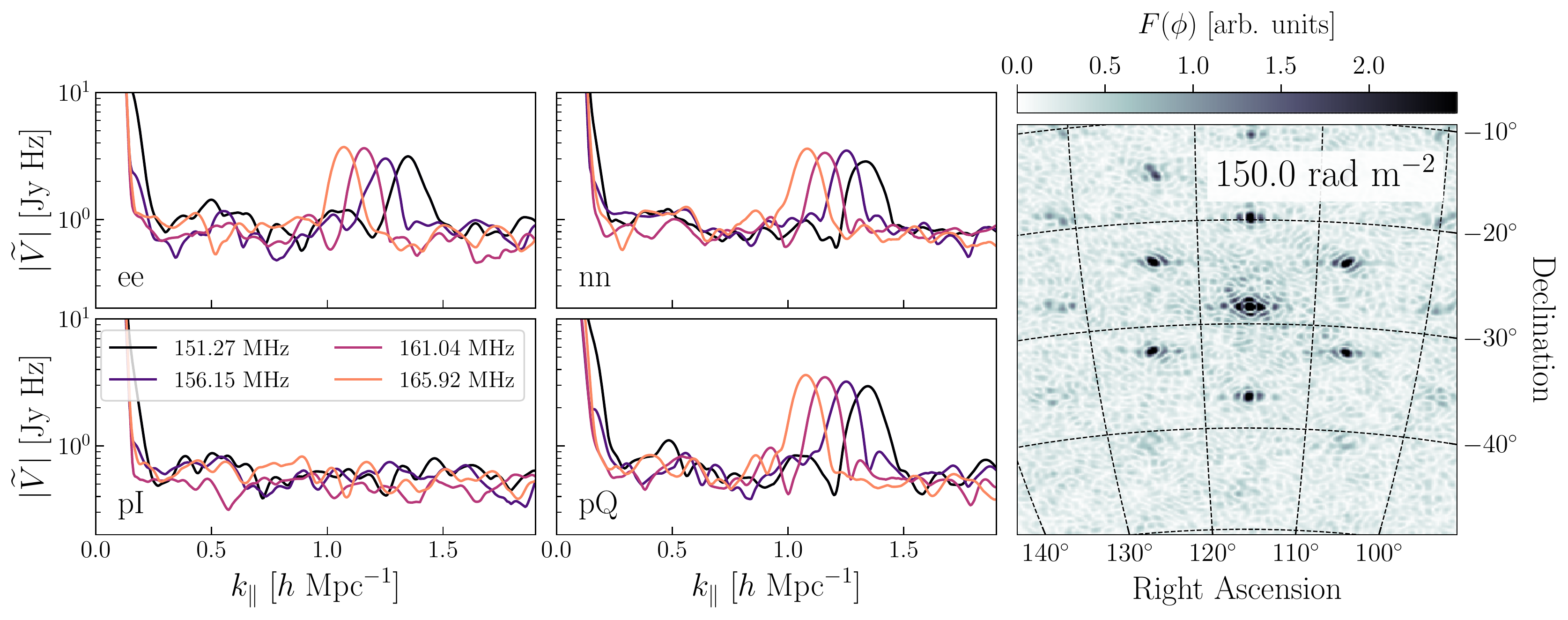}
\label{fig:pulsar}
\vspace{-5mm}
\caption{Faraday rotated foreground emission from a pulsar. The four-panel plot shows the delay transform of the LST binned data at $\alpha=115^\circ$ for two instrumental linear polarizations (top panels) and two pseudo-Stokes polarizations (bottom panels), displaying the high $k$ features that appear to drop out in the pseudo-Stokes I visibility. The different colors represent four different spectral windows for the delay transform with a fixed bandwidth but with increasing central frequency, showing the frequency dependence of the systematic that is suggestive of Faraday rotation. The right plot shows a dirty rotation measure synthesis image from 150 -- 170 MHz, clearly demonstrating a strong point source with a high rotation measure at image-center. The artifacts around the central source are the grating lobes of the point spread function.}
\end{figure*}

The most prominent source of systematics observed in Phase I data is the presence of high delay features in the visibilities that are nearly constant in time, thus occupying the fringe-rate 0 Hz mode, and roughly symmetric at negative and positive delays \citep{Kern2020a}.
For baselines with a projected East-West separation $\ge14$ meters, this systematic can be effectively filtered out from the data without significantly reducing the measured EoR power \citep{Kern2019}.
However, this does lead to a small amount of signal loss on the cosmological signal that we correct for in the power spectra (\autoref{sec:loss}).
Overall, the combination of reflection calibration and cross-coupling filtering leads to roughly two orders of magnitude of suppression in instrumental systematics spanning $0.1 < k < 0.8\ h\ {\rm Mpc}^{-1}$.
If not enacted properly, this step can lead to non-negligible amounts of cosmological signal loss, and is therefore vetted in our validation pipeline to ensure it performs as we expect it to \citep{Aguirre2021}.

\subsubsection{Faraday Rotated Foreground Emission}
\label{sec:pulsars}

After removing instrumental systematics from the data, we found low-level excesses in the power spectrum at high delays (beyond the foreground horizon) that transited on a beam-crossing timescale.
These excesses decreased their characteristic delay at increasing frequency, suggestive of Faraday rotation.
We investigated these systematics both in instrumental and pseudo-Stokes polarization visibilities, the latter of which are simply defined as a linear combination of the instrumental polarization visibilities as
\begin{align}
\label{eq:pstokes}
\left(\begin{array}{c}V_{pI} \\ V_{pQ} \\ V_{pU} \\V_{pV}\end{array}\right) = \frac{1}{2} \left(\begin{array}{cccc}1 & 0 & 0 & 1\\ 1 & 0 & 0 & -1\\ 0 & 1 & 1 & 0\\ 0 & i & -i & 0\\ \end{array}\right)\left(\begin{array}{c}V_{EE}\\ V_{EN} \\ V_{NE} \\ V_{NN}\end{array}\right),
\end{align}
where the East-East (or EE) instrumental polarization is an East-facing feed correlated with another East-facing feed, and likewise for the North-North (or NN) polarization.

We are able to identify these systematics as Faraday-rotated foreground emission through a few pieces of evidence.
First, the characteristic period of the frequency oscillations (i.e. the delay or $k_\parallel$ mode of the systematic) decreases at higher frequencies.
This is shown in \autoref{fig:pulsar}, where the four-panel plot shows the delay transform of the instrumental and pseudo-Stokes visibilities over four different spectral windows with increasing central frequency (but fixed bandwidth of 10 MHz), showing a notable decrease in the $k_\parallel$ mode of the systematic at higher frequencies.
In probing the relationship between the systematic's peak $k_\parallel$ and the central frequency of the spectral window, we find a tight relationship that follows the theoretical expectation of Faraday-rotated foregrounds found in Equation 3 of \citet{Moore2017} \citep[see also][]{Jelic2010, Asad2015}.
Finally, we can perform a direct rotation measure synthesis of the data to look for known sources of foreground emission with high rotation measures in the field-of-view.
Note that we make the assumption here that the inferred rotation measure is equal to the Faraday depth, which amounts to assuming that the rotation measure is locally compact along the line of sight.
We do this by forming a spectral cube of dirty Stokes Q and U images and then take their rotation measure synthesis for each pixel on the sky.
This yields the Faraday depth distribution, written as
\begin{align}
F(\phi) = \frac{1}{\pi}\int \left[Q(\lambda^2) + iU(\lambda^2)\right]e^{-2i\phi\lambda^2}d\lambda^2,
\end{align}
where Q and U are the Stokes Q \& U maps, $\lambda$ is the observing wavelength, and $\phi$ is the Faraday depth in radians meter$^{-2}$ \citep{Brentjens2005, Kim2016}.
Note we do not actually form the true Stokes Q and U maps, but simply image the pseudo-Stokes Q and U visibilities, which is a good approximation of the Stokes parameters near the center of the FoV.
Doing this and scanning through $\phi$ reveals a bright point source at the center of the field (right panel of \autoref{fig:pulsar}), which aligns exactly in location and in the inferred rotation measure with a known pulsar \citep[PSR J0742-2822;][]{Lenc2017}.
There are few other cases in the data where we see a high delay excess and can connect it to a known pulsar in the Faraday depth maps; however, they are seen at fairly low SNR.

While these systematics appear above the noise in the EE, NN, pseudo-Stokes Q and U visibilities, they do not appear appreciably in the pseudo-Stokes I visibility (\autoref{fig:pulsar}).
While this is expected, it is also true that intrinsic Q$\rightarrow$I leakage from primary beam asymmetry between feeds can cause these systematics to appear at suppressed levels in the Stokes I power spectrum \citep{Moore2013, Asad2016, Asad2018}.
Feed and LST jacknife tests could help to determine if future, deeper data sets are beginning to detect these systematics in Stokes I.

\subsection{Coherent Time Averaging}
\label{sec:integration}

While HERA observations are taken in drift-scan mode, integrations taken at nearby LSTs can be coherently averaged if they are phased to a common pointing center \citep{Parsons2016}. 
However, summing integrations that are separated by too much in time will begin to decohere the averaged signal due to drift in the primary beam weighting, in addition to the inability to properly re-align the fringes across the entire sky with a single fringe-stopping procedure.
We evaluated the amount of loss induced by time averaging by generating an ensemble set of mock EoR visibilities and averaging them with increasingly large time windows.
We settled on a maximum time window of 428 seconds, which leads to an average of $\sim$1\% decoherence of EoR power that is scale independent.
In \citet{Aguirre2021}, this test is repeated with a different kind of EoR model and report a similar finding of $\sim$1\% decoherence in power.
Recall that our power spectrum estimator cross multiplies adjacent time bins to avoid a noise bias.
In order to ensure that data separated by more than 428 seconds are not cross-correlated, we actually halve the coherent averaging time window to 214 seconds.

\subsection{Decoherence due to Non-Redundancies}
\label{sec:decoherence}

HERA's array layout has a large number of instantaneously redundant baselines.
Averaging the visibilities directly rather than their power spectra provides an extra boost in sensitivity by a factor of $\sqrt{N_{\rm baselines}}$, and is a key factor in HERA's overall sensitivity.
The instrument is not perfectly redundant however -- variations in baseline length and orientation, antenna primary beams, and even the calibration itself are expected at some level \citep{Orosz2019}. Deviations from perfect redundancy will cause the signal to decohere to some extent under redundant averaging, and if these deviations are large enough, some degree of signal loss will be sustained.

Here we develop an empirical metric to quantify the amount of signal loss incurred due to decoherence when averaging slightly non-redundant visibilities.
The impact of non-redundancy-based decoherence can be estimated by comparing the foreground power spectrum amplitude derived using coherent redundant averaging compared to incoherent redundant averaging.
The incoherently averaged power spectrum, $P_{\rm inc}$, is constructed by first estimating power spectra for each baseline-pair in the redundant group and then averaging them together such that
\begin{align}
P_{\rm inc} = \frac{1}{N}\left(|\widetilde{V}_1|^2 + |\widetilde{V}_2|^2 + |\widetilde{V}_3|^2 + \ldots\right),
\end{align}
where $V_1, V_2, \ldots$ indexes individual baselines in a group, $\widetilde{V}$ is the Fourier transformed visibility, and $N$ is the number of baselines in a group.
Since phase information is discarded by the initial power spectrum estimation step, phase errors cannot combine destructively when the individual spectra are subsequently averaged.
Thus, $P_{\rm inc}$ should be less susceptible to phase non-redundancies (e.g. caused by miscalibration or primary beam variations).

A cross coherent power spectrum is constructed by cross multiplying all redundant baseline pairs and then taking their average, which does allow phase errors to combine destructively.
Note that this is very similar to averaging the redundant visibilities and then cross multiplying them, except for the explicit exclusion of baseline terms paired with themselves.
In other words, we form the cross coherent power spectrum as
\begin{align}
P_{\rm coh} = \frac{1}{M}\left(\widetilde{V}_1\widetilde{V}_2^\ast + \widetilde{V}_1\widetilde{V}_3^\ast + \widetilde{V}_2\widetilde{V}_3^\ast + \ldots\right),
\end{align}
where $M$ is the number of baseline-pairs in a group.
In the ideal limit of perfect redundancy there should be no decoherence effect as each baseline is an exact replication of the signal.
Thus the coherent and incoherent spectra should measure the same value, with the exception of the different integrated thermal noise level which, as we explain later, is a negligible difference for the delay modes of interest.
Note that this is not the estimator used to estimate the 21\,cm power spectrum is later sections, this is only used as an approximate estimator for the purposes of assessign signal loss.

To measure the amount of decoherence induced by the coherent average we compute the fractional loss in power,
\begin{align}
\Delta\chi(t, \tau) = \frac{P_{\rm coh}(t, \tau) - P_{\rm inc}(t, \tau)}{\langle P_{\rm inc}(t, \tau)\rangle},
\end{align}
where $t$ denotes LST, $\tau$ is delay, and $\langle\ldots\rangle$ denotes a time (LST) average.
The time average is necessary to provide a stable baseline that limits spikes in the ratio when $P_{\rm inc}$ goes through a null.
Note that the delay of the visibility, $\tau$, is simply the direct Fourier dual to frequency in the Fourier transform, with units of 1/seconds.

What we are interested in is determing how the EoR signal is suppressed in the coherently averaged power spectrum due to non-redundant decoherence.
Being an isotropic cosmological signal, the EoR field is statistically invariant across the sky, meaning our sensitivity to it comes predominately from the main lobe of the primary beam.
In order to use the measured foreground emission, which is not statistically isotropic, as a proxy for the amount of EoR signal loss we will sharpen the metric in two ways.
First, we only inspect the $\tau=0$ ns mode, which isolates smooth spectrum foreground emission to a strip intersecting the main field of view of the primary beam.
And second, we choose to evalute the metric at times where diffuse foregrounds fill the main lobe of the primary beam, thus upweighting the parts of the beam where we expect the EoR signal to be most sensitively measured.
For HERA, this occurs most prominently at the transit of the galactic anti-center.
The behavior of this metric has been explored in an accompanying paper \citep{Choudhuri2020}.
In this study, numerical simulations of a small HERA-like array were performed with different types of primary beam non-redundancies and calibration non-redundancies.
The $\Delta\chi$ foreground metric was found to accurately reflect the degree of non-redundancy, signal loss and modulation effects in the recovered EoR power spectrum.
Its behavior in the presence of baseline non-redundancies was not studied.

\begin{figure}
\centering
\includegraphics[width=\columnwidth]{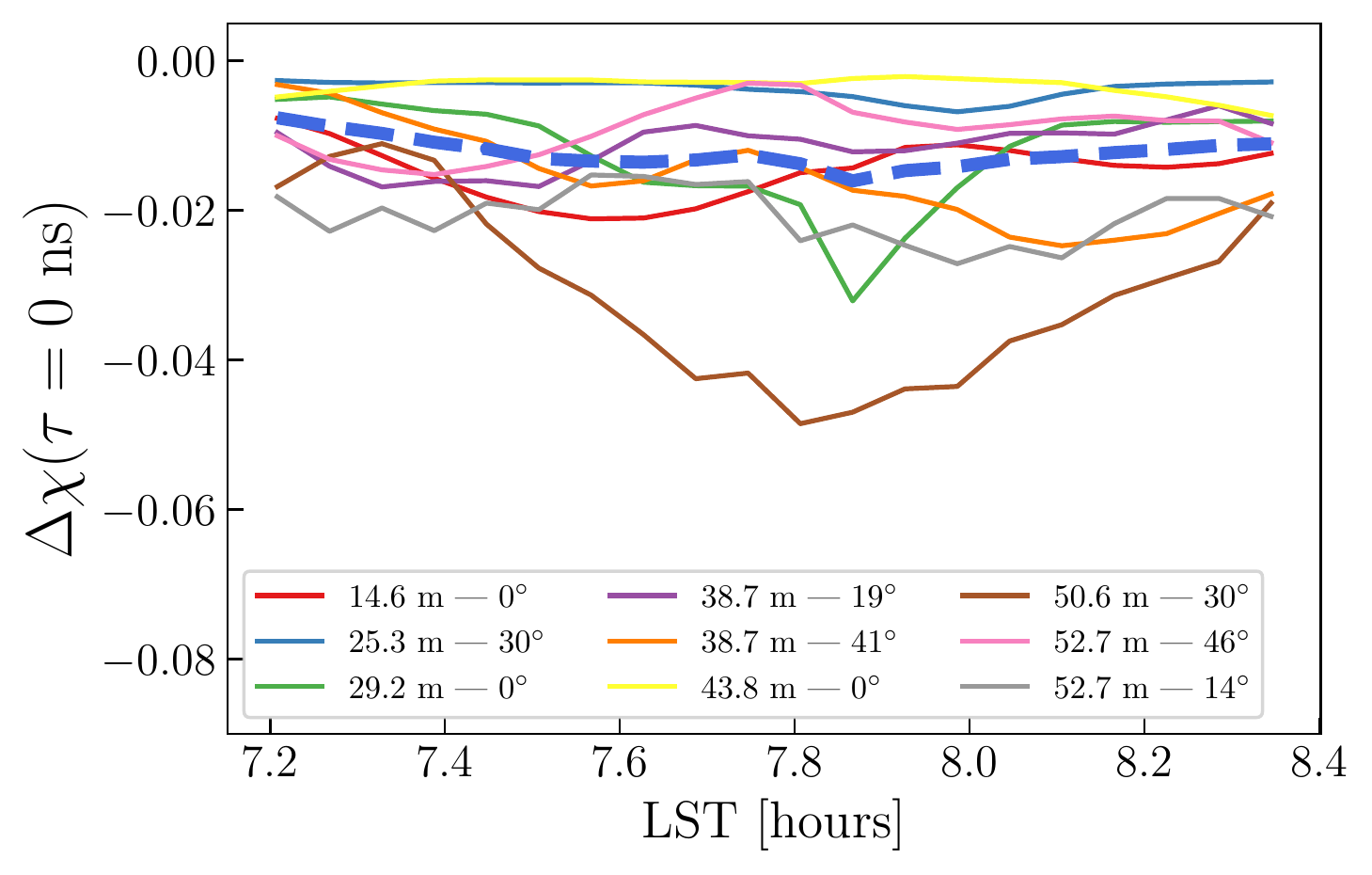}
\vspace{-5mm}
\caption{Redundancy decoherence test for 9 redundant groups, marked by the baseline length and angle in local array XYZ coordinates (blue dashed is their average).
This plots the difference of the coherently and incoherently averaged power spectrum normalized by the time-average of the latter.
We show the metric for the $\tau=0$ Fourier mode at LSTs when bright, diffuse foregrounds fill the primary beam.
On average, we see roughly 1\% power loss, suggesting that while our final set of redundant visibilities are not perfect, they are redundant enough to retain the vast majority of sky power in the main lobe of the primary beam when forming baseline-to-baseline cross power spectra.
}
\label{fig:red_avg}
\end{figure}

\autoref{fig:red_avg} shows the $\Delta\chi$ metric at $\tau = 0$ ns for several redundant baseline groups.
We see an average of $\sim$1\% power loss, suggesting that the real non-redundancies between baselines within a redundant group do not significantly decohere the sky signal.
The worst case is the $b=50.6$ m baseline, which sees an average decoherence of about 3\%.
Inspecting the visibilities of this baseline we see that this is due in part because this baseline experiences a particularly strong null in power compared to the other baselines, which systematically brings down $\langle P_{\rm inc}\rangle$ in the denominator of the metric, causing the overall ratio to slightly inflate.

\subsection{Other forms of Signal Loss in the Pipeline}
\label{sec:loss}

Here we tabulate all of the identified forms of systematic signal loss that arise from the analysis pipeline.
Recall that the philosphy of the analysis employed here is one that is generally meant to minimize the inherent loss of the analysis pipeline, therefore, corrections to the loss that have been identified are generally a small fraction of the total measured power.

Some forms of loss are entirely expected, like from the time averaging of drift-scan visibilities, and other forms were not expected but found in the process of pipeline validation \citep{Aguirre2021}, the most notable being the absolute calibration bias.
This bias was due to the logarithmic linearization of the gain amplitude parameter in the process of absolute gain calibration, which when employed on visibilities with low signal-to-noise can result in a bias in the recovered amplitude \citep{Boonstra2003}.
Because we calibrate HERA's drift-scan observations every 10.7 seconds, there are some fields where this bias can be non-negligible, and reaches into the percent level on the gains.
After gain smoothing, this bias is shown to be largely time independent but does have a slight frequency dependence \citep{Aguirre2021}.
We estimate this bias as a single number for each of the two power spectrum bands (\autoref{sec:pspec_params}) by making multi-frequency synthesis images of the calibrated and LST binned visibilities and comparing them to images of the initial absolute flux density model.
The ratio of a zenith-located point source in the data and model is used as the correction factor.

All forms of loss measured here are EoR model independnet and scale independent (i.e. flat across $k$).
In some cases we have physically motivated reasons to expect this (e.g. an overall bias in absolute calibration is scale and EoR model independent) and in other cases we have explicitely tested this (as is the case for coherent time averaging and the non-redundancy decoherence discussed above).
These are tabulated in \autoref{tab:signal_loss} and are each corrected for at the end stage by correcting the power spectra.

\begin{deluxetable}{lc} 
\tabletypesize{\footnotesize} 
\tablewidth{0pt} 
\tablecaption{
Systematic Loss in Analysis Pipeline
\label{tab:signal_loss}
}
\tablehead{Analysis Step & Fractional Power Loss}
\startdata
Absolute Calibration & 9\% (8\%) \\ [.1cm]
Cross-Coupling Filtering & 3\% (1\%) \\[.1cm]
LST Time Averaging & 1\% (1\%) \\[.1cm]
Redundant Averaging & 1\% (1\%) \\[.1cm]
\enddata
%\vspace{-.3cm} 
\tablecomments{Percentage loss in power for Band 1 (Band 2), which are corrected for after forming the power spectrum. These figures are partially derived from the validation analysis \citep{Aguirre2021}, and from this analysis (\autoref{fig:red_avg}).}
\end{deluxetable}

%%%%%%%%%%%%%%%%%%%%%%%%%%%%%%
%%%% Power Spectrum Limits %%%
%%%%%%%%%%%%%%%%%%%%%%%%%%%%%%

\section{HERA Phase I Power Spectrum Limits}
\label{sec:pspec}

In this section we outline our power spectrum estimator, giving a brief overview of quadratic estimators (QE), our power spectrum pipeline, and present our derived power spectrum limits from the analysis discussed in this work.
Our power spectrum estimator codebase, \texttt{hera\_pspec},\footnote{\url{https://github.com/HERA-Team/hera_pspec}} is publicly available software.
Throughout this work, we adopt a $\Lambda$CDM cosmology \citep{Planck2016} with $\Omega_\Lambda=0.6844$, $\Omega_b=0.04911$, $\Omega_c=0.26442$, and $H_0=67.27\,\textrm{km}/\textrm{s}/\textrm{Mpc}$.

\subsection{The Quadratic Estimator Formalism}
\label{sec:qe}
The 21\,cm power spectrum is the square of the three-dimensional spatial Fourier transform of the 21\,cm brightness temperature field. 
Given that in the flat-sky limit the interferometric visibility is the 2D transverse spatial Fourier transform of the temperature field, we can estimate of 3D 21\,cm power spectrum by taking the Fourier transform across frequency,
\begin{align}
\label{eq:dtransform}
\widetilde{V}(\tau) = \int V(\nu) e^{2\pi i \nu\tau}d\nu,
\end{align}
where the delay domain ($\tau$) is the Fourier dual of frequency.
Note that delay modes are not a direct mapping to the line-of-sight $\kpara$ wavevector; however, for short baselines they are nearly the same and approximating $\tau$ modes as $k_\parallel$ modes is known as the delay approximation \citep{Parsons2012a, Liu2014a}.
This is written as
\begin{align}
\label{eq:dspec}
\hat{P}(\kperp, \kpara) = \frac{X^2Y}{\Omega_{pp}B}\widetilde{V}_1(u, \tau)\widetilde{V}_2^\ast(u, \tau),
\end{align}
where $\hat{P}$ is our estimate of the power spectrum, $X$ and $Y$ are scalars mapping sky angles and frequency to cosmological distances, respectively, $\Omega_{pp}$ is the sky integral of the squared primary beam response and $B$ is the Fourier transform bandwidth.
Furthermore, the subscript on our visibilities, $\widetilde{V}_1$ and $\widetilde{V}_2$, indicates that we are cross-multiplying visibilities with independent noise realizations, meaning that the estimated power spectrum has no noise bias.
As written, the power spectrum $P$ has units ${\rm mK^2}\ h^{-3}\ {\rm Mpc^3}$.
The cosmological wavevectors are related to the natural telescope units of delay and baseline length \citep{Parsons2014} via
\begin{align}
\label{eq:kvecs}
\kpara &= \frac{2\pi\tau}{X} \\
\kperp &= \frac{2\pi}{Y}\frac{b}{\lambda},
\end{align}
where $X = c(1+z)^2\nu_{21}^{-1}H(z)^{-1}$, $Y=D(z)$, $\nu_{21}=1.420$ GHz, $H(z)$ is the Hubble parameter, and $D(z)$ is the transverse comoving distance \citep{Hogg1999, Condon2018}.
We can further define the cosmologically dimensionless power spectrum,
\begin{align}
\label{eq:delsq}
\Delta^2(k) = P(k)\frac{k^3}{2\pi^2},
\end{align}
which has units of ${\rm mK}^2$.

The ``Fourier transform and square'' delay spectrum estimator can also be cast into the more general form of a quadratic estimator,
\begin{align}
\label{eq:qe}
\hat{q}_\alpha = \frac{1}{2}\x_1^\dagger\R^\dagger\Q_\alpha\R\x_2,
\end{align}
where $\hat{q}_\alpha$ is the unnormalized estimate of the $\alpha$ bandpower, $\x_1$ and $\x_2$ are the visibility vectors we will cross-correlate, $\R$ is any weighting matrix we may want to apply to the data (e.g. the inverse covariance in the optimal quadratic estimator case), and $\Q_\alpha$ is the mapping from data space to power spectrum space. In general, this mapping is given by the derivative of the covariance with respect to bandpower $\alpha$, i.e. $\Q_\alpha = \C_{,\alpha}=\tfrac{\partial\C}{\partial p_\alpha}$.
We use $\C$ to mean the true covariance matrix of the visibilities, which in general is never known exactly.
Estimating the covariance empirically is a subtle process, and if not done accurately risks biasing the power spectrum estimate \citep{Ali2015erratum, Cheng2018}. The lack of access to the true covariance also makes accurate error estimation tricky, as we will discuss in more detail later in this section. For our purposes, we simply define $\Q_\alpha=\mathbf{c}_\alpha^\dagger \mathbf{c}_\alpha$, where $\mathbf{c}_\alpha$ is a discrete Fourier transform operator that takes the (weighted) data to delay space.
We denote our data vectors as $\x_1$ and $\x_2$ because in this work we opt to form cross power spectra between data vectors with independent noise realizations, which eliminates the noise bias term that can be difficult to estimate precisely \citep{Dillon2014, Ali2015, Pober2016}.

To normalize our power spectrum estimate we gather our unnormalized bandpowers into a vector and form the quantity,
\begin{align}
\label{eq:norm_p}
\hat{\p} = \M\hat{\q},
\end{align}
where $\M$ is the normalization matrix and $\hat{\p}$ is the normalized bandpower estimates.
Our estimated power spectra are related to the true bandpowers by the window function matrix,
\begin{align}
\label{eq:window_func}
\hat{\p} = \W\p,
\end{align}
where $\p$ is a vector holding the true bandpowers.
While $\p$ is never known a priori, we can use the window function matrix to appropriately map a theoretical EoR power spectrum to the basis of the measured power spectrum.
Relating this to our normalization matrix we have
\begin{align}
\W = \M\H,
\end{align}
where $\H$ is the response matrix of the bandpowers, defined as
\begin{align}
H_{\alpha\beta} = \frac{1}{2}{\rm tr}(\R^\dagger\Q_\alpha\R\Q_\beta).
\end{align}
In the case where $\R=\Cinv$, our estimator becomes the optimal estimator, and $\H$ becomes $\F$, the bandpower Fisher matrix \citep{Tegmark1997}.

The last component of our power spectrum formalism is the bandpower covariance matrix,
\begin{align}
\mathbf{\Sigma} = \Cov[\hat{\p}] = \langle\hat{\p}\hat{\p}^\dagger\rangle - \langle\hat{\p}\rangle\langle\hat{\p}\rangle^\dagger.
\end{align}
Defining $\E^\alpha = \tfrac{1}{2}\sum_{\beta}M_{\alpha\beta}\R^\dagger \Q_\beta \R$ such that $\hat{p}_\alpha = \x_1^\dagger\E^\alpha\x_2$ and substituting this in, we get that the power spectrum covariance is
\begin{align}
\label{eq:pspec_cov}
\Sigma_{\alpha\beta} = {\rm tr}[\C\E^\alpha\C\E^\beta] + {\rm tr}[\bS\E^\alpha\bS\E^\beta],
\end{align}
where the total covariance, $\C = \langle\x\x^\dagger\rangle = \bS + \N$, can be written as a sum of the sky signal and noise covariance \citep{Dillon2014}.
While the noise covariance of our data is straightforward to quantify, the signal terms are considerably harder.
We will come back to how we estimate the bandpower errors in practice.

An unbiased estimate of the power spectrum is made by building a window function matrix whose rows sum to unity, giving the analyst some freedom on exactly how to construct $\M$, with statistical implications for $\hat{p}$ and $\mathbf{\Sigma}$ \citep{Tegmark1997, Liu2011, Ali2015}.
Choosing $\M=\H^{-1}$, for example, yields $\W=\I$, meaning the bandpowers are independent; however, we also see that the bandpower errors become quite large and correlated.
In this work we choose arguably the simplest approach, which is to set $\M$ to a diagonal matrix. This minimizes the resultant error bars at the expense of measurements that overlap in Fourier space and also have slightly correlated errors.
This returns us to the simple delay spectrum estimator, but framed in the machinery of a quadratic estimator, with the diagonal of $\M$ given by the coefficients of \autoref{eq:dspec}.

Our quadratic estimator is sub-optimal in the sense that we do not weight our data by the inverse covariance matrix; however, this also safeguards our estimator against concerns about signal loss from empirical inverse covariance weighting \citep{Cheng2018}, and the distortion of the window functions at low $k$ modes due to complicated data weighting \citep{Liu2014b, Kern2021}.
In the meantime, we defer optimal inverse covariance weighted power spectrum estimation to future work.
Nonetheless, we still need to account for the fact that, without foreground removal, there exists a large dynamic range between the foreground signal at low $k$ modes and the EoR and noise signal at higher $k$ modes.
To limit foreground spectral leakage in the discrete Fourier transform, we apply a tapering (or apodization) function along the diagonal of $\R$.
We use a Blackman-Harris function \citep{Blackman1958}, which achieves 50 decibels of sidelobe suppression in Fourier space.

\subsection{Error Bar Methodology}
\label{sec:errorbars}

\citet{Tan2021} discuss different kinds of error bars one can use to quantify uncertainty on the measured bandpowers.
In summary, they find that all of the error bar methodologies explored are in general agreement with each other and with the true sampling distribution of the bandpowers in the ensemble limit; however, they suggest two specific error bars that: 1. encapsulate the full thermal noise contribution to the bandpower uncertainty, and 2. can be computed relatively straightforwardly without requiring a model of the signal covariance.
We will briefly summarize these here.

\begin{figure}
\centering
\includegraphics[width=\linewidth]{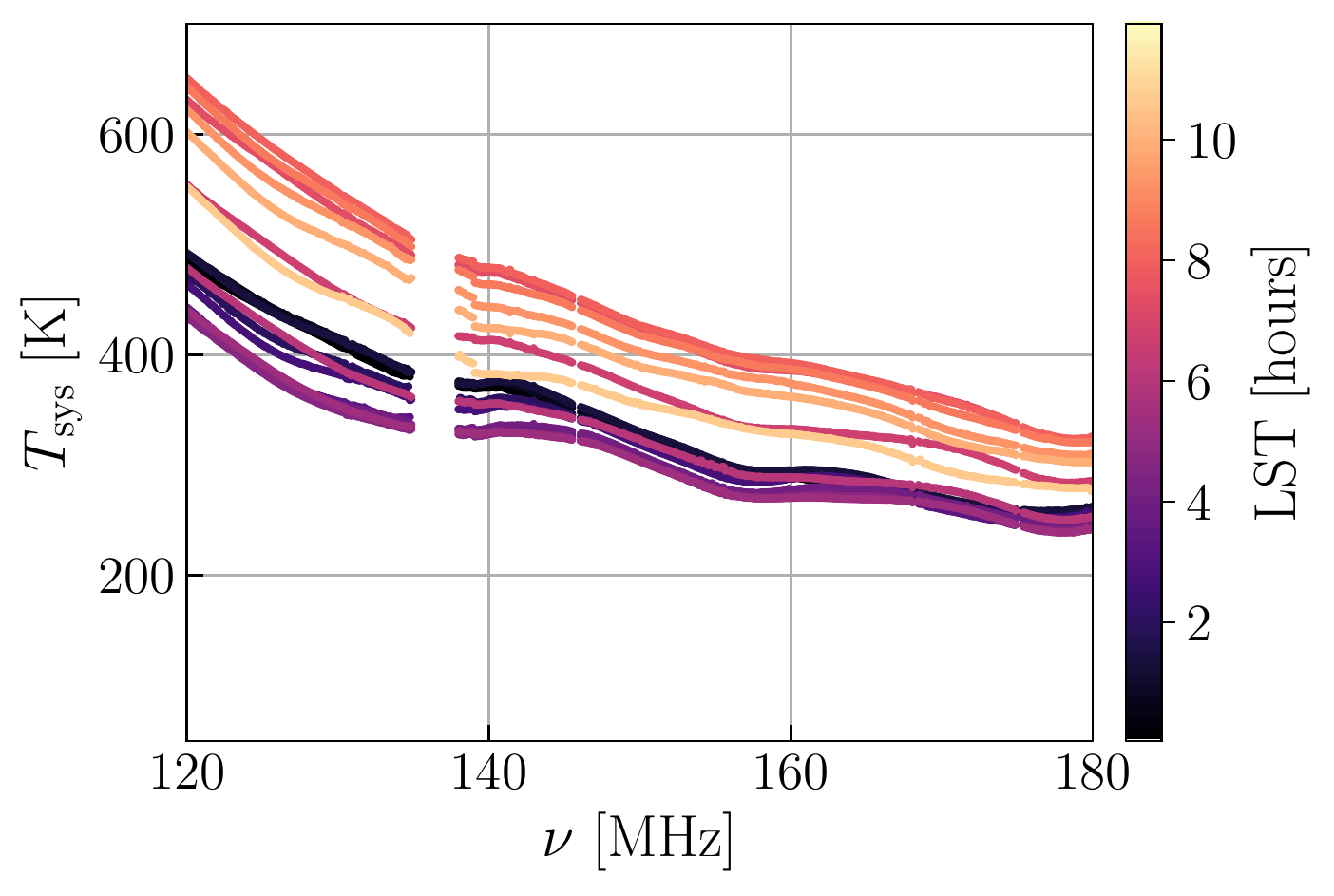}
\caption{A frequency and LST dependent model for the system temperature of a particular antenna that is used to compute power spectrum error bars, and also to generate realistic thermal noise realizations of the data.}
\label{fig:Tsys}
\end{figure}

At minimum, one needs to encapsulate the fundamental thermal noise uncertainty of the fully averaged power spectrum, which is the root-mean-square (RMS) of the bandpowers in the limit that they are noise dominated.
However, as \citet{Tan2021} point out, there is an additional source of noise variance on the bandpowers in the limit of a non-neglible coherent signal in the data.\footnote{Note that this is entirely separate from cosmic variance uncertainty, and also by signal we mean any coherent signal in the data, be it the EoR, foregrounds, or systematics.}
For weak, marginally detectable signals the additional variance is a small correction, however, when there is a significant detection of the signal this correction dominates the thermal noise uncertainty on the bandpowers, and thus should be accounted for.

An analytic expression for the root-mean-square (RMS) of noise-limited power spectra after a given amount of averaging can be written as
\begin{align}
\label{eq:PN}
P_{N} = \frac{X^2Y\Omega_{\rm eff}T^2_{\rm sys}}{t_{\rm int}N_{\rm coherent}\sqrt{2N_{\rm incoherent}}},
\end{align}
where $t_{\rm int}$ is the integration time of the data, $N_{\rm coherent}$ is the number of coherent averages of the data (i.e. visibility averages), $N_{\rm incoherent}$ is the number of incoherent averages (i.e. averaging after forming the power spectra), and $T_{\rm sys}$ is the system temperature of the signal chain, which is the sum of the sky and receiver temperatures. 
Lastly, $\Omega_{\rm eff}$ is the effective beam area, defined as $\Omega_{\rm eff} = \Omega_p^2 / \Omega_{pp}$, where $\Omega_{p}$ is the sky integral of the primary beam and $\Omega_{pp}$ is the sky integral of the squared beam \citep{Pober2013a, Parsons2014, Cheng2018}.
A frequency and LST dependent model for $T_{\rm sys}$ is derived from the auto-correlation visibilities of each antenna to produce a baseline, time, and frequency dependent noise model \citep[e.g.][]{Kern2020b, Dillon2020, Tan2021}.
\autoref{fig:Tsys} shows that HERA measures a system temperature in the range of 200 -- 400 Kelvin at 160 MHz depending on the LST.
Note that the $T_{\rm sys}$ models are used not only to compute the error bars on the power spectrum, but also to generate realistic noise simulations of the data that are used for null testing.

The second error bar that \citet{Tan2021} investigate accounts for the additional sources of noise in the power spectrum that arise in the presence of a signal, also derived in \citet{Kolopanis2019}.
These extra terms come simply from the signal-noise cross terms when squaring the visibilities to form the power spectrum.
This error bar is written as
\begin{align}
\label{eq:PSN}
P_{\rm SN} = \sqrt{\sqrt{2}P_SP_N + P_N^2},
\end{align}
where $P_S$ is the power spectrum of the signal in the data.
Note that, like $P_N$, $P_{SN}$ can be thought of as the 1$\sigma$ uncertainty on the measured bandpower.
What is readily apparent is that to compute this quantity we need the power spectrum of the signal, which we do not have any more than we have the covariance of the signal.
In that case, we can estimate it directly from the data, using $\hat{P}$ in its place, which \citet{Tan2021} show is a good approximation.
However, for noise-limited bandpowers where $P_S<<P_N$ we see that by substituting $P_S$ for $\hat{P}$ we are actually overestimating the error bar (in a sense we are double counting the noise-noise contribution).
Nevertheless, \citet{Tan2021} show that this can be corrected by constructing a modified error bar estimator
\begin{align}
\label{eq:mod_PSN}
\tilde{P}_{\rm SN} = \sqrt{\sqrt{2}P_SP_N + P_N^2} - (\sqrt{1 / \sqrt{\pi} + 1} - 1) P_N.
\end{align}

One downside to \autoref{eq:PN}, \autoref{eq:PSN}, and \autoref{eq:mod_PSN} is that they make certain simplifying assumptions.
Mainly, they assume that the bandpowers are uncorrelated.
In other words, they only account for the diagonal of the bandpower covariance.
However, we expect the off-diagonal components to possibly have non-negligible covariance (specifically due to the apodization function) and it should be accounted for.
\citet{Tan2021} show how the error bar estimators presented above can also be derived by propagating a combination of the data and the frequency-frequency noise covariance through the quadratic estimator, which conveniently yields the full bandpower covariance matrix, not just its diagonal.
Therefore, to quantify the off-diagonal components in the bandpower covariance matrix, we use the frequency-frequency noise covariance of the data propagated through the QE formalism to $\mathbf{\Sigma}$.

When quoting upper limits on the power spectrum or presenting power spectrum measurements, we will exclusively use the modified $\tilde{P}_{SN}$ error bar as the $1\sigma$ uncertainty, which is also the 68\% confidence interval given that we expect the bandpowers to be Gaussian distributed \citep[c.f.][]{Tan2021}.
We will also generally plot the $P_N$ limit as a reference for whether the data are consistent with the thermal noise floor.
Lastly, we also use $P_N$ and the bandpower noise covariance matrix when evaluating statistical null tests in \autoref{sec:validation}.

\subsection{Forming Power Spectra}
\label{sec:pspec_params}

\begin{figure}
\centering
\includegraphics[scale=0.6]{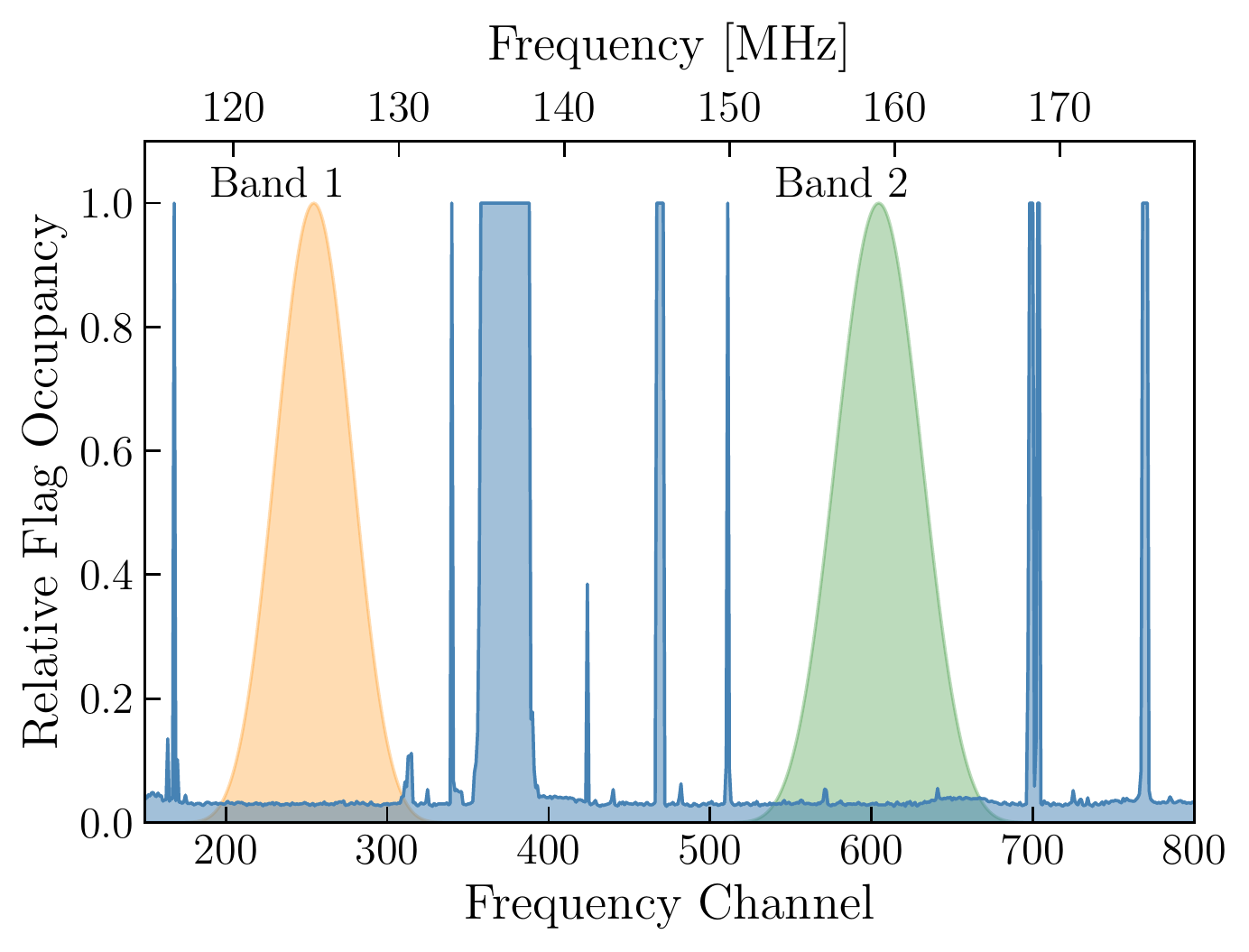}
\caption{The average flag occupancy of each frequency channel across LST (blue shaded), showing the two spectral windows used in the power spectrum analysis. The orange and green shaded regions show the tapering function applied across each spectral window for Band 1 and Band 2, which have central frequencies of 124.8 and 159.0 MHz, corresponding to redshifts of 10.4 and 7.9, respectively.}
\label{fig:spw_selection}
\end{figure}

We form power spectra from the pseudo-Stokes I visibilities (\autoref{eq:pstokes}) after LST binning, systematics treatment, and coherent time averaging.
We choose two spectral windows over which to form power spectra spanning 117.1--132.6\,MHz and 150.3--167.8\,MHz, corresponding to redshifts of 10.4 and 7.9, respectively.
We refer to these as Band 1 and Band 2, shown in \autoref{fig:spw_selection}, which were selected based on their relatively low flag occupancy of a few percent.
The shaded curves shows the extent of the Blackman-Harris tapering function applied along the spectral window of each band.
Evolution of the cosmological signal will begin to affect the power spectrum over large line-of-sight bandwidths, also known as lightcone effects.
Although this effect is technically EoR model dependent, it has been found that for $\Delta\nu<10$ MHz, lightcone effects on the power spectrum are kept below 10\% for $k<0.5\ h\ {\rm Mpc}^{-1}$ and $7 < z < 10$ \citep{Datta2014}.
Note that because we apply a Blackman-Harris tapering function across each of our spectral windows, their effective bandwidths are 7.75 MHz and 8.75 MHz for Band 1 and Band 2, respectively.

To form power spectra following \autoref{eq:qe}, we cross multiply adjacent time integrations \citep{Pober2013b} separated by 214 seconds after time averaging between all baselines in a redundant set.
A redundant set is defined by all physical baselines that share a common length and orientation.
Note that we compute the power spectrum of all baseline permutations (i.e. we compute $V_1\times V_2$ and $V_2\times V_1$, where $V_1$ and $V_2$ are visibilities in the same redundant set); however, we do not compute the power spectrum of a baseline paired with itself (i.e. $V_1\times V_1$), which reduces the impact of baseline-dependent systematics.
This achieves nearly the full sensitivity of a coherent visibility average across redundant baselines, thus the need to account for signal loss from a coherent redundant average (\autoref{sec:loss}).

Next we motivate the choice of the three unique ``fields'' in LST over which we form power spectra.
Tracking arrays like the GMRT, the MWA and LOFAR can pick out specific parts of the sky where foregrounds emission is minimal \citep{Ghosh2012, Trott2020, Mertens2020}.
However, HERA is not afforded this luxury because it is a static-array that observes a uniform stripe across the sky.
Nonetheless, we can pick out certain LSTs where foreground emission is relatively less bright.
This is important because we know that systematics like RFI and residual instrumental gains act to leak foregrounds modes in the power spectrum to higher $k_\parallel$, thus partially contaminated the EoR window.
Furthermore, as noted in \autoref{sec:reduction}, we do not perform any kind of foreground subtraction in this work.
Therefore, picking LSTs that avoid the brightest foregrounds can help to minimize the impact of residual systematics.
Specifically, we pick three fields, each spanning $\sim2$ hours in LST that avoid the extended radio galaxy Fornax A at an RA of 3.36 hours and the galactic anticenter at $\sim7.5$ hours (\autoref{fig:hera_stripe}).
They also give a 0.5 hour buffer at the ends of the total LST range to limit edge effects in our fringe-rate filtering.
The chosen fields span an LST range of 1.25--2.7 hours, 4.5--6.5 hours, and 8.5--10.75 hours for fields 1, 2, and 3, respectively.

\begin{figure}
\centering
\includegraphics[width=\linewidth]{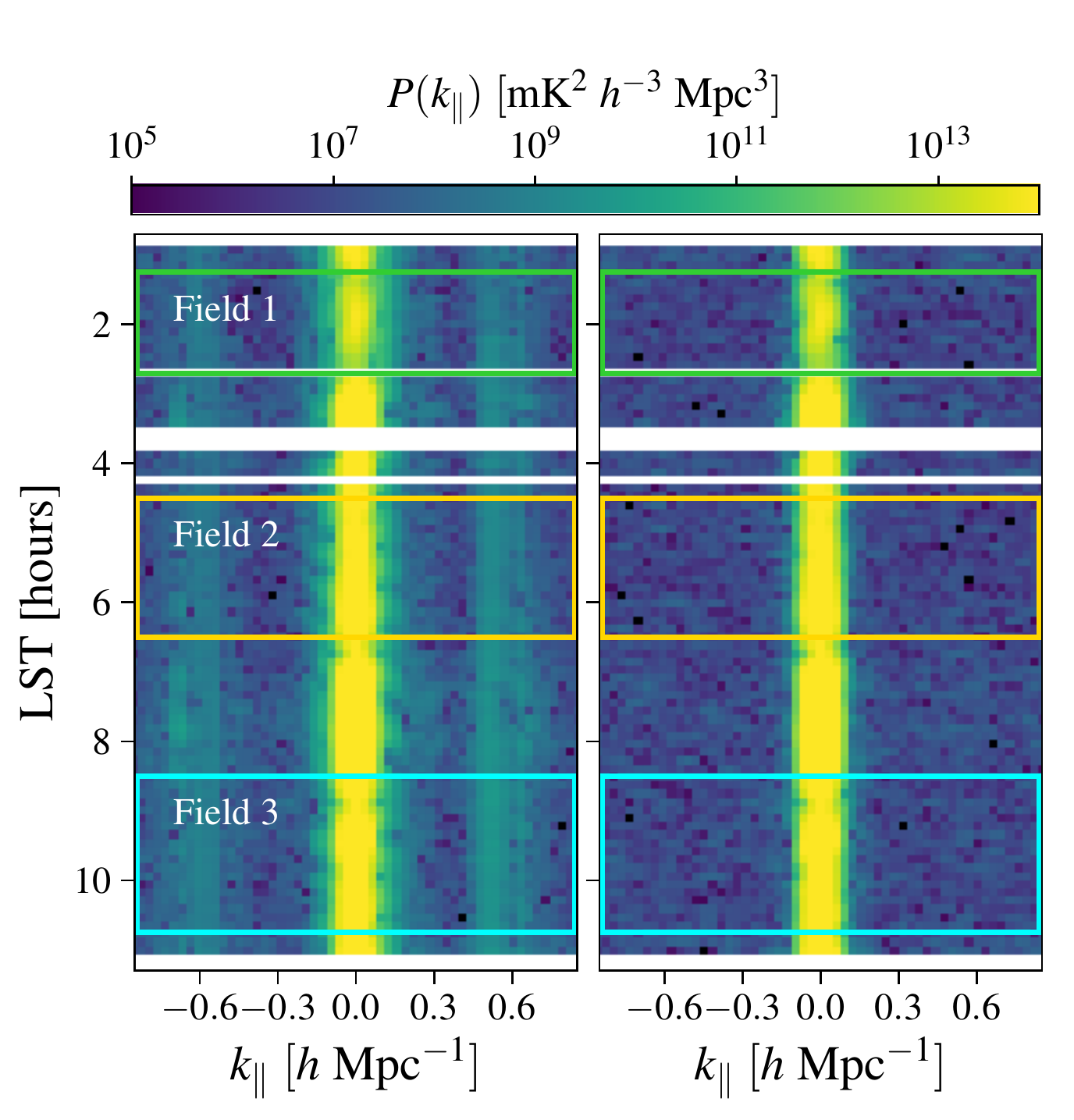}
\caption{A waterfall of a redundantly averaged power spectrum for a single baseline group without systematic treatment (left) and with systematic treatment (right). The LST cuts for each of the three fields are shown in blue, orange and blue, which are chosen to avoid the transiting of bright foreground sources like Fornax A at 3.36 hours, the galactic anticenter at $\sim$7.5 hours, and to give a buffer of $\sim$30 minutes from both the beginning and end of the full LST range.}
\label{fig:pspec_waterfall}
\end{figure}

\autoref{fig:pspec_waterfall} shows a redundantly-averaged power spectrum waterfall for a single redundant set with and without systematics treatment (\autoref{sec:systematics}) over the full LST range, with the three fields marked.
Based on the amplitude of the $k_\parallel=0\ h\ {\rm Mpc}^{-1}$ mode, we can see that Field 1 has the least amount of overall foreground power, making it an ideal candidate for power spectrum analysis.
\autoref{fig:pspec_waterfall} also demonstrates the $\sim2$ orders of magnitude of systematic suppression achieved by our pipeline, which would otherwise contaminate a large portion of the EoR window modes.

\begin{figure*}
\centering
\includegraphics[width=\linewidth]{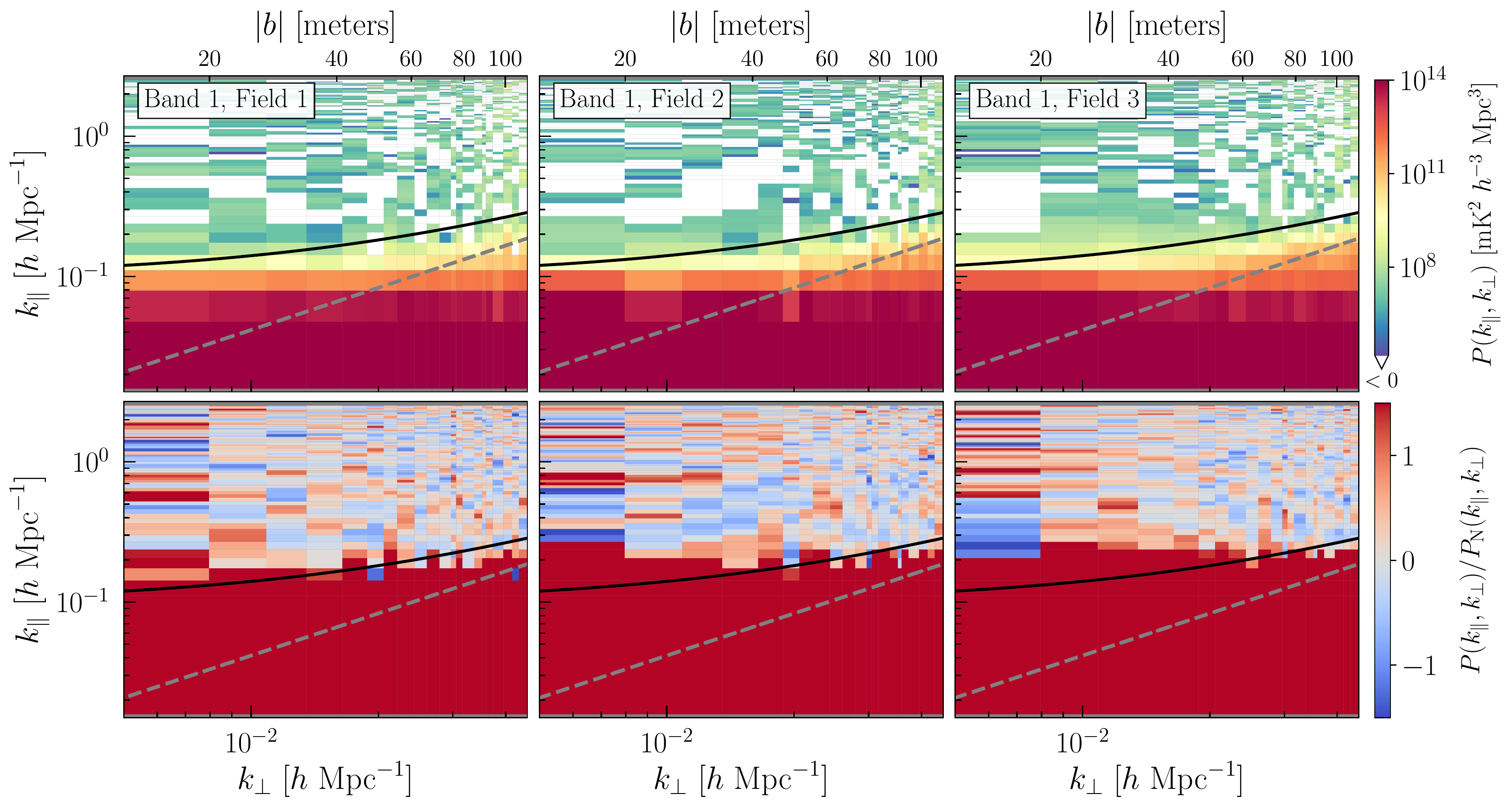}
\caption{The 2D power spectra $P(k_\parallel, k_\perp)$ for Band 1 (z=10.4) at each field (top panels), and their ratio with the $1\sigma$ thermal noise floor (bottom panels). We also plot the horizon wedge (dashed) and the horizon buffer used when spherically binning (solid).}
\label{fig:band1_wedge}
\end{figure*}

\begin{figure*}
\centering
\includegraphics[width=\linewidth]{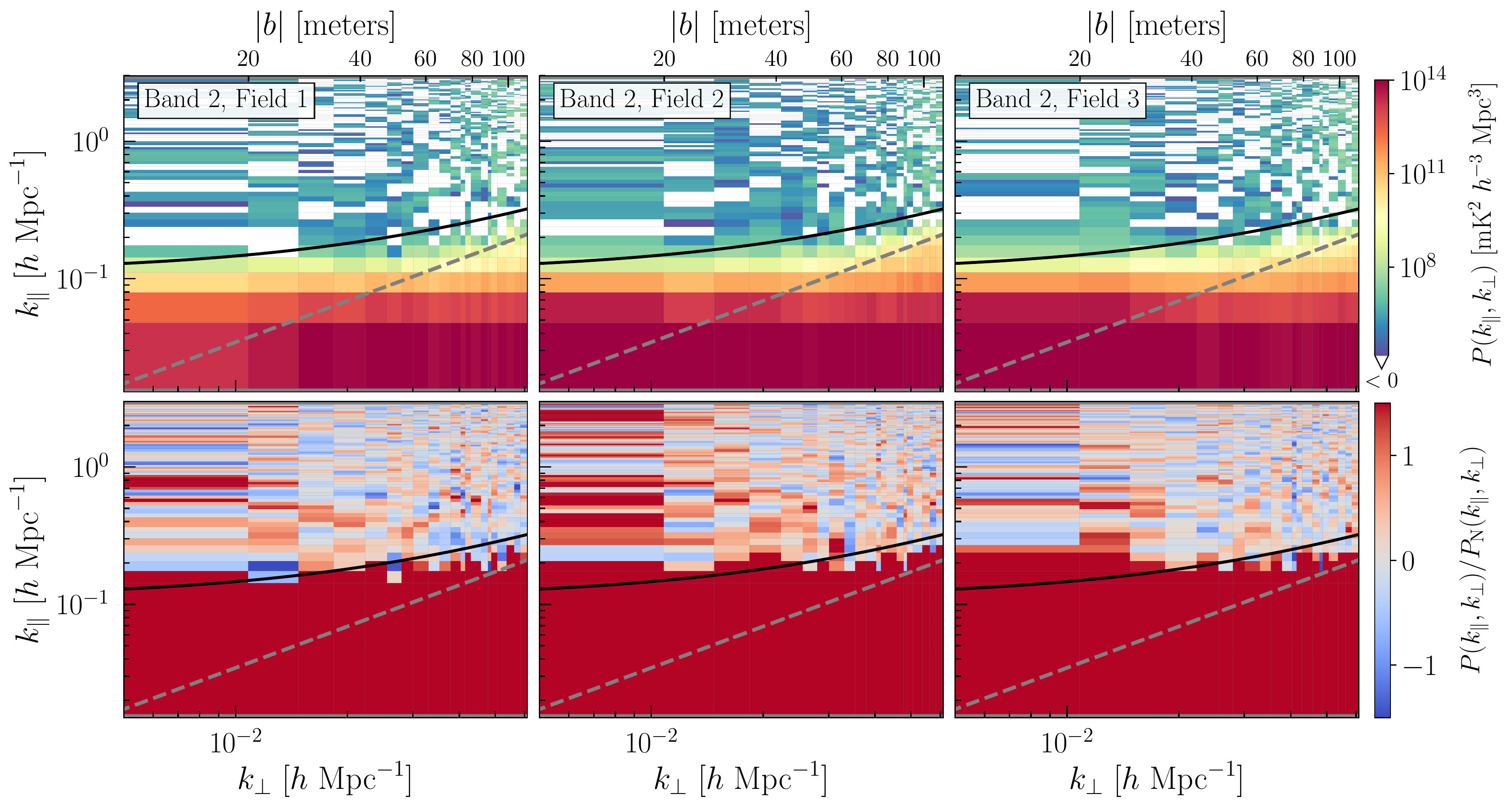}
\vspace{-5mm}
\caption{The same 2D power spectra as \autoref{fig:band1_wedge} but for Band 2 ($z=7.9$).}
\label{fig:band2_wedge}
\end{figure*}

Next we incoherently average the power spectra over the redundant baseline axis and across the remaining time bins in each field.
This is equivalent to a cylindrical binning in $\mathbf{k}$ space onto a $k_\parallel$ and $k_\perp$ plane.
All averaging is weighted by the squared inverse of the computed thermal noise floor, $P_N$, which is both a function of baseline-pair and time but is independent of $k_\parallel$.
Simultaneously, we also propagate the bandpower covariance and the $\tilde{P}_{SN}$ error bar through the averaging.
This leaves us with a 2D power spectrum for each band and field, shown in \autoref{fig:band1_wedge} for Band 1 and \autoref{fig:band2_wedge} for Band 2.
In each figure the top panels show the real component of the power spectrum, with negative measured bandpowers shown in white, while the bottom panels show the ratio of the power spectrum with the thermal noise floor.
Even though the power spectrum of a coherent signal is non-negative by construction, the measured power spectrum can fluctuate negative due to the fact that we have cross-correlated data with independent noise realizations, which is a mean-zero process with non-zero variance.

The root-mean-square of the 2D power spectra divided by their thermal noise floors for $k_\parallel>0.2\ h\ {\rm Mpc}^{-1}$ is consistent with one to within $\sim$3\%, which means that (1) we are able to estimate the noise variance in the real data at a precision of a few percent, and (2), as we will also demonstrate later, the data at high $k$ modes are largely noise dominated.
The dashed line plots the foreground horizon limit, which is the theoretical maximum $k_\parallel$ for smooth spectrum foregrounds, and the solid line plots the horizon limit plus a buffer of 200 nanoseconds ($\Delta k_\parallel=0.11\ h\ {\rm Mpc}^{-1}$ for $z=7.9$ and $\Delta k_\parallel=0.10\ h\ {\rm Mpc}^{-1}$ for $z=10.4$).
As we will discuss below, this cut at the horizon limit plus a buffer helps to reduce foreground contamination in the final spherically averaged power spectrum.
The exact buffer was chosen to mitigate foreground leakage beyond the horizon limit (particularly at large $k_\perp$ modes) while keeping some non-zero weight at the lower $k_\parallel$ modes, even though there is some observed leakage beyond this buffer at low $k_\perp$ (\autoref{fig:band1_wedge} \& \autoref{fig:band2_wedge}).

\begin{figure}
\centering
\includegraphics[width=\linewidth]{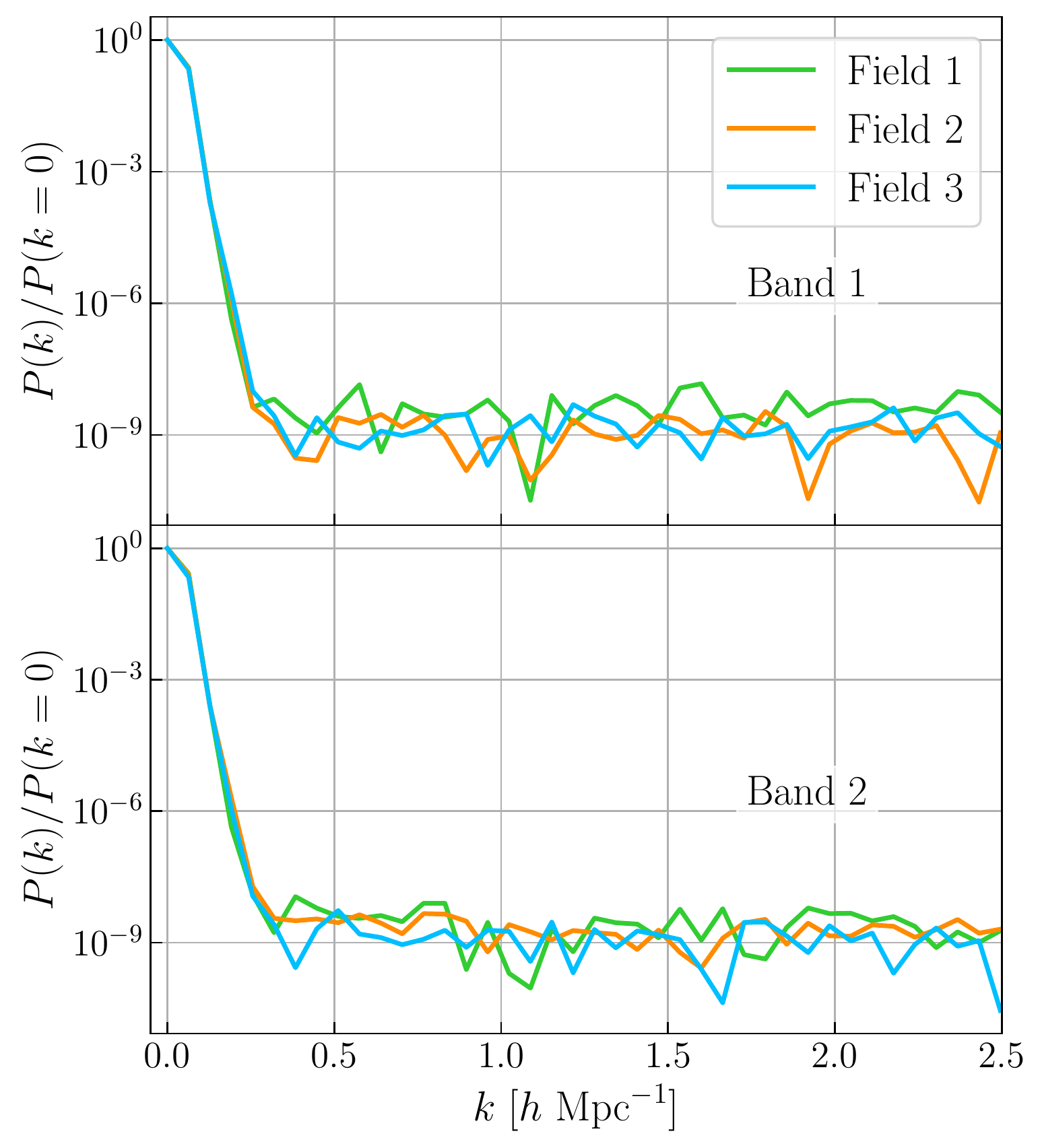}
\caption{The amplitude of the spherically binned power spectrum without enacting a minimum $k_\parallel$ cut in the binning, normalized to the foreground power at $k=0\ h\ {\rm Mpc}^{-1}$. This demonstrates that our analysis pipeline achieves a dynamic range of $\sim10^9$ with respect to the peak foreground power.}
\label{fig:dynamic_range}
\end{figure}

The fact that the data are in rough agreement with the expected noise at high $k_\parallel$, especially given that we have performed no foreground removal, is a testament to both the instrument design as well as the data reduction algorithms, which have been able to effectively contain foreground emission to the lowest $k_\parallel$ modes within the foreground wedge.
\autoref{fig:dynamic_range} shows this more clearly, showing the spherically averaged power spectrum (discussed below) without enacting a low $k_\parallel$ cut and normalizing by the peak foreground power at $k=0\ h\ {\rm Mpc}^{-1}$.
This shows that the full process of measuring the sky brightness, including the instrument itself and our analysis pipeline, achieves a dynamic range of $10^9$ in power between the peak foreground emission and the thermal noise floor at high $k$, which is a necessary but not quite sufficient criterion for an eventual 21\,cm signal detection.
Note that a complementary analysis of Phase I data using the bispectrum phase also achieve a dynamic range of $10^8$ between peak foreground power and the recovered noise floor \citep{Thyagarajan2020}, but used less data than what is presented here.

However, while \autoref{fig:band1_wedge} and \autoref{fig:band2_wedge} show decent agreement with the noise floor at high $k_\parallel$ (a point we come back to in \autoref{sec:validation}), we also clearly see evidence of foreground emission at $k_\parallel$ beyond that determined by the horizon delay (termed supra-horizon emission or foreground leakage), particularly at low $k_\perp$.
Some of this is simply due to the frequency tapering function's footprint in $k_\parallel$ space, which pushes $k_\parallel\approx0$ foreground emission to higher $k_\parallel$; however, that cannot explain the full extent of the supra-horizon emission.
\citet{Kern2020a} speculate that this could be due to residual instrumental cross-coupling, which is both generally most prominent and hardest to filter off at the low $k_\perp$ modes.

\begin{figure*}[ht]
\centering
\includegraphics[width=\linewidth]{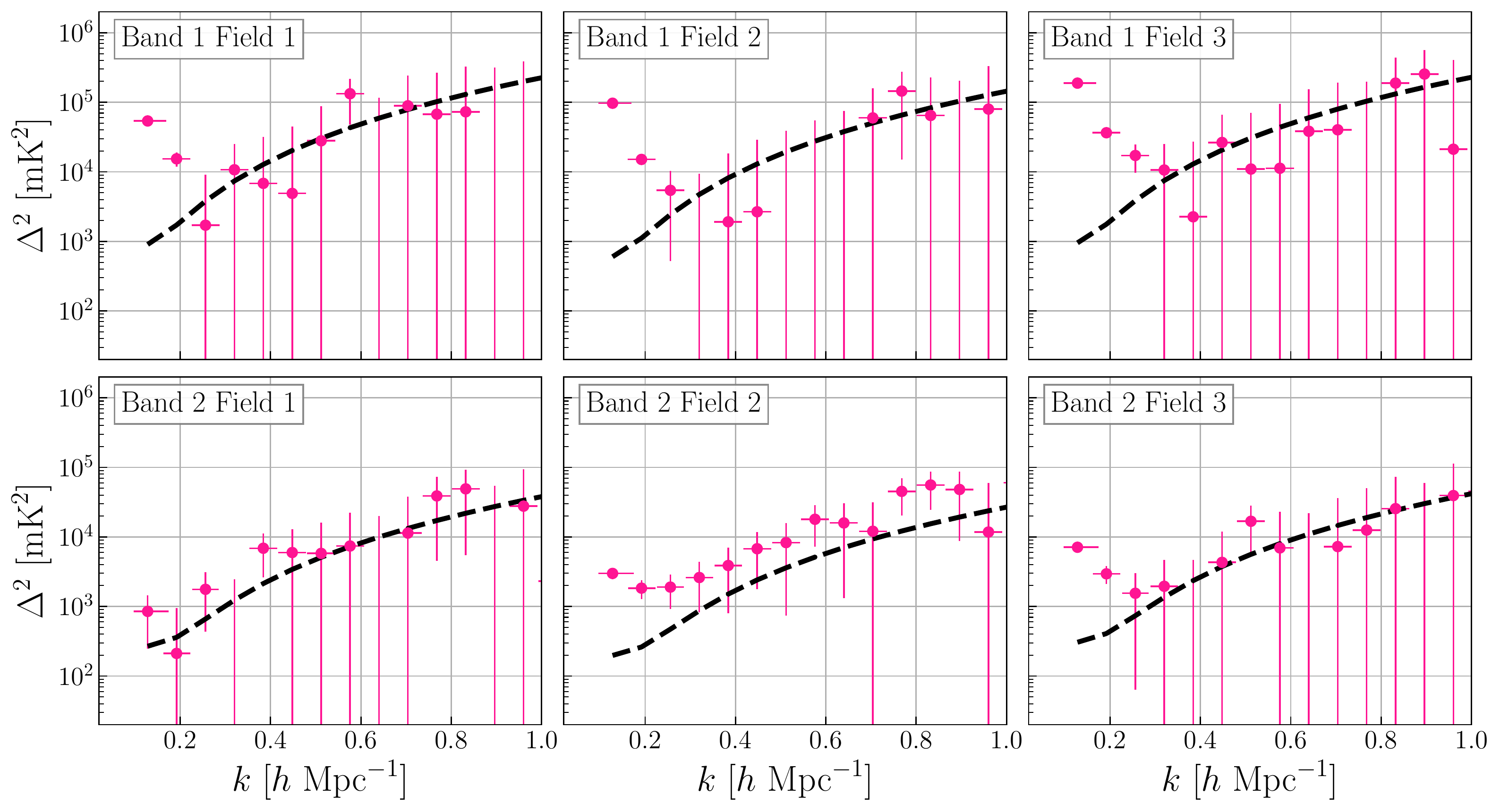}
\vspace{-5mm}
\caption{Spherically averaged, dimensionless power spectra for Band 1 ($z=10.4$) and Band 2 ($z=7.9$) at each of the three fields. The vertical error bars are $2\sigma$, while the horizontal error bars are the 16 and 84th percentiles of the window functions. The theoretical thermal noise floor (dashed) is also shown, which represents the $1\sigma$ noise floor. Bandpowers that are meausured to be negative are set to zero. The most sensitive limits come from Field 1 (for both Band 1 and Band 2), which is largely due to the fact that Field 1 has the least amount of foreground power in the field of view.}
\label{fig:integrated_dsq}
\end{figure*}

Looking at Band 2, Field 2, we also see evidence for a slight excess beyond the foreground horizon and buffer.
While it is unclear exactly what this is due to, one possibility is that it is low-level spectral artifact (possibly unflagged RFI) that exhibits a baseline dependence, which has been seen before at a low-level in Phase I data.
Future, more comprehensive work detailing low-level spectral artifacts in the fully averaged data, for example with the SSINS algorithm \citep{Wilensky2019}, may help us better understand these features.

\subsection{Integrated Limits on the Power Spectrum}
\label{sec:limits}

Next we spherically average the data by binning the 2D power spectra in bins of constant $|k|$, assuming statistical isotropy of the cosmologial signal, and again weighting the average by the squared inverse of $P_N(k_\parallel, k_\perp)$ in each pixel.
We compute the spherically averaged power spectrum and propagate our measures of uncertainty using Equations 33 -- 38 of \citet{Dillon2014}.
Recall that we give zero weight to all pixels with $k_\parallel < k_{\rm horizon} + \Delta k_\parallel$, where $\Delta k_\parallel$ is the foreground horizon buffer.
This leaves us with three spherically averaged power spectra at each field for Band 1 and Band 2, which we show in \autoref{fig:integrated_dsq}.
The vertical error bars are the $2\sigma$ $P_{SN}$ error bar discussed in \autoref{sec:errorbars} and \citet{Tan2021}, and the horizontal error bars are the 16 and 84th percentile of the window functions.
We also plot the theoretical $1\sigma$ thermal noise floor (dashed), which is simply the $P_N$ error bar discussed in \autoref{sec:errorbars}.
Bandpowers that are measured as negative are set to zero.
While coherent sky signal manifests as purely positive signal in the power spectrum, thermal noise and other uncorrelated systematics are mean-zero with non-zero variance, which can manifest as negative values in the power spectrum. 
Note that in the limit that the data are nearly noise-dominated, which is the most of the data shown in \autoref{fig:integrated_dsq}, the $1\sigma$ errorbar is the same as the $1\sigma\ P_N$ dashed line.

%\begin{figure}
%\centering
%\includegraphics[width=\linewidth]{imgs/wedge_hist.pdf}
%\caption{Histogram of the real component of the two dimensional power spectra divided by the thermal noise RMS for all $k$ bins where $k_\parallel \ge k_{\rm horizon}$ plus a suprahorizon buffer of $k_\parallel=0.2\ h\ {\rm Mpc}^{-1}$. The inset shows the distribution of the robust standard deviation (median absolute deviation from \autoref{eq:modz}) of each histogram, having resampled with replacement to compute their distributions. We observe deviations from $\sigma_{P/P_{N}}\sim1$ at the the level of $\sim3\%$, except for Band 2 Field 2 which shows evidence for a positive outlier distribution.}
%\label{fig:wedge_hist}
%\end{figure}

\begin{figure}
\centering
\includegraphics[width=\linewidth]{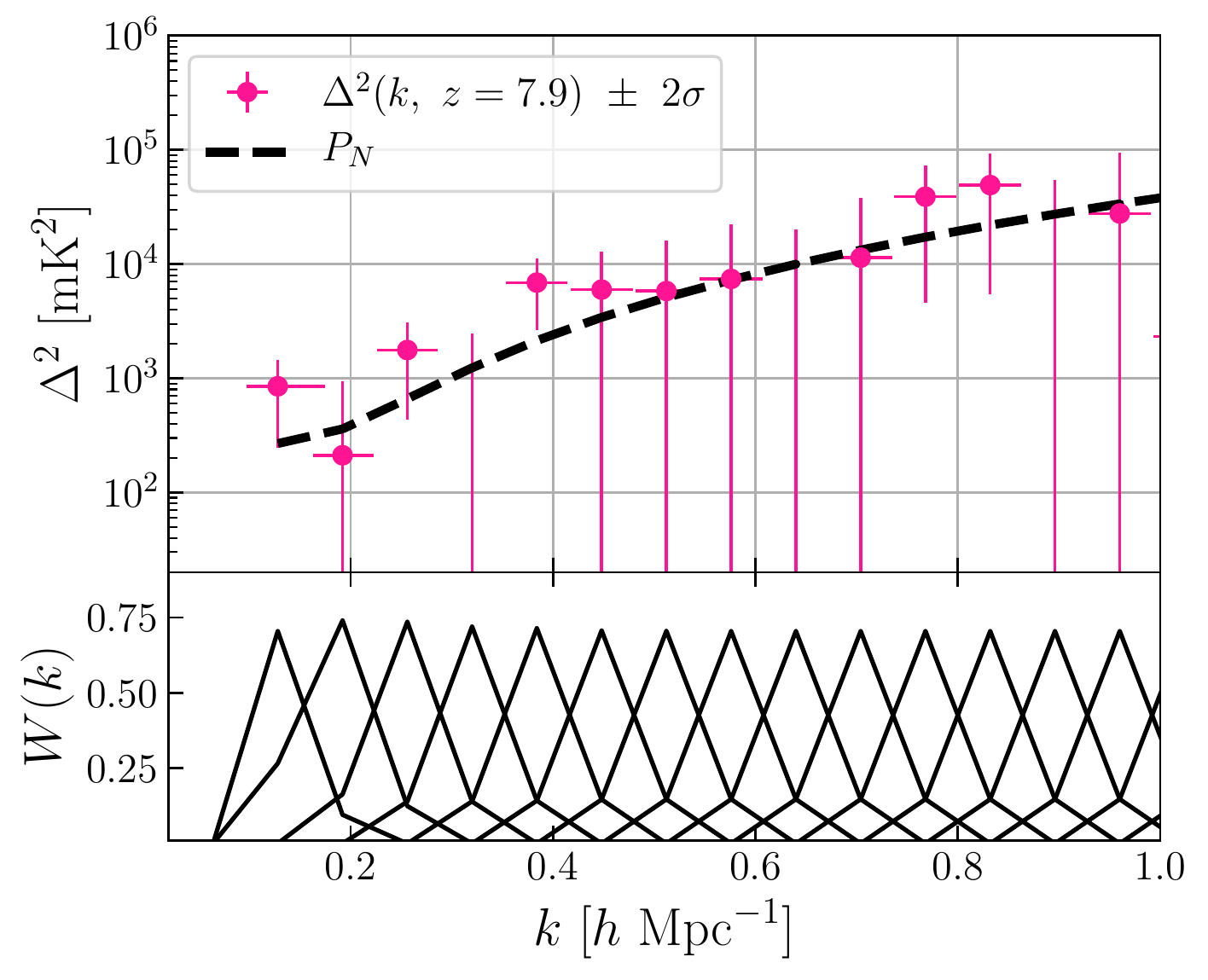}
\caption{The spherically averaged, dimensionless power spectrum from Band 2 and Field 1, showing 2$\sigma$ vertical error bars, their associated window functions (lower panel), and the theoretical thermal noise floor (dashed). The horizontal error bar is the distance of the 16 and 84th percentile of the window function from its 50th percentile, equal to $\Delta k=0.031\ h\ {\rm Mpc}^{-1}$ for all the bandpowers except the lowest $k$ mode, which is slightly asymmetric due to the horizon buffer's effect on spherical binning.}
\label{fig:band2_wf}
\end{figure}

Broadly, we can evaluate whether the data are consistent with thermal noise fluctuations if (1) the vertical error bars reach $\Delta^2=0\ {\rm mK}^2$ for the majority of the data points, (2) the measured bandpowers are roughly consistent with the $P_N$ curve, and (3) we measure a roughly equal number of positive and negative band powers.
However, all of this is complicated by the fact that the fully averaged products reduce us to a regime of low number statistics, in addition to the fact there will be some amount of correlation between the bandpowers in $k$, in part due to the frequency tapering function (the bandpowers are independent across bands and fields).
With just a qualitative assessment for the time being, however, we can see general agreement of the data with the expected thermal noise level at high $k$ modes, and increasing discrepancy at low $k$ modes.

The discrepancy at low $k$ is easily explained as residual foreground leakage that can be directly observed in \autoref{fig:band1_wedge} and \autoref{fig:band2_wedge}.
The fact that this leakage is stronger for Band 1 is largely due to the fact that the foregrounds are brighter at low frequencies.
Furthermore, we see that Field 1 exhibits the least amount of foreground leakage at the low $k$ modes, which is also largely due to the fact that it sees relatively dimmer foregrounds than the other fields (\autoref{fig:hera_stripe} \& \autoref{fig:pspec_waterfall}).
Furthermore, we also see that Band 2, Field 2 sees a systematic excess in power compared to thermal noise expectation, which is likely due to a low-level time and frequency dependent systematic.
We speculate that this could be the effect of low-level radio frequency interference, due in part to its spectral and temporal transience; however, more comprehensive work identifying low-level RFI is required to understand this at a deeper level.

The band powers in \autoref{fig:integrated_dsq} are sampled at a cadence of $\Delta k=0.64\ h\ {\rm Mpc}^{-1}$, which is twice the native spacing of the Fourier modes given our choice of spectral windows.
This helps to reduce the correlations between neighboring bandpowers, especially given that we applied a tapering function across frequency before estimating the power spectra.
This tapering impacts both the resultant bandpower window functions as well as the bandpower covariance matrix.
\autoref{fig:band2_wf} shows the Band 2, Field 1 power spectrum with its associated window functions (bottom panel), which shows that they are well-behaved even without decorrelation; however, they do show low-level overlap between neighboring modes, which in principle should be taken into account when comparing the data to astrophysical models.
The window functions are nearly identical for all measured modes,\footnote{The lowest $k$ mode in \autoref{fig:band2_wf} has no response below $k=0.128\ h\ {\rm Mpc}^{-1}$ due to the horizon buffer enacted in spherical binning.} and are also nearly identical between Band 1 and Band 2.
The 16 and 84th percentile of the window functions are $\Delta k = 0.031\ h\ {\rm Mpc}^{-1}$ away from their 50th percentile, which is a rough approximation of the $1\sigma$ horizontal error bar of the bandpowers.

Upper limits on the 21\,cm power spectrum (at a 95\% confidence level) are constructed by taking the measured dimensionless power spectrum, $\Delta^2$, and adding the $2\sigma$ error bar.
In the case where the measured bandpower is negative, we set it to zero, which is justified by our assertion that the cosmological signal (or any coherent signal in the visibilities for that matter) is a purely real and positive quantity in the power spectrum by definition.
In this case, the upper limit simply becomes the $2\sigma$ error bar.

The measured bandpowers, their uncertainties, and the derived upper limits are tabulated in \autoref{tab:band1_limits} and \autoref{tab:band2_limits}.
The most stringent upper limits at both Band 1 and 2 are achieved from Field 1, yielding an upper limit of $(95.74)^2\ {\rm mK}^2$ at $z=10.4$ and $k=0.256\ h\ {\rm Mpc}^{-1}$ and $(30.76)^2\ {\rm mK}^2$ at $z=7.9$ $k=0.192\ h\ {\rm Mpc}^{-1}$.
This is the most sensitive upper limit on the EoR 21\,cm power spectum at $z\sim8$ by over an order of magnitude, achieved with only a fraction of the full sensitivity of the HERA array.
HERA's nominal sensitivty over a year-long observing campaign could be pushed two orders of magnitude deeper, thus reaching the fiducial signal amplitudes of standard EoR models; however, low-level systematics will need to be modeled and appropriately resolved in order to reach those sensitivities.

\begin{table*}
\centering
\caption{Band 1 power spectra, errors, and upper limits on $\Delta^2_{21}(z=10.4)$ from \autoref{fig:integrated_dsq}. Ther upper limit is the measurement plus the $2\sigma$ error bar. In the process, $\Delta^2$ is set to zero where it is measured to be negative.}
\vspace{-2mm}
\label{tab:band1_limits}
\begin{tabular}{c | ccc | ccc | ccc}
\toprule
\multicolumn{1}{c}{} &
\multicolumn{3}{c}{Field 1}    &
\multicolumn{3}{c}{Field 2}    &
\multicolumn{3}{c}{Field 3}    \\
\cmidrule(lr){2-4}
\cmidrule(lr){5-7}
\cmidrule(lr){8-10}
\multicolumn{1}{c}{$k$} &
\multicolumn{1}{c}{$\Delta^2$}     &
\multicolumn{1}{c}{$1\sigma$} &
\multicolumn{1}{c}{$\Delta^2_{\rm UL}$} &
\multicolumn{1}{c}{$\Delta^2$}     &
\multicolumn{1}{c}{$1\sigma$} &
\multicolumn{1}{c}{$\Delta^2_{\rm UL}$} &
\multicolumn{1}{c}{$\Delta^2$}     &
\multicolumn{1}{c}{$1\sigma$} &
\multicolumn{1}{c}{$\Delta^2_{\rm UL}$} \\
\multicolumn{1}{c}{$h\ {\rm Mpc}^{-1}$} &
\multicolumn{1}{c}{$({\rm mK})^2$}     &
\multicolumn{1}{c}{$({\rm mK})^2$} &
\multicolumn{1}{c}{$({\rm mK})^2$} &
\multicolumn{1}{c}{$({\rm mK})^2$}     &
\multicolumn{1}{c}{$({\rm mK})^2$} &
\multicolumn{1}{c}{$({\rm mK})^2$} &
\multicolumn{1}{c}{$({\rm mK})^2$}     &
\multicolumn{1}{c}{$({\rm mK})^2$} &
\multicolumn{1}{c}{$({\rm mK})^2$} \\
\midrule
$0.128$ & $(232.14)^2$ & $(36.90)^2$ & $(237.93)^2$& $(311.36)^2$ & $(36.84)^2$ & $(315.69)^2$& $(434.21)^2$ & $(48.03)^2$ & $(439.49)^2$\\ 
$0.192$ & $(124.01)^2$ & $(42.26)^2$ & $(137.66)^2$& $(122.81)^2$ & $(34.88)^2$ & $(132.34)^2$& $(191.00)^2$ & $(43.14)^2$ & $(200.50)^2$\\ 
$0.256$ & $(41.34)^2$ & $(61.07)^2$ & $(95.74)^2$& $(73.63)^2$ & $(49.50)^2$ & $(101.60)^2$& $(131.11)^2$ & $(61.15)^2$ & $(157.06)^2$\\ 
$0.320$ & $(103.46)^2$ & $(84.89)^2$ & $(158.49)^2$& $-(81.56)^2$ & $(68.63)^2$ & $(97.06)^2$& $(103.08)^2$ & $(84.82)^2$ & $(158.16)^2$\\ 
$0.384$ & $(82.64)^2$ & $(111.66)^2$ & $(178.23)^2$& $(43.73)^2$ & $(90.51)^2$ & $(135.27)^2$& $(47.52)^2$ & $(111.84)^2$ & $(165.15)^2$\\ 
$0.448$ & $(69.98)^2$ & $(141.55)^2$ & $(212.06)^2$& $(51.61)^2$ & $(114.92)^2$ & $(170.52)^2$& $(162.61)^2$ & $(141.37)^2$ & $(257.70)^2$\\ 
$0.512$ & $(167.19)^2$ & $(172.97)^2$ & $(296.30)^2$& $-(195.21)^2$ & $(139.86)^2$ & $(197.80)^2$& $(104.72)^2$ & $(172.32)^2$ & $(265.24)^2$\\ 
$0.576$ & $(364.27)^2$ & $(206.10)^2$ & $(466.52)^2$& $-(200.33)^2$ & $(166.17)^2$ & $(235.00)^2$& $(106.11)^2$ & $(205.11)^2$ & $(308.86)^2$\\ 
$0.640$ & $-(73.05)^2$ & $(240.38)^2$ & $(339.95)^2$& $-(296.41)^2$ & $(194.00)^2$ & $(274.36)^2$& $(195.72)^2$ & $(239.46)^2$ & $(391.14)^2$\\ 
$0.704$ & $(298.47)^2$ & $(277.29)^2$ & $(492.81)^2$& $(244.19)^2$ & $(223.98)^2$ & $(399.95)^2$& $(200.57)^2$ & $(275.83)^2$ & $(438.63)^2$\\ 
$0.768$ & $(259.69)^2$ & $(314.80)^2$ & $(515.40)^2$& $(380.60)^2$ & $(254.70)^2$ & $(524.02)^2$& $-(264.99)^2$ & $(314.01)^2$ & $(444.07)^2$\\ 
$0.832$ & $(270.03)^2$ & $(354.73)^2$ & $(569.72)^2$& $(254.18)^2$ & $(286.82)^2$ & $(478.69)^2$& $(434.40)^2$ & $(354.47)^2$ & $(663.33)^2$\\ 
$0.896$ & $-(326.77)^2$ & $(396.45)^2$ & $(560.67)^2$& $-(111.95)^2$ & $(320.48)^2$ & $(453.23)^2$& $(504.58)^2$ & $(395.78)^2$ & $(753.58)^2$\\ 
$0.960$ & $-(523.27)^2$ & $(439.41)^2$ & $(621.42)^2$& $(282.52)^2$ & $(355.29)^2$ & $(576.44)^2$& $(145.43)^2$ & $(438.38)^2$ & $(636.80)^2$\\ 
\bottomrule
\end{tabular}
\end{table*}

\begin{table*}
\centering
\caption{Band 2 power spectra, errors, and upper limits on $\Delta^2_{21}(z=7.9)$ from \autoref{fig:integrated_dsq}. The same procedure as \autoref{tab:band1_limits} is used to construct the upper limits.}
\vspace{-2mm}
\label{tab:band2_limits}
\begin{tabular}{c | ccc | ccc | ccc}
\toprule
\multicolumn{1}{c}{} &
\multicolumn{3}{c}{Field 1}    &
\multicolumn{3}{c}{Field 2}    &
\multicolumn{3}{c}{Field 3}    \\
\cmidrule(lr){2-4}
\cmidrule(lr){5-7}
\cmidrule(lr){8-10}
\multicolumn{1}{c}{$k$} &
\multicolumn{1}{c}{$\Delta^2$}     &
\multicolumn{1}{c}{$1\sigma$} &
\multicolumn{1}{c}{$\Delta^2_{\rm UL}$} &
\multicolumn{1}{c}{$\Delta^2$}     &
\multicolumn{1}{c}{$1\sigma$} &
\multicolumn{1}{c}{$\Delta^2_{\rm UL}$} &
\multicolumn{1}{c}{$\Delta^2$}     &
\multicolumn{1}{c}{$1\sigma$} &
\multicolumn{1}{c}{$\Delta^2_{\rm UL}$} \\
\multicolumn{1}{c}{$h\ {\rm Mpc}^{-1}$} &
\multicolumn{1}{c}{$({\rm mK})^2$}     &
\multicolumn{1}{c}{$({\rm mK})^2$} &
\multicolumn{1}{c}{$({\rm mK})^2$} &
\multicolumn{1}{c}{$({\rm mK})^2$}     &
\multicolumn{1}{c}{$({\rm mK})^2$} &
\multicolumn{1}{c}{$({\rm mK})^2$} &
\multicolumn{1}{c}{$({\rm mK})^2$}     &
\multicolumn{1}{c}{$({\rm mK})^2$} &
\multicolumn{1}{c}{$({\rm mK})^2$} \\
\midrule
$0.128$ & $(29.17)^2$ & $(17.39)^2$ & $(38.16)^2$& $(54.66)^2$ & $(16.01)^2$ & $(59.17)^2$& $(84.32)^2$ & $(19.49)^2$ & $(88.71)^2$\\ 
$0.192$ & $(14.55)^2$ & $(19.17)^2$ & $(30.76)^2$& $(42.85)^2$ & $(16.74)^2$ & $(48.95)^2$& $(54.33)^2$ & $(20.64)^2$ & $(61.67)^2$\\ 
$0.256$ & $(42.04)^2$ & $(25.84)^2$ & $(55.70)^2$& $(43.60)^2$ & $(22.08)^2$ & $(53.63)^2$& $(39.36)^2$ & $(27.25)^2$ & $(55.09)^2$\\ 
$0.320$ & $-(22.72)^2$ & $(35.07)^2$ & $(49.60)^2$& $(51.11)^2$ & $(29.87)^2$ & $(66.31)^2$& $(44.19)^2$ & $(36.84)^2$ & $(68.31)^2$\\ 
$0.384$ & $(82.98)^2$ & $(46.17)^2$ & $(105.59)^2$& $(62.34)^2$ & $(39.28)^2$ & $(83.50)^2$& $-(18.72)^2$ & $(48.42)^2$ & $(68.48)^2$\\ 
$0.448$ & $(77.29)^2$ & $(58.66)^2$ & $(113.39)^2$& $(82.31)^2$ & $(49.99)^2$ & $(108.51)^2$& $(65.69)^2$ & $(61.53)^2$ & $(109.02)^2$\\ 
$0.512$ & $(76.19)^2$ & $(71.90)^2$ & $(127.06)^2$& $(91.07)^2$ & $(61.47)^2$ & $(125.90)^2$& $(129.66)^2$ & $(75.54)^2$ & $(168.00)^2$\\ 
$0.576$ & $(86.08)^2$ & $(85.87)^2$ & $(148.85)^2$& $(133.99)^2$ & $(73.20)^2$ & $(169.32)^2$& $(83.39)^2$ & $(90.01)^2$ & $(152.18)^2$\\ 
$0.640$ & $-(108.73)^2$ & $(100.34)^2$ & $(141.90)^2$& $(126.10)^2$ & $(85.39)^2$ & $(174.59)^2$& $-(89.07)^2$ & $(104.92)^2$ & $(148.38)^2$\\ 
$0.704$ & $(106.66)^2$ & $(115.28)^2$ & $(194.83)^2$& $(109.95)^2$ & $(98.06)^2$ & $(176.98)^2$& $(85.27)^2$ & $(120.75)^2$ & $(190.87)^2$\\ 
$0.768$ & $(197.25)^2$ & $(131.05)^2$ & $(270.66)^2$& $(212.69)^2$ & $(111.44)^2$ & $(264.71)^2$& $(112.00)^2$ & $(137.38)^2$ & $(224.26)^2$\\ 
$0.832$ & $(221.71)^2$ & $(147.79)^2$ & $(304.69)^2$& $(236.41)^2$ & $(125.53)^2$ & $(295.64)^2$& $(159.84)^2$ & $(154.98)^2$ & $(271.27)^2$\\ 
$0.896$ & $-(43.63)^2$ & $(164.64)^2$ & $(232.84)^2$& $(219.42)^2$ & $(140.29)^2$ & $(295.82)^2$& $-(114.21)^2$ & $(172.95)^2$ & $(244.58)^2$\\ 
$0.960$ & $(166.53)^2$ & $(182.58)^2$ & $(307.25)^2$& $(108.48)^2$ & $(155.43)^2$ & $(245.12)^2$& $(198.43)^2$ & $(191.65)^2$ & $(335.91)^2$\\ 
\bottomrule
\end{tabular}
\end{table*}

%%%%%%%%%%%%%%%%%%%%%%%%%%%
%%%%%%%% Validation %%%%%%%
%%%%%%%%%%%%%%%%%%%%%%%%%%%
\section{Validation of Upper Limits}
\label{sec:validation}

21\,cm analysis pipelines are becoming increasingly sophisticated in order to beat down a host of systematic effects and reach the level of precision necessary to detect the cosmological 21\,cm signal.
A key concern is that these analysis choices could cause non-negligible amounts of signal loss, whether from calibration \citep{Mouri2019}, power spectrum estimation \citep{Cheng2018}, or other steps.
The extent of any possible signal loss needs to be quantified in order to build confidence that our results are not prone to signal loss.
In this analysis we use a combination of three complementary approaches to build confidence in our analysis and the robustness of upper limits on the power spectrum: simulation-based checks of the pipeline, a comparative study of uncertainty quantification, and statistical null tests on data subsets

In a companion paper, \citet{Aguirre2021} present a validation of the Phase I analysis and power spectrum pipeline, using data simulations with realistic sky, instrument, and systematic models.
In addition to tests on individual components of the pipelines, such as calibration, systematic modeling, and power spectrum estimation, they also present a near end-to-end pipeline test, demonstrating the pipeline's ability to accurately recover a realistic EoR model for $k\ge0.192\ h\ {\rm Mpc}^{-1}$ at both of the spectral windows considered in this work.
This end-to-end study is particularly important, as different steps of the pipeline could potentially interact in non-linear ways that introduce signal loss.
While the instrument model used in that work are complex enough to model the most prominent systematics, there are some areas where the models can be made more sophisticated in future work to test more subtle features of the instrument, which may be necessary for validating an analysis with a putative EoR detection.
In addition to testing the Phase I pipeline with simulated datsets, \citet{Aguirre2021} also show a semi-blinded analysis of the simulated data with an independent power spectrum pipeline as a consistency test with the Phase I power spectrum pipeline.
In this exercise, a small group of people were excluded from discussions on the construction of the simulated data and systematics, and applied an independent, simplified power spectrum pipeline to the data products to test for experimenter bias in the derived limits.
Their resultant power spectra are in good agreement with the output of the Phase I pipeline, in that they show unbiased detection of the simulated EoR signal for $k\gtrsim0.2\ h\ {\rm Mpc}^{-1}$.

\begin{figure*}[htbp]
\centering
\includegraphics[width=\linewidth]{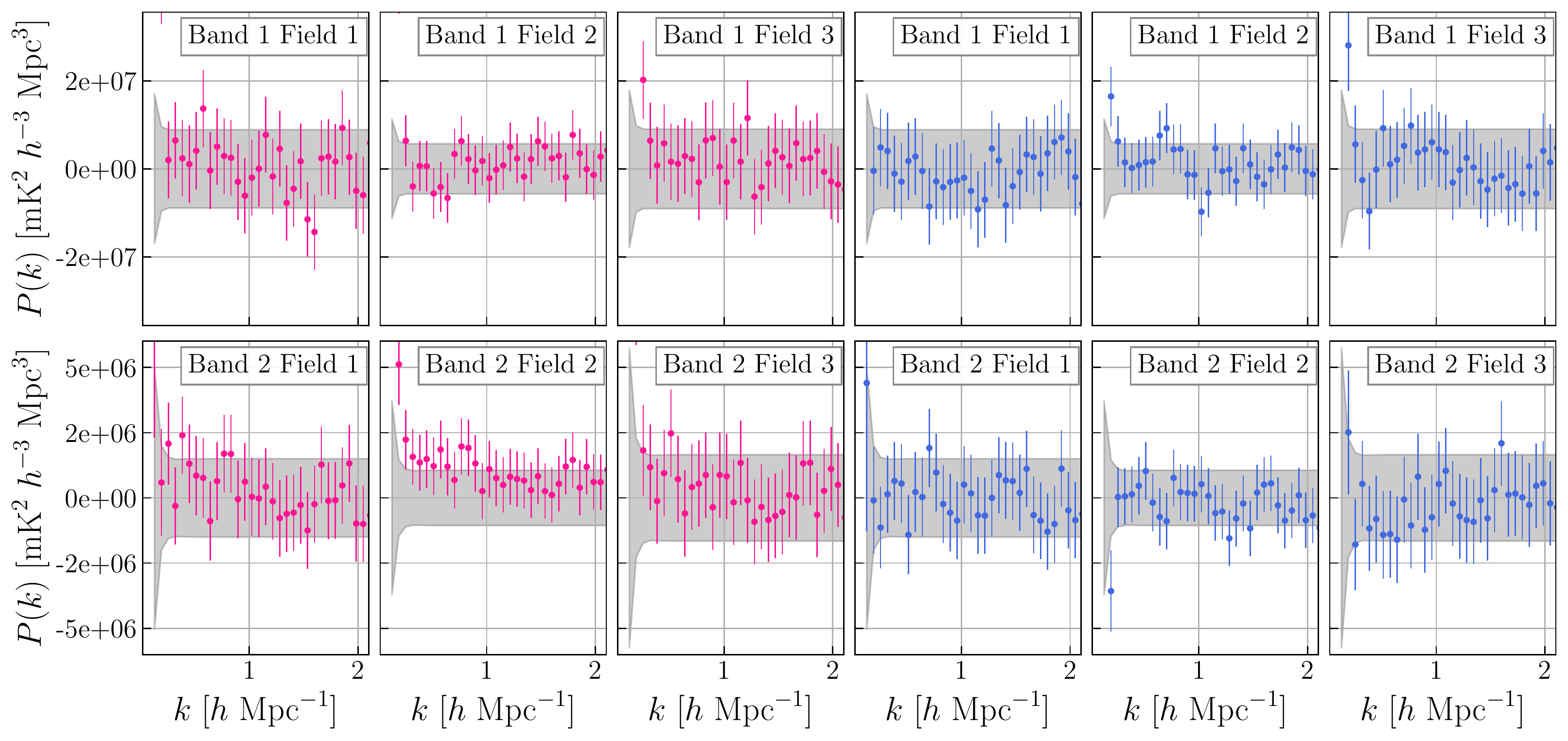}
\vspace{-5mm}
\caption{
The real and imaginary components of the spherically averaged power spectrum $P(k)$.
The six plots on the left show the real component (pink dots), while the six plots on the right show the imaginary component (blue dots).
The error bars are $2\sigma$ and the shaded region is the $P_N$ theoretical noise expectation.
Significant power in the imaginary component could indicate phase calibration errors resulting from unmodeled systematics or baseline non-redundancies.
We quantify the consistency of the data with noise via $p$-value tests, tabulated in \autoref{tab:pvalues}.}
\label{fig:integrated_sph}
\end{figure*}

In addition to the simulated pipeline validation from \citet{Aguirre2021}, \citet{Tan2021} present a comparative study of uncertainty quantification in the power spectrum pipeline.
They highlight a number of techniques used in the literature for estimating uncertainty on the 21\,cm power spectrum, both analytic and semi-empirical, and demonstrate that they all show relatively good agreement with each other when applied to both real and simulated HERA Phase I power spectra.
Specifically, as discussed in \autoref{sec:qe}, we use the $P_{N}$ and $P_{SN}$ error bars that quantify the root-mean-square (RMS) of the power spectra due soley to thermal noise ($P_N$), and the RMS due to thermal noise and its coupling to residual signal terms in the data ($P_{SN}$).
The final error bars quoted in \autoref{tab:band1_limits} and \autoref{tab:band2_limits} come from this latter metric.

In the rest of this section, we discuss a series of statistical consistency tests to better understand residual systematics observed in the data.
Some of these tests are formulated as true null tests, where the null hypothesis is that the spherically averaged power spectrum is consistent with noise-only fluctuations, while others are better described as consistency checks that seek to verify a particular scaling law.
Such tests can help to quantify the parts of the data that are systematic or noise limited \citep[e.g.][]{Kolopanis2019}.

\subsection{Real and imaginary parts of the power spectrum}
\label{sec:real_imag_null}

\begin{figure}[htbp]
\centering
\includegraphics[width=\linewidth]{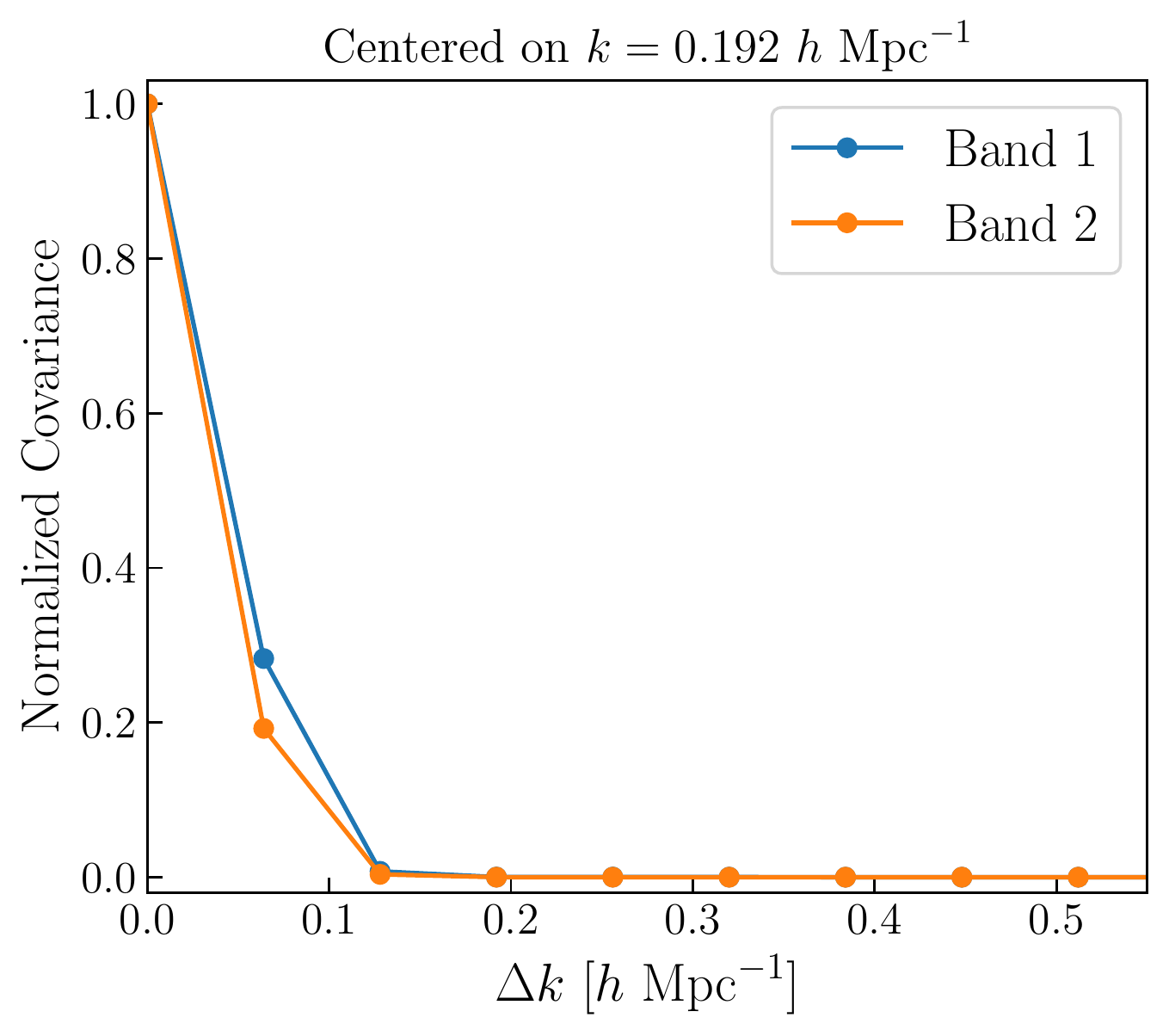}
\caption{The normalized covariance between the $k=0.192\ h\ {\rm Mpc}^{-1}$ mode and neighboring modes used in the statistical tests of \autoref{tab:pvalues}. This shows there is a small but non-zero covariance between neighboring modes, and effectively no covariance at larger separations.}
\label{fig:noise_cov}
\end{figure}

This test seeks to assess whether the final spherically averaged power spectra are consistent with thermal noise.
In the absence of residual systematics, we expect the power spectra to consist of foreground emission, EoR emission, and thermal noise.
Given that we have masked the foreground horizon upon spherical binning, any excess power above the predicted thermal noise variance is likely due to residual systematics, which can take may forms.
Generally, these systematics leak foreground emission to $k$ modes beyond the foreground horizon modes, thereby contaminating the lowest $k$ modes in the EoR window.
We assess whether the data are consistent with noise by evaluating the $p$-value of the power spectrum $\chi^2$ statistic for each band and field combination.
The statistic is defined as
\begin{align}
\label{eq:chisq_for_p_value}
\chi^2 = \bd^T \bSigma^{-1} \bd,
\end{align}
where $\bd$ is the power spectrum data vector and $\bSigma$ is its noise covariance.
The noise covariance is nearly diagonal, with small but non-zero neighbor-to-neighbor covariance (\autoref{fig:noise_cov}).
The $p$-value for a particular data vector is the area under the $\chi^2$ sampling distribution that exceeds the $\chi^2$ of the data.
The $\chi^2$ sampling distribution is derived in a Monte Carlo fashion by drawing a large number ($N=10^6$) of random noise realizations and evaluating their $\chi^2$.
For $p$-values below 0.001 we can only quote an upper limit on the $p$-value given the discrete number of random draws, however, this is a sufficiently small value to enable an unambiguous rejection of the null hypothesis.
Note that we do not enact a hard cut on the $p$-value to determine whether the data reject the null hypothesis, but rather use the collection of $p$-values for a given band and field in both the real and imaginary components to ascertain whether the data seem consistent with thermal noise.

\begin{table}[ht]
\caption{Statistical $p$-value tests on the real and imaginary components of the spherical $P(k)$ (\autoref{fig:integrated_sph}).}
\vspace{-2mm}
\label{tab:pvalues}
\centering
\begin{tabular}{l c c c}
\toprule
\multicolumn{1}{l}{} &
\multicolumn{3}{c}{$p$-value} \\
\cmidrule(lr){2-4}
\multicolumn{1}{c}{Data Selection} & 
\multicolumn{1}{c}{$k\ge0.192$} &
\multicolumn{1}{c}{$k\ge0.5$}  &
\multicolumn{1}{c}{$k\ge1.0$}  \\
\midrule
Re[$P$], Band 1, Field 1 & $< 0.001$ & $< 0.001$ & $0.003$ \\ 
Im[$P$], Band 1, Field 1 & $0.605$ & $0.477$ & $0.345$ \\ 
\midrule
Re[$P$], Band 1, Field 2 & $< 0.001$ & $0.044$ & $0.203$ \\ 
Im[$P$], Band 1, Field 2 & $< 0.001$ & $0.002$ & $0.011$ \\ 
\midrule
Re[$P$], Band 1, Field 3 & $< 0.001$ & $0.107$ & $0.050$ \\ 
Im[$P$], Band 1, Field 3 & $< 0.001$ & $0.358$ & $0.523$ \\ 
\midrule
Re[$P$], Band 2, Field 1 & $0.020$ & $0.372$ & $0.716$ \\ 
Im[$P$], Band 2, Field 1 & $0.232$ & $0.216$ & $0.504$ \\ 
\midrule
Re[$P$], Band 2, Field 2 & $< 0.001$ & $< 0.001$ & $0.007$ \\ 
Im[$P$], Band 2, Field 2 & $< 0.001$ & $0.064$ & $0.111$ \\ 
\midrule
Re[$P$], Band 2, Field 3 & $< 0.001$ & $0.439$ & $0.717$ \\ 
Im[$P$], Band 2, Field 3 & $0.036$ & $0.292$ & $0.733$ \\ 
\bottomrule
\end{tabular}
\tablecomments{The null hypothesis is that the power spectrum is consistent with mean-zero fluctuations drawn from the propagated bandpower noise covariance (\autoref{fig:noise_cov}). Upper limits on the $p$-value are quoted where it is sufficiently small. The $k$ cuts are in $h\ {\rm Mpc}^{-1}$.}
\end{table}

Based on the estimator presented in \autoref{sec:qe}, the power spectrum of a noise-only dataset will be mean-zero process in both the real and imaginary components with equal variance.
Coherent signals in the data, whether they are sky signals or instrumental systematics, are confined to the real component of the power spectrum.
However, systematics with phase differences between cross-multiplied visibilties can cause excess variance in the imaginary component of the power spectrum \citep{Kolopanis2019}.
\autoref{fig:integrated_sph} shows the real component (pink panels) and the imaginary component (blue panels) of the power spectra in $P(k)$, with their $p$-values tabulated in \autoref{tab:pvalues} showing the response of the $p$-value with an increasingly large $k$ cut.
We see that for both bands, most fields are inconsistent with the noise distribution at the low $k$ modes, evidenced by their consistently low p values in both the real and imaginary components.
Reassuringly, the $p$-values for the imaginary component are generally more consistent with noise than the $p$-values for the real component, even when there is strong evidence for systematics, like Band 2, Field 2.
The interesting exception to the rule is Band 1, Field 2, whose imaginary component is less consistent with the noise than its real component, possibly indicative of a low level, baseline dependent systematic.
Overall, the quantification provided by the $p$-value tests on the real and imaginary components supports the notion that the data are largely consistent with thermal noise at high $k$ modes, while confirming our intuition that residual systematics are beginning to be detected at the lowest $k$ modes in the EoR window.

\begin{figure}
\centering
\includegraphics[width=\linewidth]{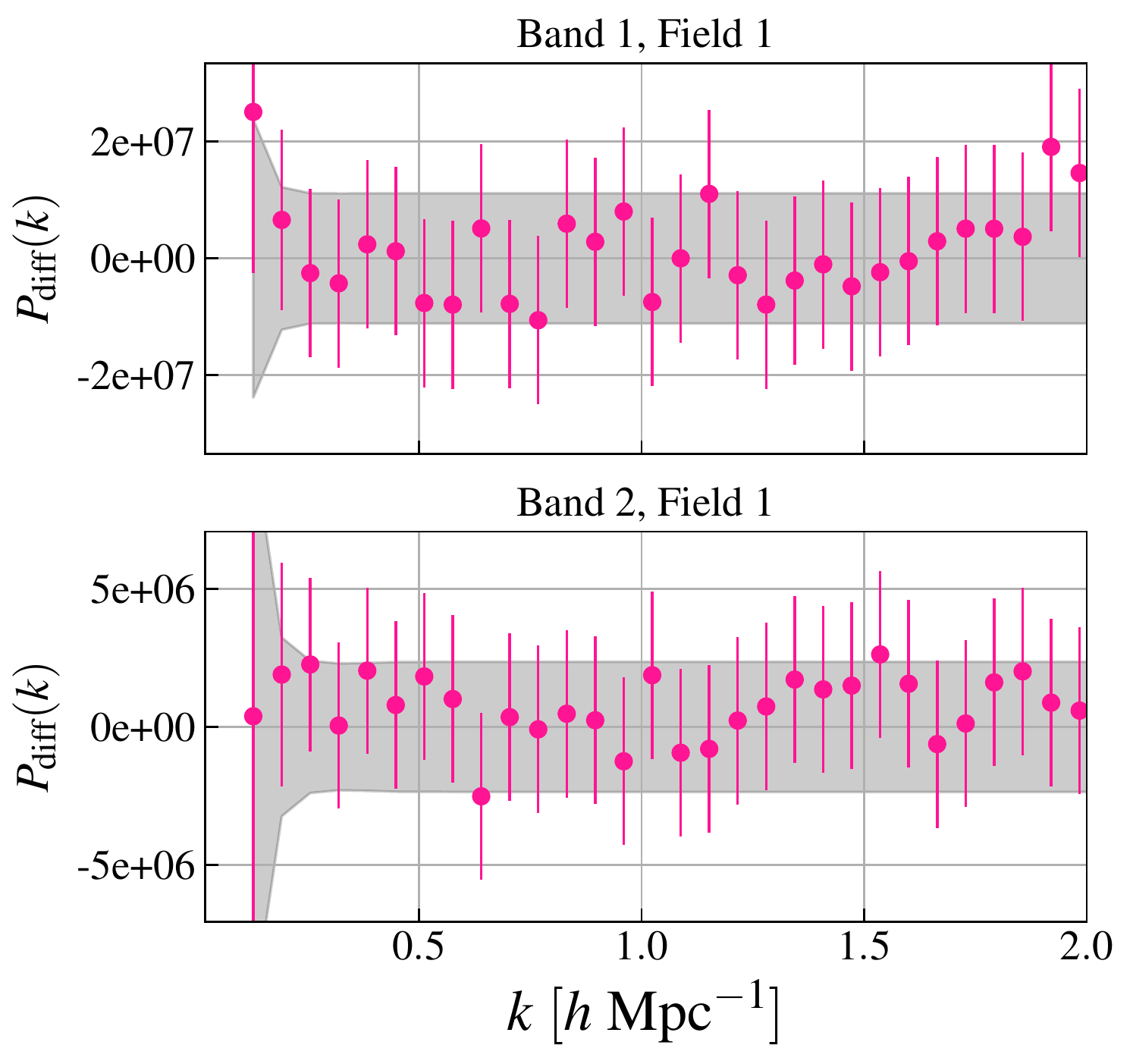}
\caption{The real component of the power spectrum computed from the difference of interleaved nights for Field 1. Vertical error bars are $2\sigma$, and the grey shaded region is $\pm2P_N$. The $p$-values for these power spectra with $k$ cuts of $\ge0.192$, $\ge0.5$, and $\ge1.0\ h\ {\rm Mpc}^{-1}$ are 0.651, 0.491, and 0.643 for Band 1 and 0.585, 0.584, and 0.619 for Band 2, respectively. These are consistent with the null hypothesis that the data are drawn from the thermal noise covariance.}
\label{fig:even_odd_split}
\end{figure}

\subsection{Night-to-night binning split}
\label{sec:even_odd_null}

A check on long-term temporal stability can be made by taking the nightly data from the dataset and separating them into two groups, differencing them, and then forming the power spectra of the differenced visibilities.
To do this we take every other night in the 18 night dataset and form a set of ``even'' nights and a set of ``odd'' nights in terms of their Julian date.
We then perform the standard night-to-night LST binning described in \autoref{sec:lstbin} on both the even and odd sets independently.
Afterwards, we pass both data sets through the standard systematics treatment and then difference the resultant visibilities before estimating the power spectrum.
This test is sensitive to systematics that exhibit non-negligible variation from night-to-night.

\autoref{fig:even_odd_split} shows the real component of the differenced power spectrum for Field 1.
Computing the $p$-values of these spectra in the same manner as \autoref{sec:real_imag_null}, we find that the spectra for both Band 1 and Band 2 are consistent with thermal noise for $k\ge0.192\ h\ {\rm Mpc}^{-1}$, which was not observed for Field 1 in the undifferenced power spectrum (\autoref{tab:pvalues}).
This suggests that the low-$k$ residual systematics seen in the data are fairly stable from night-to-night.

\begin{figure*}[tbp]
\begin{center}
\includegraphics[width=\linewidth]{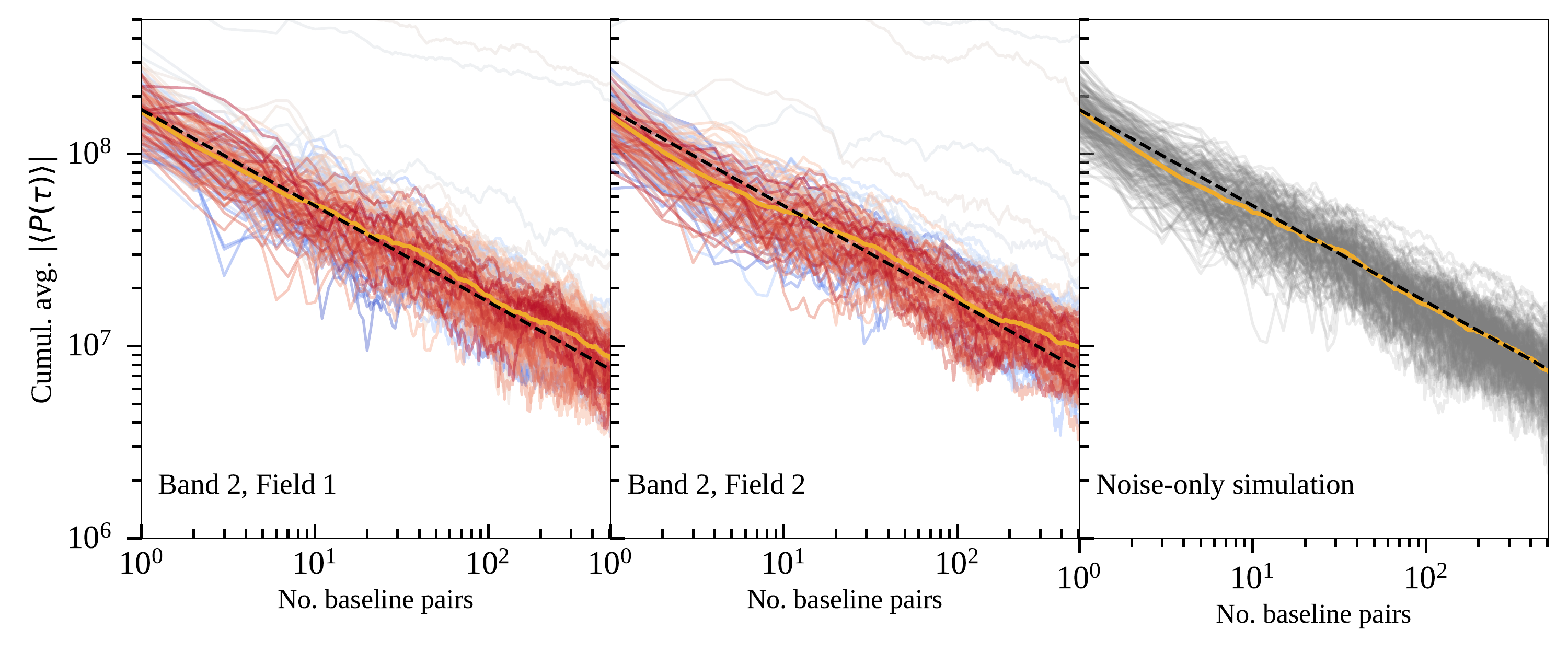}
\caption{Cumulative incoherent averaging of the delay spectra in arbitrary units within a single redundant baseline group (all 14.6 meter, East-West baselines) for Band 2 of Field 1 (left panel), Field 2 (middle panel), and a corresponding thermal noise-only realization of Band 2, Field 2 (right panel). Each line denotes a different bandpower that has been cumulatively-averaged over the baseline-pairs within the redundant group. To reduce the noisiness of the statistic, the absolute value of the real part was then taken and averaged over all LST samples in the field. Red lines denote negative delays while blue lines denote positive delays. Note that low delay modes of $|\tau|\lesssim200$ ns are not shown in the plot. The black dashed line shows a theoretical $1/\sqrt{N_{\rm baseline-pair}}$ scaling, with the same arbitrary normalization in all panels. The thick orange line shows the mean of the bandpowers for $|\tau|>500$ ns ($k_\parallel\sim0.25\ h\ {\rm Mpc}^{-1}$), showing a slight detection of a systematic at deep integration levels, particularly for Band 2, Field 2.}
\label{fig:cumulavg}    
\end{center}   
\end{figure*}

\subsection{Scaling of bandpowers under cumulative averaging}
\label{sec:allan_variance}

If delay modes outside the foreground horizon are dominated by thermal noise, as we would expect in the absence of residual systematics or an EoR signal detection, the measured power in those modes should integrate down in a characteristic way as more time samples and baseline-pairs are averaged over. In this section, we study the scaling of the measured delay spectrum bandpowers when cumulatively averaging individual baseline-pairs in a redundant baseline group, which should in principle measure the same sky signal but have independent noise contributions.
Note that this test is performed higher up in the power spectrum pipeline before cylindrical or spherical binning of the power spectra, which is necessary because we need a large number of independent samples for computing the cumulative average.
We specifically target the East-West 14.6 meter baseline group for this test as it has the largest number of independent baselines.

The left and center panel of \autoref{fig:cumulavg} show the scaling of the delay spectrum band powers for Band 2, fields 1 and 2, having cumulatively averaged over the redundant baseline pairs.
Each line shows the a different band power, ranging from negative delays to positive delays, while the black dashed line shows a representative $1/\sqrt{N}$ scaling, where $N$ is the number of averaged baseline-pairs.
To reduce the scatter in this statistic, we take the absolute value of the real part of the band power and average across the remaining LST samples associated with Field 1.

While low delay modes ($|\tau|\lesssim200$ ns) are not plotted, we see that the high delay modes are generally consistent with the expected $1/\sqrt{N}$ scaling of noise.
As a demonstration, the right panel of \autoref{fig:cumulavg} shows the same procedure applied to a pure noise simulation for Band 2, Field 2.

The thick orange line shows the average of the band powers across delays for $|\tau| > 500$ ns (or $|k_\parallel|\gtrsim0.25\ h\ {\rm Mpc}^{-1}$).
This more clearly shows the marginal detection of a systematic at deep integration levels, which is more pronounced for Band 2, Field 2, as we would expect given our conclusions from \autoref{fig:integrated_sph}.
Nevertheless, this test shows that the low-level systematics do partially integrate when averaging across different baselines, as the mean of the band powers (i.e. the orange line) still shows a downward trend.
Integrating more data will help to understand at what dynamic range these systematics hit a plateau, if at all.

\begin{figure}
\centering
\includegraphics[width=\linewidth]{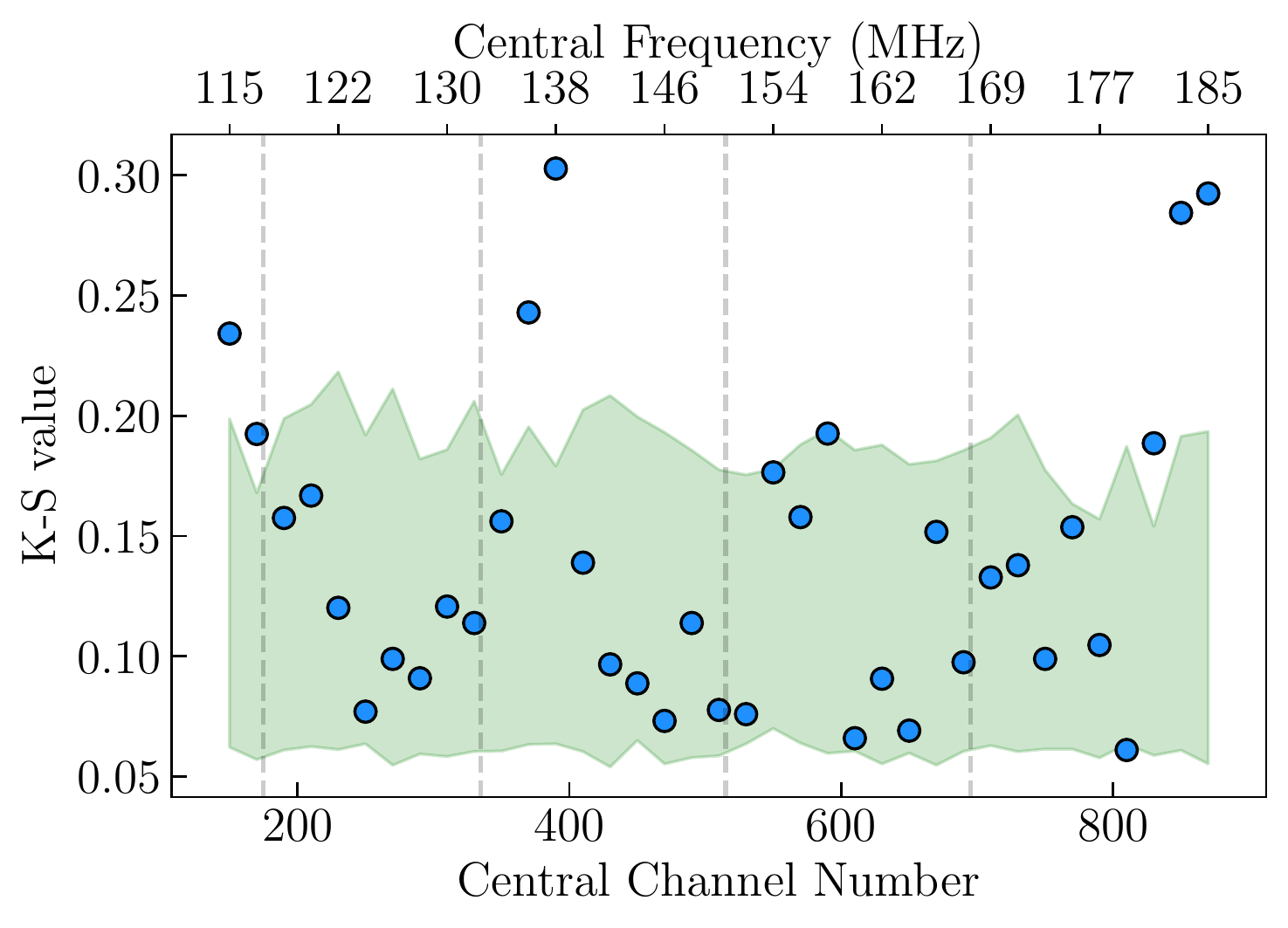}
\caption{Blue points show the K-S value between data and noise-only simulations. The green regions are 5th-95th percentile range of the K-S values from a set of noise-only simulations. The vertical dashed lines show the  spectral window chosen for the power spectrum analysis (\autoref{sec:qe}). }
\label{fig:spw_stability}
\centering
\end{figure}

\subsection{Stability with respect to frequency selection}
\label{sec:spw_stability}

Here we seek to assess the overall containment of foreground structure within the foreground horizon as a function of frequency.
To do this, we compute the delay spectrum of the data over a subband with fixed bandwidth and then iteratively shift the center of the subband through the full frequency range of the data, thereby probing for spectral discontinuities in the data at particular locations in frequency.
In order to test the consistency of the measurements with noise, we simulate mock visibilities consisting of Gaussian random noise with the same sampling and flagging patterns as the real data.
As a metric, we look at the mean band power within a delay range of $2000<|\tau|<4000$ nanoseconds for both the real data and simulated noise data sets.
By picking out the high delay regime, we are constructing a test that is sensitive to spurious spectral structure that would cause foreground power to leak significantly beyond the foreground horizon.
Similar to \autoref{sec:allan_variance}, we perform this test with the East-West 14.6 meter redundant baseline group to enable a large sample size, and focus on the LST ranges associated with Field 1.
Note that this test is performed after delay-based inpainting, where the data have been reconstructed in previously flagged channels.

We use two-sample Kolmogorov–Smirnov (K-S) test \citep{2s_ks} to examine if the underlying probability distribution of the measured band powers from the data and noise-only simulations are different.
Again, this process is repeated as we shift the spectral window up in frequency, where the window has a width of 100 channels, and is moved up 20 channels every iteration.
We compute the allowed range of the metric by taking the 5th and 95th percentiles of the two-sample KS test, having paired two independent noise simulations, and then repeating for all pair combinations of the noise simulations and averaging the result.
This interval is shown as the green shaded region in \autoref{fig:spw_stability}, which also shows the two-sample KS test between the data and the noise simulation (blue points). This shows that for most of the total bandwidth, the measured band powers at high delay are consistent with that of pure noise fluctuations.
The Band 1 and Band 2 spectral windows are shown as grey dashed lines.

The notable exception are the band edges, which are known to be sub-optimal to due tapering effects in the process of calibration, as well as the ORBCOMM band at $\sim138$ MHz, where a large swath of channels are routinely flagged, which is known to degrade the ability of the delay-based inpainting algorithm.

\begin{figure*}
\centering
\includegraphics[width=\linewidth]{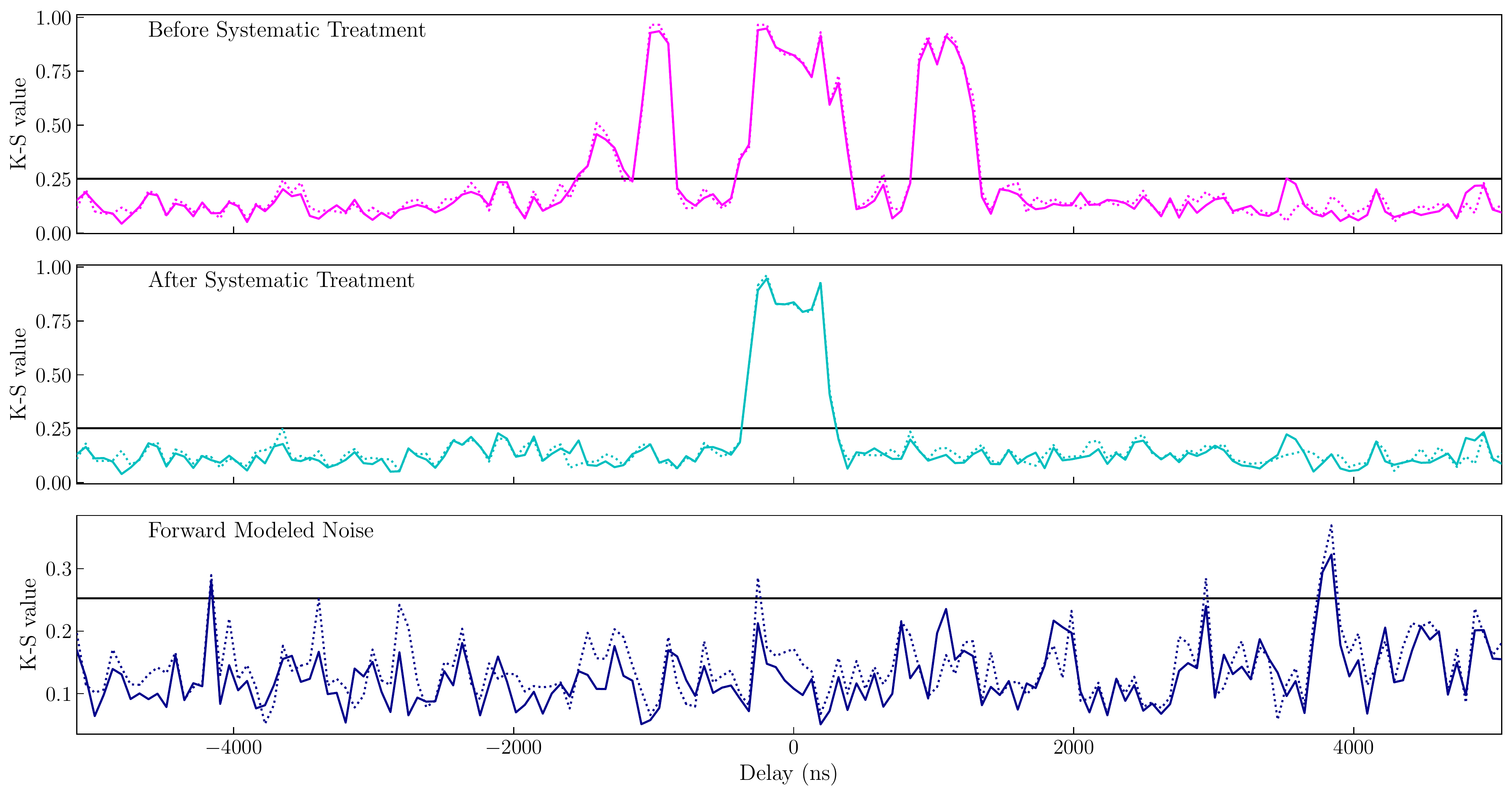}
\vspace{-5mm}
\caption{
One-sample K-S test at different delays, comparing CDFs of the bandpowers from the data, and the mean bootstrapped CDF with that expected from Gaussian random noise. The data without systematic subtraction (top panel) shows two clear regions, foreground and systematic dominated, where K-S values are clearly higher than the threshold, marked with the black horizontal line. Systematic subtraction (middle panel) continues to have high K-S values for foreground dominated region, while all other delay ranges seem consistent with thermal noise. As a reference, bottom panel shows the K-S values from the simulated dataset that consists of Gaussian random noise with the same variance as that of the data. Naturally, in this case, the K-S values are consistent with the reference distribution across all delays. Solid and dashed lines represent the results from K-S test on original band powers and mean of bootstrapped band powers respectively.}
\label{fig:systematics_null}
\end{figure*}

\subsection{Impact of systematic treatment}
\label{sec:systematics_null}

This null test is aimed at examining the consistency of the band powers with thermal noise before and after systematic treatment, discussed in more detail in \citet{DibbleeBarkman2021}.
We compute the delay power spectra for a single 14.6 meter, East-West oriented baseline this time spanning an LST range of 5 -- 5.4 hours overlapping with Field 2.
We estimate the cumulative probability distribution (CDF) of the band powers at different delays, constructing the CDF across the 30 time bins in the LST range studied here.
We then perform a one-sample KS test between the band power at each delay and the expected analytic distribution of the band powers under the assumption that the visibilities are drawn from random Gaussian noise.
Additionally, we also construct band power CDFs having resampled the time bins with replacement, which gives us a sense for the inherent variation of the CDF throughout the LST range.
We compute the one-sample KS test for each of the resampled CDFs, and then take their average.

The result is shown in \autoref{fig:systematics_null}, where the solid line shows the KS statistic from the un-resampled CDF as a function of delay mode and the dashed line shows the average KS statistic from the resampled CDF, showing good agreement between the two, suggesting that variation of the KS statistic as a function of time is minimal.
We repeat this test on the data before systematics treatment (to panel), after systematics treatment (middle panel), and also compute it on a pure-noise simulation for validation purposes (bottom panel).
The horizontal black line with a KS value of $\sim0.25$ is the analytically-computed 95th percentile of the KS values assuming the CDF is drawn from pure noise, which is validated by the results of the bottom panel showing the vast majority of the KS values falling below the solid black line.
What we see from the top and middle panels of \autoref{fig:systematics_null} shows a clear discrepancy between the data and the assumed noise distribution at intermediate delay modes before performign systematics treatment (top panel), which is not surprising given that we know those modes to be systematics dominated.
After systematics treatment (middle panel), these delay modes show good agreement with the noise model, except for at very low delays where we expect intrinsic foreground emission to dominate.

%%%%%%%%%%%%%%%%%%%%%%%%%%%%%
%%%%%%%%%%%% Summary %%%%%%%%%
%%%%%%%%%%%%%%%%%%%%%%%%%%%%%

\section{Summary}
\label{sec:summary}

In this work we have presented upper limits on the 21\,cm power spectrum from Phase I observations of the 50-element Hydrogen Epoch of Reionization Array (HERA) at redshifts 7.9 and 10.4.
The most sensitive $2\sigma$ limits achieved are $(30.76)^2$ mK$^2$ at $z=7.9$ and $k=0.192\ h\ {\rm Mpc}^{-1}$ and $(95.74)^2$ mK$^2$ at $z=10.4$ and $k=0.256\ h\ {\rm Mpc}^{-1}$.
At $z=7.9$, the limits presented here are the most sensitive to-date within the literature by roughly an order of magnitude.
Along with a series of paper describing the Phase I analysis pipeline, including redundant calibration \citep{Dillon2020}, absolute calibration \citep{Kern2020b}, systematic modeling \citep{Kern2019, Kern2020a}, uncertainty quantification \citep{Tan2021}, and pipeline validation \citep{Aguirre2021}, this work details the Phase I analysis and power spectrum pipelines, and discusses a series of statistical tests that show that the data are largely thermal noise limited for $k\ge0.5\ h\ {\rm Mpc}^{-1}$, whereas at lower $k$ modes the data show evidence for low-level systematics.
We speculate that these residual systematics could be due to radio frequency intereference, as well as possibly residual gain errors or residual instrumental coupling effects.
Future work exploring better RFI identification, as well as more comprehensive systematic models, will help to better understand the origin of these systematics.

Note that no explicit foreground subtraction or filtering was performed in the anlaysis.
Instead, our analysis emphasized strong control of spectral systematics in order to keep foreground structure largely contained within the foreground horizon.
Ultimately, this enabled the Phase I analysis pipeline achieved a dynamic range between the thermal noise floor and the peak foreground power of $10^9$ in power.
Future work will focus both on a more comprehensive analysis of low-level systematics, and the modeling and subtraction of foreground emission.
Hardware upgrades in the HERA Phase II system will potentially mitigate some of the observed Phase I systematics in the field, such as cross-coupling and reflection contamination.

The overall systematic uncertainties in this analysis come predominately from the uncertainty on the absolute flux density scale that is known to an accuracy of $\sim$10\%, which is pinned to the GLEAM catalogue \citep{Hurley-Walker2017}.
Furthermore, nightly drift in the gain amplitude is not corrected for in this work, which \citet{Kern2020b} estimate to be a roughly 3\% effect in the gains.
Signal loss arising from the analysis pipeline are explicitely derived and corrected for, but in total this amounts to less than a 15\% correction of the power spectrum amplitude.
Cosmic variance uncertainty on the 21\,cm power spectrum limits is expected to be 5.5\% ($1\sigma$) given the $uv$ sampling of the instrument for a 2 hour integration across LST \citep{Aguirre2021}.

Going forward, future Phase I analyses may include a $\sim$100-night dataset that theoretically stands to improve the sensitivity of our limits here by over a factor of 5, assuming the low $k$ systematics observed in this work can be further mitigated.
Additionally, Phase II commissioning and observations have already commenced, with a new front-end receiver spanning an increased bandwidth from 50 -- 250 MHz.
Overall, the analysis pipeline presented in this work has laid a framework for future analyses of HERA data as construction and commissioning is completed.

\section*{Acknowledgements}

This material is based upon work supported by the National Science Foundation under Grant Nos. 1636646 and 1836019 and institutional support from the HERA collaboration partners.
This research is funded in part by the Gordon and Betty Moore Foundation.
HERA is hosted by the South African Radio Astronomy Observatory, which is a facility of the National Research Foundation, an agency of the Department of Science and Innovation.
Parts of this research were supported by the Australian Research Council Centre of Excellence for All Sky Astrophysics in 3 Dimensions (ASTRO 3D), through project number CE170100013.
G.~Bernardi acknowledges funding from the INAF PRIN-SKA 2017 project 1.05.01.88.04 (FORECaST), support from the Ministero degli Affari Esteri della Cooperazione Internazionale - Direzione Generale per la Promozione del Sistema Paese Progetto di Grande Rilevanza ZA18GR02 and the National Research Foundation of South Africa (Grant Number 113121) as part of the ISARP RADIOSKY2020 Joint Research Scheme, from the Royal Society and the Newton Fund under grant NA150184 and from the National Research Foundation of South Africa (grant No. 103424).
P.~Bull acknowledges funding for part of this research from the European Research Council (ERC) under the European Union's Horizon 2020 research and innovation programme (Grant agreement No. 948764), and from STFC Grant ST/T000341/1.
E. ~de Lera Acedo acknowledges the funding support of the UKRI Science and Technology Facilities Council SKA grant.
J.S.~Dillon gratefully acknowledges the support of the NSF AAPF award \#1701536.
N.~Kern acknowledges support from the MIT Pappalardo fellowship.
A.~Liu acknowledges support from the New Frontiers in Research Fund Exploration grant program, the Canadian Institute for Advanced Research (CIFAR) Azrieli Global Scholars program, a Natural Sciences and Engineering Research Council of Canada (NSERC) Discovery Grant and a Discovery Launch Supplement, the Sloan Research Fellowship, and the William Dawson Scholarship at McGill.
We gratefully acknowledge an anonymous referee whose feedback improved the clarity of this work.

%%%%%%%%%%%%%%%%%%%%%%%%%%%%%
%%%%%%%%% Appendix A %%%%%%%%
%%%%%%%%%%%%%%%%%%%%%%%%%%%%%

\section*{Data}
\label{appendix:data}
The data from \autoref{tab:band1_limits}, \autoref{tab:band2_limits}, \autoref{fig:integrated_dsq}, and \autoref{fig:band2_wf} are publicly available and can be accessed at \url{https://reionization.org/science/public-data-release-1/}.

\section*{Software}
\label{appendix:software}
This analysis utilized custom-built software by the HERA Collaboration (\url{https://github.com/hera-team}) in addition to software built by both HERA members and collaborators (\url{https://github.com/RadioAstronomySoftwareGroup}).
This analysis also relied on publicly-accessible and open-sourced software, including \texttt{numpy} \citep{2020NumPy-Array}, \texttt{scipy} \citep{scipy2020}, and \texttt{astropy} \citep{astropy:2018}.

%% Bibliography %%
\bibliographystyle{aasjournal}
\bibliography{idr2_pspec}

\end{document}

%% file: author-list.tex
\correspondingauthor{Nicholas Kern}
\email{nkern@mit.edu}

\author{Zara  Abdurashidova}
\affiliation{Department of Astronomy, University of California, Berkeley, CA}

\author{James E. Aguirre}
\affiliation{Department of Physics and Astronomy, University of Pennsylvania, Philadelphia, PA}

\author{Paul  Alexander}
\affiliation{Cavendish Astrophysics, University of Cambridge, Cambridge, UK}

\author{Zaki S. Ali}    
\affiliation{Department of Astronomy, University of California, Berkeley, CA}

\author{Yanga  Balfour}
\affiliation{South African Radio Astronomy Observatory, Cape Town, South Africa}

\author{Adam P. Beardsley}
\altaffiliation{NSF Astronomy and Astrophysics Postdoctoral Fellow}
\affiliation{Department of Physics, Winona State University, Winona, MN}
\affiliation{School of Earth and Space Exploration, Arizona State University, Tempe, AZ}

\author{Gianni  Bernardi}
\affiliation{Department of Physics and Electronics, Rhodes University, PO Box 94, Grahamstown, 6140, South Africa}
\affiliation{INAF-Istituto di Radioastronomia, via Gobetti 101, 40129 Bologna, Italy}
\affiliation{South African Radio Astronomy Observatory, Cape Town, South Africa}

\author{Tashalee S. Billings}
\affiliation{Department of Physics and Astronomy, University of Pennsylvania, Philadelphia, PA}

\author{Judd D. Bowman}
\affiliation{School of Earth and Space Exploration, Arizona State University, Tempe, AZ}

\author{Richard F. Bradley}
\affiliation{National Radio Astronomy Observatory, Charlottesville, VA}

\author{Philip  Bull}
\affiliation{School of Physics \& Astronomy, Queen Mary University of London, London, UK}
\affiliation{Department of Physics and Astronomy, University of Western Cape, Cape Town 7535, South Africa}

\author{Jacob  Burba}
\affiliation{Department of Physics, Brown University, Providence, RI}

\author{Steve  Carey}
\affiliation{Cavendish Astrophysics, University of Cambridge, Cambridge, UK}

\author{Chris L. Carilli}
\affiliation{National Radio Astronomy Observatory, Socorro, NM}

\author{Carina  Cheng}
\affiliation{Department of Astronomy, University of California, Berkeley, CA}

\author{David R. DeBoer}
\affiliation{Department of Astronomy, University of California, Berkeley, CA}

\author{Matt  Dexter}
\affiliation{Department of Astronomy, University of California, Berkeley, CA}

\author{Eloy  de~Lera~Acedo}
\affiliation{Cavendish Astrophysics, University of Cambridge, Cambridge, UK}

\author{Taylor Dibblee-Barkman}
\affiliation{Department of Physics and McGill Space Institute, McGill University, 3600 University Street, Montreal, QC H3A 2T8, Canada}

\author{Joshua S. Dillon}
\altaffiliation{NSF Astronomy and Astrophysics Postdoctoral Fellow}
\affiliation{Department of Astronomy, University of California, Berkeley, CA}

\author{John  Ely}
\affiliation{Cavendish Astrophysics, University of Cambridge, Cambridge, UK}

\author{Aaron  Ewall-Wice}
\affiliation{Department of Astronomy, University of California, Berkeley, CA}

\author{Nicolas  Fagnoni}
\affiliation{Cavendish Astrophysics, University of Cambridge, Cambridge, UK}

\author{Randall  Fritz}
\affiliation{South African Radio Astronomy Observatory, Cape Town, South Africa}

\author{Steven R. Furlanetto}
\affiliation{Department of Physics and Astronomy, University of California, Los Angeles, CA}

\author{Kingsley  Gale-Sides}
\affiliation{Cavendish Astrophysics, University of Cambridge, Cambridge, UK}

\author{Brian  Glendenning}
\affiliation{National Radio Astronomy Observatory, Socorro, NM}

\author{Deepthi  Gorthi}
\affiliation{Department of Astronomy, University of California, Berkeley, CA}

\author{Bradley  Greig}
\affiliation{School of Physics, University of Melbourne, Parkville, VIC 3010, Australia}

\author{Jasper  Grobbelaar}
\affiliation{South African Radio Astronomy ObservatoryA, Cape Town, South Africa}

\author{Ziyaad  Halday}
\affiliation{South African Radio Astronomy Observatory, Cape Town, South Africa}

\author{Bryna J. Hazelton}
\affiliation{Department of Physics, University of Washington, Seattle, WA}
\affiliation{eScience Institute, University of Washington, Seattle, WA}

\author{Jacqueline N. Hewitt}
\affiliation{Department of Physics and MIT Kavli Institute, Massachusetts Institute of Technology, Cambridge, MA}

\author{Jack  Hickish}
\affiliation{Department of Astronomy, University of California, Berkeley, CA}

\author{Daniel C. Jacobs}
\affiliation{School of Earth and Space Exploration, Arizona State University, Tempe, AZ}

\author{Austin  Julius}
\affiliation{South African Radio Astronomy Observatory, Cape Town, South Africa}

\author{Nicholas S. Kern}
\affiliation{Department of Physics, Massachusetts Institute of Technology, Cambridge, MA}
\affiliation{Department of Astronomy, University of California, Berkeley, CA}

\author{Joshua  Kerrigan}
\affiliation{Department of Physics, Brown University, Providence, RI}

\author{Piyanat  Kittiwisit}
\affiliation{Department of Physics and Astronomy, University of Western Cape, Cape Town 7535, South Africa}

\author{Saul A. Kohn}
\affiliation{Department of Physics and Astronomy, University of Pennsylvania, Philadelphia, PA}

\author{Matthew  Kolopanis}
\affiliation{School of Earth and Space Exploration, Arizona State University, Tempe, AZ}

\author{Adam  Lanman}
\affiliation{Department of Physics, Brown University, Providence, RI}

\author{Paul  La~Plante}
\affiliation{Department of Astronomy, University of California, Berkeley, CA}
\affiliation{Department of Physics and Astronomy, University of Pennsylvania, Philadelphia, PA}

\author{Telalo  Lekalake}
\affiliation{South African Radio Astronomy Observatory, Cape Town, South Africa}

\author{David  Lewis}
\affiliation{School of Earth and Space Exploration, Arizona State University, Tempe, AZ}

\author{Adrian  Liu}
\affiliation{Department of Physics and McGill Space Institute, McGill University, 3600 University Street, Montreal, QC H3A 2T8, Canada}

\author{David  MacMahon}
\affiliation{Department of Astronomy, University of California, Berkeley, CA}

\author{Lourence  Malan}
\affiliation{South African Radio Astronomy Observatory, Cape Town, South Africa}

\author{Cresshim  Malgas}
\affiliation{South African Radio Astronomy Observatory, Cape Town, South Africa}

\author{Matthys  Maree}
\affiliation{South African Radio Astronomy Observatory, Cape Town, South Africa}

\author{Zachary E. Martinot}
\affiliation{Department of Physics and Astronomy, University of Pennsylvania, Philadelphia, PA}

\author{Eunice  Matsetela}
\affiliation{South African Radio Astronomy Observatory, Cape Town, South Africa}

\author{Andrei  Mesinger}
\affiliation{Scuola Normale Superiore, 56126 Pisa, PI, Italy}

\author{Mathakane  Molewa}
\affiliation{South African Radio Astronomy Observatory, Cape Town, South Africa}

\author{Miguel F. Morales}
\affiliation{Department of Physics, University of Washington, Seattle, WA}

\author{Tshegofalang  Mosiane}
\affiliation{South African Radio Astronomy Observatory, Cape Town, South Africa}

\author{Steven G. Murray}
\affiliation{School of Earth and Space Exploration, Arizona State University, Tempe, AZ}

\author{Abraham R. Neben}
\affiliation{Department of Physics, Massachusetts Institute of Technology, Cambridge, MA}

\author{Bojan  Nikolic}
\affiliation{Cavendish Astrophysics, University of Cambridge, Cambridge, UK}

\author{Chuneeta D. Nunhokee}
\affiliation{Department of Astronomy, University of California, Berkeley, CA}

\author{Aaron R. Parsons}
\affiliation{Department of Astronomy, University of California, Berkeley, CA}

\author{Nipanjana  Patra}
\affiliation{Department of Astronomy, University of California, Berkeley, CA}

\author{Robert  Pascua}
\affiliation{Department of Physics and McGill Space Institute, McGill University, 3600 University Street, Montreal, QC H3A 2T8, Canada}
\affiliation{Department of Astronomy, University of California, Berkeley, CA}

\author{Samantha  Pieterse}
\affiliation{South African Radio Astronomy Observatory, Cape Town, South Africa}

\author{Jonathan C. Pober}
\affiliation{Department of Physics, Brown University, Providence, RI}

\author{Nima  Razavi-Ghods}
\affiliation{Cavendish Astrophysics, University of Cambridge, Cambridge, UK}

\author{Jon  Ringuette}
\affiliation{Department of Physics, University of Washington, Seattle, WA}

\author{James  Robnett}
\affiliation{National Radio Astronomy Observatory, Socorro, NM}

\author{Kathryn  Rosie}
\affiliation{South African Radio Astronomy Observatory, Cape Town, South Africa}

\author{Peter  Sims}
\affiliation{Department of Physics, Brown University, Providence, RI}

\author{Saurabh Singh}
\affiliation{Department of Physics and McGill Space Institute, McGill University, 3600 University Street, Montreal, QC H3A 2T8, Canada}

\author{Craig  Smith}
\affiliation{South African Radio Astronomy Observatory, Cape Town, South Africa}

\author{Angelo  Syce}
\affiliation{South African Radio Astronomy Observatory, Cape Town, South Africa}

\author{Nithyanandan  Thyagarajan}
\altaffiliation{Jansky Fellow of the National Radio Astronomy Observatory}
\affiliation{National Radio Astronomy Observatory, Socorro, NM}
\affiliation{School of Earth and Space Exploration, Arizona State University, Tempe, AZ}

\author{Peter K.~G. Williams}
\affiliation{Center for Astrophysics | Harvard \& Smithsonian, Cambridge, MA}
\affiliation{American Astronomical Society, Washington, DC}

\author{Haoxuan  Zheng}
\affiliation{Department of Physics, Massachusetts Institute of Technology, Cambridge, MA}

%% file: idr2_pspec.bbl
\begin{thebibliography}{}
\expandafter\ifx\csname natexlab\endcsname\relax\def\natexlab#1{#1}\fi
\providecommand{\url}[1]{\href{#1}{#1}}
\providecommand{\dodoi}[1]{doi:~\href{http://doi.org/#1}{\nolinkurl{#1}}}
\providecommand{\doeprint}[1]{\href{http://ascl.net/#1}{\nolinkurl{http://ascl.net/#1}}}
\providecommand{\doarXiv}[1]{\href{https://arxiv.org/abs/#1}{\nolinkurl{https://arxiv.org/abs/#1}}}

\bibitem[{{Aguirre} {et~al.}(in prep.){Aguirre}, {Ali}, \& the
  HERA~Collaboration}]{Aguirre2021}
{Aguirre}, J., {Ali}, Z.~S., \& the HERA~Collaboration. in prep.

\bibitem[{Ali {et~al.}(2015)Ali, Parsons, Zheng, Pober, Liu, Aguirre, Bradley,
  Bernardi, Carilli, Cheng, DeBoer, Dexter, Grobbelaar, Horrell, Jacobs, Klima,
  MacMahon, Maree, Moore, Razavi, Stefan, Walbrugh, \& Walker}]{Ali2015}
Ali, Z., Parsons, A., Zheng, H., {et~al.} 2015, \apj, 809, 61,
  \dodoi{10.1088/0004-637X/809/1/61}

\bibitem[{Ali {et~al.}(2018)Ali, Parsons, Zheng, Pober, Liu, Aguirre, Bradley,
  Bernardi, Carilli, Cheng, DeBoer, Dexter, Grobbelaar, Horrell, Jacobs, Klima,
  MacMahon, Maree, Moore, Razavi, Stefan, Walbrugh, \& Walker}]{Ali2015erratum}
---. 2018, ApJ (Erratum), 863, 201, \dodoi{10.3847/1538-4357/aad7b4}

\bibitem[{Asad {et~al.}(2015)Asad, Koopmans, Jeli{\'{c}}, Pandey, Ghosh,
  Abdalla, Bernardi, Brentjens, de~Bruyn, Bus, Ciardi, Chapman, Daiboo,
  Fernandez, Harker, Iliev, Jensen, Martinez-Rubi, Mellema, Mevius, Offringa,
  Patil, Schaye, Thomas, van~der Tol, Vedantham, Yatawatta, \&
  Zaroubi}]{Asad2015}
Asad, K., Koopmans, L., Jeli{\'{c}}, V., {et~al.} 2015, \mnras, 451, 3709,
  \dodoi{10.1093/mnras/stv1107}

\bibitem[{{Asad} {et~al.}(2018){Asad}, {Koopmans}, {Jeli{\'c}}, {de Bruyn},
  {Pandey}, \& {Gehlot}}]{Asad2018}
{Asad}, K.~M.~B., {Koopmans}, L.~V.~E., {Jeli{\'c}}, V., {et~al.} 2018, \mnras,
  476, 3051, \dodoi{10.1093/mnras/sty258}

\bibitem[{{Asad} {et~al.}(2016){Asad}, {Koopmans}, {Jeli{\'c}}, {Ghosh},
  {Abdalla}, {Brentjens}, {de Bruyn}, {Ciardi}, {Gehlot}, {Iliev}, {Mevius},
  {Pandey}, {Yatawatta}, \& {Zaroubi}}]{Asad2016}
---. 2016, \mnras, 462, 4482, \dodoi{10.1093/mnras/stw1863}

\bibitem[{{Astropy Collaboration} {et~al.}(2018){Astropy Collaboration},
  {Price-Whelan}, {Sip{H{o}}cz}, {G{\"u}nther}, {Lim}, {Crawford}, {Conseil},
  {Shupe}, {Craig}, {Dencheva}, {Ginsburg}, {Vand erPlas}, {Bradley},
  {P{\'e}rez-Su{\'a}rez}, {de Val-Borro}, {Aldcroft}, {Cruz}, {Robitaille},
  {Tollerud}, {Ardelean}, {Babej}, {Bach}, {Bachetti}, {Bakanov}, {Bamford},
  {Barentsen}, {Barmby}, {Baumbach}, {Berry}, {Biscani}, {Boquien}, {Bostroem},
  {Bouma}, {Brammer}, {Bray}, {Breytenbach}, {Buddelmeijer}, {Burke},
  {Calderone}, {Cano Rodr{\'\i}guez}, {Cara}, {Cardoso}, {Cheedella}, {Copin},
  {Corrales}, {Crichton}, {D'Avella}, {Deil}, {Depagne}, {Dietrich}, {Donath},
  {Droettboom}, {Earl}, {Erben}, {Fabbro}, {Ferreira}, {Finethy}, {Fox},
  {Garrison}, {Gibbons}, {Goldstein}, {Gommers}, {Greco}, {Greenfield},
  {Groener}, {Grollier}, {Hagen}, {Hirst}, {Homeier}, {Horton}, {Hosseinzadeh},
  {Hu}, {Hunkeler}, {Ivezi{\'c}}, {Jain}, {Jenness}, {Kanarek}, {Kendrew},
  {Kern}, {Kerzendorf}, {Khvalko}, {King}, {Kirkby}, {Kulkarni}, {Kumar},
  {Lee}, {Lenz}, {Littlefair}, {Ma}, {Macleod}, {Mastropietro}, {McCully},
  {Montagnac}, {Morris}, {Mueller}, {Mumford}, {Muna}, {Murphy}, {Nelson},
  {Nguyen}, {Ninan}, {N{\"o}the}, {Ogaz}, {Oh}, {Parejko}, {Parley}, {Pascual},
  {Patil}, {Patil}, {Plunkett}, {Prochaska}, {Rastogi}, {Reddy Janga},
  {Sabater}, {Sakurikar}, {Seifert}, {Sherbert}, {Sherwood-Taylor}, {Shih},
  {Sick}, {Silbiger}, {Singanamalla}, {Singer}, {Sladen}, {Sooley},
  {Sornarajah}, {Streicher}, {Teuben}, {Thomas}, {Tremblay}, {Turner},
  {Terr{\'o}n}, {van Kerkwijk}, {de la Vega}, {Watkins}, {Weaver}, {Whitmore},
  {Woillez}, {Zabalza}, \& {Astropy Contributors}}]{astropy:2018}
{Astropy Collaboration}, {Price-Whelan}, A.~M., {Sip{H{o}}cz}, B.~M., {et~al.}
  2018, \aj, 156, 123, \dodoi{10.3847/1538-3881/aabc4f}

\bibitem[{{Barry} {et~al.}(2019{\natexlab{a}}){Barry}, {Beardsley}, {Byrne},
  {Hazelton}, {Morales}, {Pober}, \& {Sullivan}}]{Barry2019a}
{Barry}, N., {Beardsley}, A.~P., {Byrne}, R., {et~al.} 2019{\natexlab{a}},
  \pasa, 36, e026, \dodoi{10.1017/pasa.2019.21}

\bibitem[{{Barry} {et~al.}(2019{\natexlab{b}}){Barry}, {Wilensky}, {Trott},
  {Pindor}, {Beardsley}, {Hazelton}, {Sullivan}, {Morales}, {Pober}, {Line},
  {Greig}, {Byrne}, {Lanman}, {Li}, {Jordan}, {Joseph}, {McKinley}, {Rahimi},
  {Yoshiura}, {Bowman}, {Gaensler}, {Hewitt}, {Jacobs}, {Mitchell}, {Udaya
  Shankar}, {Sethi}, {Subrahmanyan}, {Tingay}, {Webster}, \&
  {Wyithe}}]{Barry2019b}
{Barry}, N., {Wilensky}, M., {Trott}, C.~M., {et~al.} 2019{\natexlab{b}}, \apj,
  884, 1, \dodoi{10.3847/1538-4357/ab40a8}

\bibitem[{Beardsley {et~al.}(2016)Beardsley, Hazelton, Sullivan, Carroll,
  Barry, Rahimi, Pindor, Trott, Line, Jacobs, Morales, Pober, Bernardi, Bowman,
  Busch, Briggs, Cappallo, Corey, de~Oliveira-Costa, Dillon, Emrich,
  Ewall-Wice, Feng, Gaensler, Goeke, Greenhill, Hewitt, Hurley-Walker,
  Johnston-Hollitt, Kaplan, Kasper, Kim, Kratzenberg, Lenc, Loeb, Lonsdale,
  Lynch, McKinley, McWhirter, Mitchell, Morgan, Neben, Thyagarajan, Oberoi,
  Offringa, Ord, Paul, Prabu, Procopio, Riding, Rogers, Roshi, {Udaya Shankar},
  Sethi, Srivani, Subrahmanyan, Tegmark, Tingay, Waterson, Wayth, Webster,
  Whitney, Williams, Williams, Wu, \& Wyithe}]{Beardsley2016}
Beardsley, A., Hazelton, B., Sullivan, I., {et~al.} 2016, \apj, 833, 102,
  \dodoi{10.3847/1538-4357/833/1/102}

\bibitem[{{Becker} {et~al.}(2015){Becker}, {Bolton}, {Madau}, {Pettini},
  {Ryan-Weber}, \& {Venemans}}]{Becker15}
{Becker}, G.~D., {Bolton}, J.~S., {Madau}, P., {et~al.} 2015, \mnras, 447,
  3402, \dodoi{10.1093/mnras/stu2646}

\bibitem[{{Bernardi} {et~al.}(2016){Bernardi}, {Zwart}, {Price}, {Greenhill},
  {Mesinger}, {Dowell}, {Eftekhari}, {Ellingson}, {Kocz}, \&
  {Schinzel}}]{Bernardi2016}
{Bernardi}, G., {Zwart}, J.~T.~L., {Price}, D., {et~al.} 2016, \mnras, 461,
  2847, \dodoi{10.1093/mnras/stw1499}

\bibitem[{Blackman \& Tukey(1958)}]{Blackman1958}
Blackman, R.~B., \& Tukey, J.~W. 1958, Bell System Technical Journal, 37, 185,
  \dodoi{10.1002/j.1538-7305.1958.tb03874.x}

\bibitem[{{Bolton} {et~al.}(2011){Bolton}, {Haehnelt}, {Warren}, {Hewett},
  {Mortlock}, {Venemans}, {McMahon}, \& {Simpson}}]{Bolton11}
{Bolton}, J.~S., {Haehnelt}, M.~G., {Warren}, S.~J., {et~al.} 2011, \mnras,
  416, L70, \dodoi{10.1111/j.1745-3933.2011.01100.x}

\bibitem[{{Boonstra} \& {van der Veen}(2003)}]{Boonstra2003}
{Boonstra}, A., \& {van der Veen}, A. 2003, IEEE Transactions on Signal
  Processing, 51, 25, \dodoi{10.1109/TSP.2002.806588}

\bibitem[{{Bosman} {et~al.}(2018){Bosman}, {Fan}, {Jiang}, {Reed}, {Matsuoka},
  {Becker}, \& {Haehnelt}}]{Bosman18}
{Bosman}, S. E.~I., {Fan}, X., {Jiang}, L., {et~al.} 2018, \mnras, 479, 1055,
  \dodoi{10.1093/mnras/sty1344}

\bibitem[{{Bowman} {et~al.}(2018){Bowman}, {Rogers}, {Monsalve}, {Mozdzen}, \&
  {Mahesh}}]{Bowman2018}
{Bowman}, J.~D., {Rogers}, A.~E.~E., {Monsalve}, R.~A., {Mozdzen}, T.~J., \&
  {Mahesh}, N. 2018, \nat, 555, 67, \dodoi{10.1038/nature25792}

\bibitem[{{Bradley} {et~al.}(2019){Bradley}, {Tauscher}, {Rapetti}, \&
  {Burns}}]{Bradley2019}
{Bradley}, R.~F., {Tauscher}, K., {Rapetti}, D., \& {Burns}, J.~O. 2019, \apj,
  874, 153, \dodoi{10.3847/1538-4357/ab0d8b}

\bibitem[{{Brentjens} \& {de Bruyn}(2005)}]{Brentjens2005}
{Brentjens}, M.~A., \& {de Bruyn}, A.~G. 2005, \aap, 441, 1217,
  \dodoi{10.1051/0004-6361:20052990}

\bibitem[{{Byrne} {et~al.}(2019){Byrne}, {Morales}, {Hazelton}, {Li}, {Barry},
  {Beardsley}, {Joseph}, {Pober}, {Sullivan}, \& {Trott}}]{Byrne2019}
{Byrne}, R., {Morales}, M.~F., {Hazelton}, B., {et~al.} 2019, \apj, 875, 70,
  \dodoi{10.3847/1538-4357/ab107d}

\bibitem[{{Carilli} {et~al.}(2018){Carilli}, {Nikolic}, {Thyagarayan}, \&
  {Gale-Sides}}]{Carilli2019}
{Carilli}, C.~L., {Nikolic}, B., {Thyagarayan}, N., \& {Gale-Sides}, K. 2018,
  Radio Science, 53, 845, \dodoi{10.1029/2018RS006537}

\bibitem[{{Caruana} {et~al.}(2014){Caruana}, {Bunker}, {Wilkins}, {Stanway},
  {Lorenzoni}, {Jarvis}, \& {Ebert}}]{Caruana14}
{Caruana}, J., {Bunker}, A.~J., {Wilkins}, S.~M., {et~al.} 2014, \mnras, 443,
  2831, \dodoi{10.1093/mnras/stu1341}

\bibitem[{{Cheng} {et~al.}(2018){Cheng}, {Parsons}, {Kolopanis}, {Jacobs},
  {Liu}, {Kohn}, {Aguirre}, {Pober}, {Ali}, {Bernardi}, {Bradley}, {Carilli},
  {DeBoer}, {Dexter}, {Dillon}, {Klima}, {MacMahon}, {Moore}, {Nunhokee},
  {Walbrugh}, \& {Walker}}]{Cheng2018}
{Cheng}, C., {Parsons}, A.~R., {Kolopanis}, M., {et~al.} 2018, \apj, 868, 26,
  \dodoi{10.3847/1538-4357/aae833}

\bibitem[{{Choudhuri} {et~al.}(2021){Choudhuri}, {Bull}, \&
  {Garsden}}]{Choudhuri2020}
{Choudhuri}, S., {Bull}, P., \& {Garsden}, H. 2021, arXiv e-prints,
  arXiv:2101.02684.
\newblock \doarXiv{2101.02684}

\bibitem[{{Choudhury} {et~al.}(2021){Choudhury}, {Paranjape}, \&
  {Bosman}}]{Choudhury20}
{Choudhury}, T.~R., {Paranjape}, A., \& {Bosman}, S. E.~I. 2021, \mnras,
  \dodoi{10.1093/mnras/stab045}

\bibitem[{{Ciardi} \& {Ferrara}(2005)}]{Ciardi2005}
{Ciardi}, B., \& {Ferrara}, A. 2005, \ssr, 116, 625,
  \dodoi{10.1007/s11214-005-3592-0}

\bibitem[{{Condon} \& {Matthews}(2018)}]{Condon2018}
{Condon}, J.~J., \& {Matthews}, A.~M. 2018, \pasp, 130, 073001,
  \dodoi{10.1088/1538-3873/aac1b2}

\bibitem[{{Datta} {et~al.}(2014){Datta}, {Jensen}, {Majumdar}, {Mellema},
  {Iliev}, {Mao}, {Shapiro}, \& {Ahn}}]{Datta2014}
{Datta}, K.~K., {Jensen}, H., {Majumdar}, S., {et~al.} 2014, \mnras, 442, 1491,
  \dodoi{10.1093/mnras/stu927}

\bibitem[{{Davies} {et~al.}(2018){Davies}, {Hennawi}, {Ba{\~n}ados},
  {Luki{\'c}}, {Decarli}, {Fan}, {Farina}, {Mazzucchelli}, {Rix}, {Venemans},
  {Walter}, {Wang}, \& {Yang}}]{davies18}
{Davies}, F.~B., {Hennawi}, J.~F., {Ba{\~n}ados}, E., {et~al.} 2018, \apj, 864,
  142, \dodoi{10.3847/1538-4357/aad6dc}

\bibitem[{{de Oliveira-Costa} {et~al.}(2008){de Oliveira-Costa}, {Tegmark},
  {Gaensler}, {Jonas}, {Landecker}, \& {Reich}}]{Oliveira2008}
{de Oliveira-Costa}, A., {Tegmark}, M., {Gaensler}, B.~M., {et~al.} 2008,
  \mnras, 388, 247, \dodoi{10.1111/j.1365-2966.2008.13376.x}

\bibitem[{DeBoer {et~al.}(2017)DeBoer, Parsons, Aguirre, Alexander, Ali,
  Beardsley, Bernardi, Bowman, Bradley, Carilli, Cheng, {de Lera Acedo},
  Dillon, Ewall-Wice, Fadana, Fagnoni, Fritz, Furlanetto, Glendenning, Greig,
  Grobbelaar, Hazelton, Hewitt, Hickish, Jacobs, Julius, Kariseb, Kohn,
  Lekalake, Liu, Loots, MacMahon, Malan, Malgas, Maree, Martinot, Mathison,
  Matsetela, Mesinger, Morales, Neben, Patra, Pieterse, Pober, Razavi-Ghods,
  Ringuette, Robnett, Rosie, Sell, Smith, Syce, Tegmark, Thyagarajan, Williams,
  \& Zheng}]{DeBoer2017}
DeBoer, D., Parsons, A., Aguirre, J., {et~al.} 2017, \pasp, 129, 45001,
  \dodoi{10.1088/1538-3873/129/974/045001}

\bibitem[{{Dibblee-Barkman} \& {Singh}(2021)}]{DibbleeBarkman2021}
{Dibblee-Barkman}, T., \& {Singh}, S. 2021, {KS Testing on HERA Data}, Tech.
  rep.

\bibitem[{Dillon \& Parsons(2016)}]{Dillon2016}
Dillon, J., \& Parsons, A. 2016, \apj, 826, 181,
  \dodoi{10.3847/0004-637X/826/2/181}

\bibitem[{Dillon {et~al.}(2014)Dillon, Liu, Williams, Hewitt, Tegmark, Morgan,
  Levine, Morales, Tingay, Bernardi, Bowman, Briggs, Cappallo, Emrich,
  Mitchell, Oberoi, Prabu, Wayth, \& Webster}]{Dillon2014}
Dillon, J., Liu, A., Williams, C., {et~al.} 2014, \prd, 89, 23002,
  \dodoi{10.1103/PhysRevD.89.023002}

\bibitem[{Dillon {et~al.}(2015)Dillon, Neben, Hewitt, Tegmark, Barry,
  Beardsley, Bowman, Briggs, Carroll, de~Oliveira-Costa, Ewall-Wice, Feng,
  Greenhill, Hazelton, Hernquist, Hurley-Walker, Jacobs, Kim, Kittiwisit, Lenc,
  Line, Loeb, McKinley, Mitchell, Morales, Offringa, Paul, Pindor, Pober,
  Procopio, Riding, Sethi, Shankar, Subrahmanyan, Sullivan, Thyagarajan,
  Tingay, Trott, Wayth, Webster, Wyithe, Bernardi, Cappallo, Deshpande,
  Johnston-Hollitt, Kaplan, Lonsdale, McWhirter, Morgan, Oberoi, Ord, Prabu,
  Srivani, Williams, \& Williams}]{Dillon2015b}
Dillon, J., Neben, A., Hewitt, J., {et~al.} 2015, \prd, 91, 123011,
  \dodoi{10.1103/PhysRevD.91.123011}

\bibitem[{{Dillon} {et~al.}(2018){Dillon}, {Kohn}, {Parsons}, {Aguirre}, {Ali},
  {Bernardi}, {Kern}, {Li}, {Liu}, {Nunhokee}, \& {Pober}}]{Dillon2018}
{Dillon}, J.~S., {Kohn}, S.~A., {Parsons}, A.~R., {et~al.} 2018, \mnras, 477,
  5670, \dodoi{10.1093/mnras/sty1060}

\bibitem[{{Dillon} {et~al.}(2020){Dillon}, {Lee}, {Ali}, {Parsons}, {Orosz},
  {Devi Nunhokee}, {La Plante}, {Beardsley}, {Kern}, {Abdurashidova},
  {Aguirre}, {Alexander}, {Balfour}, {Bernardi}, {Billings}, {Bowman},
  {Bradley}, {Bull}, {Burba}, {Carey}, {Carilli}, {Cheng}, {DeBoer}, {Dexter},
  {de Lera Acedo}, {Ely}, {Ewall-Wice}, {Fagnoni}, {Fritz}, {Furlanetto},
  {Gale-Sides}, {Glendenning}, {Gorthi}, {Greig}, {Grobbelaar}, {Halday},
  {Hazelton}, {Hewitt}, {Hickish}, {Jacobs}, {Julius}, {Kerrigan},
  {Kittiwisit}, {Kohn}, {Kolopanis}, {Lanman}, {Lekalake}, {Lewis}, {Liu},
  {Ma}, {MacMahon}, {Malan}, {Malgas}, {Maree}, {Martinot}, {Matsetela},
  {Mesinger}, {Molewa}, {Morales}, {Mosiane}, {Murray}, {Neben}, {Nikolic},
  {Pascua}, {Patra}, {Pieterse}, {Pober}, {Razavi-Ghods}, {Ringuette},
  {Robnett}, {Rosie}, {Santos}, {Sims}, {Smith}, {Syce}, {Tegmark},
  {Thyagarajan}, {Williams}, \& {Zheng}}]{Dillon2020}
{Dillon}, J.~S., {Lee}, M., {Ali}, Z.~S., {et~al.} 2020, arXiv e-prints,
  arXiv:2003.08399.
\newblock \doarXiv{2003.08399}

\bibitem[{{Douspis} {et~al.}(2015){Douspis}, {Aghanim}, {Ili{\'c}}, \&
  {Langer}}]{Douspis15}
{Douspis}, M., {Aghanim}, N., {Ili{\'c}}, S., \& {Langer}, M. 2015, \aap, 580,
  L4, \dodoi{10.1051/0004-6361/201526543}

\bibitem[{{Eastwood} {et~al.}(2019){Eastwood}, {Anderson}, {Monroe},
  {Hallinan}, {Catha}, {Dowell}, {Garsden}, {Greenhill}, {Hicks}, {Kocz},
  {Price}, {Schinzel}, {Vedantham}, \& {Wang}}]{Eastwood2019}
{Eastwood}, M.~W., {Anderson}, M.~M., {Monroe}, R.~M., {et~al.} 2019, \aj, 158,
  84, \dodoi{10.3847/1538-3881/ab2629}

\bibitem[{Ewall-Wice {et~al.}(2016{\natexlab{a}})Ewall-Wice, Hewitt, Mesinger,
  Dillon, Liu, \& Pober}]{Ewall-Wice2016a}
Ewall-Wice, A., Hewitt, J., Mesinger, A., {et~al.} 2016{\natexlab{a}}, \mnras,
  458, 2710, \dodoi{10.1093/mnras/stw452}

\bibitem[{Ewall-Wice {et~al.}(2016{\natexlab{b}})Ewall-Wice, Dillon, Hewitt,
  Loeb, Mesinger, Neben, Offringa, Tegmark, Barry, Beardsley, Bernardi, Bowman,
  Briggs, Cappallo, Carroll, Corey, de~Oliveira-Costa, Emrich, Feng, Gaensler,
  Goeke, Greenhill, Hazelton, Hurley-Walker, Johnston-Hollitt, Jacobs, Kaplan,
  Kasper, Kim, Kratzenberg, Lenc, Line, Lonsdale, Lynch, McKinley, McWhirter,
  Mitchell, Morales, Morgan, Thyagarajan, Oberoi, Ord, Paul, Pindor, Pober,
  Prabu, Procopio, Riding, Rogers, Roshi, Shankar, Sethi, Srivani,
  Subrahmanyan, Sullivan, Tingay, Trott, Waterson, Wayth, Webster, Whitney,
  Williams, Williams, Wu, \& Wyithe}]{Ewall-Wice2016b}
Ewall-Wice, A., Dillon, J., Hewitt, J., {et~al.} 2016{\natexlab{b}}, \mnras,
  460, 4320, \dodoi{10.1093/mnras/stw1022}

\bibitem[{{Ewall-Wice} {et~al.}(2016){Ewall-Wice}, {Bradley}, {Deboer},
  {Hewitt}, {Parsons}, {Aguirre}, {Ali}, {Bowman}, {Cheng}, {Neben}, {Patra},
  {Thyagarajan}, {Venter}, {de Lera Acedo}, {Dillon}, {Dickenson}, {Doolittle},
  {Egan}, {Hedrick}, {Klima}, {Kohn}, {Schaffner}, {Shelton}, {Saliwanchik},
  {Taylor}, {Taylor}, {Tegmark}, \& {Wirt}}]{Ewall-Wice2016c}
{Ewall-Wice}, A., {Bradley}, R., {Deboer}, D., {et~al.} 2016, \apj, 831, 196,
  \dodoi{10.3847/0004-637X/831/2/196}

\bibitem[{{Ewall-Wice} {et~al.}(2021){Ewall-Wice}, {Kern}, {Dillon}, {Liu},
  {Parsons}, {Singh}, {Lanman}, {Plante}, {Fagnoni}, {Acedo}, {DeBoer},
  {Nunhokee}, {Bull}, {Chang}, {Lazio}, {Aguirre}, \&
  {Weinberg}}]{Ewall-Wice2021}
{Ewall-Wice}, A., {Kern}, N., {Dillon}, J.~S., {et~al.} 2021, \mnras, 500,
  5195, \dodoi{10.1093/mnras/staa3293}

\bibitem[{{Fagnoni} {et~al.}(2020){Fagnoni}, {de Lera Acedo}, {Drought},
  {DeBoer}, {Riley}, {Razavi-Ghods}, {Carey}, \& {Parsons}}]{Fagnoni2020}
{Fagnoni}, N., {de Lera Acedo}, E., {Drought}, N., {et~al.} 2020, arXiv
  e-prints, arXiv:2009.07939.
\newblock \doarXiv{2009.07939}

\bibitem[{{Fagnoni} {et~al.}(2021){Fagnoni}, {de Lera Acedo}, {DeBoer},
  {Abdurashidova}, {Aguirre}, {Alexander}, {Ali}, {Balfour}, {Beardsley},
  {Bernardi}, {Billings}, {Bowman}, {Bradley}, {Bull}, {Burba}, {Carilli},
  {Cheng}, {Dexter}, {Dillon}, {Ewall-Wice}, {Fritz}, {Furlanetto},
  {Gale-Sides}, {Glendenning}, {Gorthi}, {Greig}, {Grobbelaar}, {Halday},
  {Hazelton}, {Hewitt}, {Hickish}, {Jacobs}, {Josaitis}, {Julius}, {Kern},
  {Kerrigan}, {Kim}, {Kittiwisit}, {Kohn}, {Kolopanis}, {Lanman}, {Plante},
  {Lekalake}, {Liu}, {MacMahon}, {Malan}, {Malgas}, {Maree}, {Martinot},
  {Matsetela}, {Mena Parra}, {Mesinger}, {Molewa}, {Morales}, {Mosiane},
  {Neben}, {Nikolic}, {Parsons}, {Patra}, {Pieterse}, {Pober}, {Razavi-Ghods},
  {Robnett}, {Rosie}, {Sims}, {Smith}, {Syce}, {Thyagarajan}, {Williams}, \&
  {Zheng}}]{Fagnoni2021}
{Fagnoni}, N., {de Lera Acedo}, E., {DeBoer}, D.~R., {et~al.} 2021, \mnras,
  500, 1232, \dodoi{10.1093/mnras/staa3268}

\bibitem[{Furlanetto {et~al.}(2006)Furlanetto, Oh, \& Briggs}]{Furlanetto2006c}
Furlanetto, S., Oh, S., \& Briggs, F. 2006, \physrep, 433, 181,
  \dodoi{10.1016/j.physrep.2006.08.002}

\bibitem[{{Gehlot} {et~al.}(2019){Gehlot}, {Mertens}, {Koopmans}, {Brentjens},
  {Zaroubi}, {Ciardi}, {Ghosh}, {Hatef}, {Iliev}, {Jeli{\'c}}, {}, {Kooistra},
  {Krause}, {Mellema}, {Mevius}, {Mitra}, {Offringa}, {Pandey}, {Sardarabadi},
  {Schaye}, {Silva}, {Vedantham}, \& {Yatawatta}}]{Gehlot2019}
{Gehlot}, B.~K., {Mertens}, F.~G., {Koopmans}, L.~V.~E., {et~al.} 2019, \mnras,
  488, 4271, \dodoi{10.1093/mnras/stz1937}

\bibitem[{{Ghosh} {et~al.}(2012){Ghosh}, {Prasad}, {Bharadwaj}, {Ali}, \&
  {Chengalur}}]{Ghosh2012}
{Ghosh}, A., {Prasad}, J., {Bharadwaj}, S., {Ali}, S.~S., \& {Chengalur}, J.~N.
  2012, \mnras, 426, 3295, \dodoi{10.1111/j.1365-2966.2012.21889.x}

\bibitem[{{Ghosh} {et~al.}(2020){Ghosh}, {Mertens}, {Bernardi}, {Santos},
  {Kern}, {Carilli}, {Grobler}, {Koopmans}, {Jacobs}, {Liu}, {Parsons},
  {Morales}, {Aguirre}, {Dillon}, {Hazelton}, {Smirnov}, {Gehlot}, {Matika},
  {Alexander}, {Ali}, {Beardsley}, {Benefo}, {Billings}, {Bowman}, {Bradley},
  {Cheng}, {Chichura}, {DeBoer}, {Acedo}, {Ewall-Wice}, {Fadana}, {Fagnoni},
  {Fortino}, {Fritz}, {Furlanetto}, {Gallardo}, {Glendenning}, {Gorthi},
  {Greig}, {Grobbelaar}, {Hickish}, {Josaitis}, {Julius}, {Igarashi},
  {Kariseb}, {Kohn}, {Kolopanis}, {Lekalake}, {Loots}, {MacMahon}, {Malan},
  {Malgas}, {Maree}, {Martinot}, {Mathison}, {Matsetela}, {Mesinger}, {Neben},
  {Nikolic}, {Nunhokee}, {Patra}, {Pieterse}, {Razavi-Ghods}, {Ringuette},
  {Robnett}, {Rosie}, {Sell}, {Smith}, {Syce}, {Tegmark}, {Thyagarajan},
  {Williams}, \& {Zheng}}]{Ghosh2020}
{Ghosh}, A., {Mertens}, F., {Bernardi}, G., {et~al.} 2020, \mnras, 495, 2813,
  \dodoi{10.1093/mnras/staa1331}

\bibitem[{{Gorce} {et~al.}(2018){Gorce}, {Douspis}, {Aghanim}, \&
  {Langer}}]{Gorce18}
{Gorce}, A., {Douspis}, M., {Aghanim}, N., \& {Langer}, M. 2018, \aap, 616,
  A113, \dodoi{10.1051/0004-6361/201629661}

\bibitem[{Greig \& Mesinger(2017)}]{Greig17}
Greig, B., \& Mesinger, A. 2017, \mnras, 465, 4838,
  \dodoi{10.1093/mnras/stw3026}

\bibitem[{{Greig} {et~al.}(2017){Greig}, {Mesinger}, {Haiman}, \&
  {Simcoe}}]{Greig17b}
{Greig}, B., {Mesinger}, A., {Haiman}, Z., \& {Simcoe}, R.~A. 2017, \mnras,
  466, 4239, \dodoi{10.1093/mnras/stw3351}

\bibitem[{Greig {et~al.}(2016)Greig, Mesinger, \& Pober}]{Greig2016}
Greig, B., Mesinger, A., \& Pober, J. 2016, \mnras, 455, 4295,
  \dodoi{10.1093/mnras/stv2618}

\bibitem[{Guthrie(2020)}]{NIST_Stats}
Guthrie, W.~F. 2020, NIST/SEMATECH e-Handbook of Statistical Methods (NIST
  Handbook 151),  National Institute of Standards and Technology,
  \dodoi{10.18434/M32189}

\bibitem[{{Hamaker} {et~al.}(1996){Hamaker}, {Bregman}, \&
  {Sault}}]{Hamaker1996}
{Hamaker}, J.~P., {Bregman}, J.~D., \& {Sault}, R.~J. 1996, \aaps, 117, 137

\bibitem[{Harris {et~al.}(2020)Harris, Millman, van~der Walt, Gommers,
  Virtanen, Cournapeau, Wieser, Taylor, Berg, Smith, Kern, Picus, Hoyer, van
  Kerkwijk, Brett, Haldane, Fernández~del Río, Wiebe, Peterson,
  Gérard-Marchant, Sheppard, Reddy, Weckesser, Abbasi, Gohlke, \&
  Oliphant}]{2020NumPy-Array}
Harris, C.~R., Millman, K.~J., van~der Walt, S.~J., {et~al.} 2020, Nature, 585,
  357–362, \dodoi{10.1038/s41586-020-2649-2}

\bibitem[{{Hazra} {et~al.}(2019){Hazra}, {Paoletti}, {Finelli}, \&
  {Smoot}}]{Hazra19}
{Hazra}, D.~K., {Paoletti}, D., {Finelli}, F., \& {Smoot}, G.~F. 2019, arXiv
  e-prints, arXiv:1904.01547.
\newblock \doarXiv{1904.01547}

\bibitem[{{Hills} {et~al.}(2018){Hills}, {Kulkarni}, {Meerburg}, \&
  {Puchwein}}]{Hills2018}
{Hills}, R., {Kulkarni}, G., {Meerburg}, P.~D., \& {Puchwein}, E. 2018, \nat,
  564, E32, \dodoi{10.1038/s41586-018-0796-5}

\bibitem[{{Hodges}(1958)}]{2s_ks}
{Hodges}, J.~L. 1958, Arkiv for Matematik, 3, 469, \dodoi{10.1007/BF02589501}

\bibitem[{Hogan \& Rees(1979)}]{Hogan1979}
Hogan, C., \& Rees, M. 1979, \mnras, 188, 791, \dodoi{10.1093/mnras/188.4.791}

\bibitem[{{H{\"o}gbom}(1974)}]{Hogbom1974}
{H{\"o}gbom}, J.~A. 1974, \aaps, 15, 417

\bibitem[{{Hogg}(1999)}]{Hogg1999}
{Hogg}, D.~W. 1999, arXiv e-prints, astro.
\newblock \doarXiv{astro-ph/9905116}

\bibitem[{{Hurley-Walker} {et~al.}(2017){Hurley-Walker}, {Callingham},
  {Hancock}, {Franzen}, {Hindson}, {Kapi{\'n}ska}, {Morgan}, {Offringa},
  {Wayth}, {Wu}, {Zheng}, {Murphy}, {Bell}, {Dwarakanath}, {For}, {Gaensler},
  {Johnston-Hollitt}, {Lenc}, {Procopio}, {Staveley-Smith}, {Ekers}, {Bowman},
  {Briggs}, {Cappallo}, {Deshpande}, {Greenhill}, {Hazelton}, {Kaplan},
  {Lonsdale}, {McWhirter}, {Mitchell}, {Morales}, {Morgan}, {Oberoi}, {Ord},
  {Prabu}, {Shankar}, {Srivani}, {Subrahmanyan}, {Tingay}, {Webster},
  {Williams}, \& {Williams}}]{Hurley-Walker2017}
{Hurley-Walker}, N., {Callingham}, J.~R., {Hancock}, P.~J., {et~al.} 2017,
  \mnras, 464, 1146, \dodoi{10.1093/mnras/stw2337}

\bibitem[{{Jeli{\'c}} {et~al.}(2010){Jeli{\'c}}, {Zaroubi}, {Labropoulos},
  {Bernardi}, {de Bruyn}, \& {Koopmans}}]{Jelic2010}
{Jeli{\'c}}, V., {Zaroubi}, S., {Labropoulos}, P., {et~al.} 2010, \mnras, 409,
  1647, \dodoi{10.1111/j.1365-2966.2010.17407.x}

\bibitem[{{Jensen} {et~al.}(2013){Jensen}, {Laursen}, {Mellema}, {Iliev},
  {Sommer-Larsen}, \& {Shapiro}}]{Jensen13}
{Jensen}, H., {Laursen}, P., {Mellema}, G., {et~al.} 2013, \mnras, 428, 1366,
  \dodoi{10.1093/mnras/sts116}

\bibitem[{{Keating} {et~al.}(2020){Keating}, {Weinberger}, {Kulkarni},
  {Haehnelt}, {Chardin}, \& {Aubert}}]{keating20}
{Keating}, L.~C., {Weinberger}, L.~H., {Kulkarni}, G., {et~al.} 2020, \mnras,
  491, 1736, \dodoi{10.1093/mnras/stz3083}

\bibitem[{{Kern} \& {Liu}(2021)}]{Kern2021}
{Kern}, N.~S., \& {Liu}, A. 2021, \mnras, 501, 1463,
  \dodoi{10.1093/mnras/staa3736}

\bibitem[{{Kern} {et~al.}(2017){Kern}, {Liu}, {Parsons}, {Mesinger}, \&
  {Greig}}]{Kern2017}
{Kern}, N.~S., {Liu}, A., {Parsons}, A.~R., {Mesinger}, A., \& {Greig}, B.
  2017, \apj, 848, 23, \dodoi{10.3847/1538-4357/aa8bb4}

\bibitem[{{Kern} {et~al.}(2019){Kern}, {Parsons}, {Dillon}, {Lanman},
  {Fagnoni}, \& {de Lera Acedo}}]{Kern2019}
{Kern}, N.~S., {Parsons}, A.~R., {Dillon}, J.~S., {et~al.} 2019, \apj, 884,
  105, \dodoi{10.3847/1538-4357/ab3e73}

\bibitem[{{Kern} {et~al.}(2020{\natexlab{a}}){Kern}, {Dillon}, {Parsons},
  {Carilli}, {Bernardi}, {Abdurashidova}, {Aguirre}, {Alexander}, {Ali},
  {Balfour}, {Beardsley}, {Billings}, {Bowman}, {Bradley}, {Bull}, {Burba},
  {Carey}, {Cheng}, {DeBoer}, {Dexter}, {de Lera Acedo}, {Ely}, {Ewall-Wice},
  {Fagnoni}, {Fritz}, {Furlanetto}, {Gale-Sides}, {Glendenning}, {Gorthi},
  {Greig}, {Grobbelaar}, {Halday}, {Hazelton}, {Hewitt}, {Hickish}, {Jacobs},
  {Julius}, {Kerrigan}, {Kittiwisit}, {Kohn}, {Kolopanis}, {Lanman}, {La
  Plante}, {Lekalake}, {Liu}, {MacMahon}, {Malan}, {Malgas}, {Maree},
  {Martinot}, {Matsetela}, {Mesinger}, {Molewa}, {Morales}, {Mosiane},
  {Murray}, {Neben}, {Nikolic}, {Nunhokee}, {Patra}, {Pieterse}, {Pober},
  {Razavi-Ghods}, {Ringuette}, {Robnett}, {Rosie}, {Sims}, {Smith}, {Syce},
  {Thyagarajan}, {Williams}, \& {Zheng}}]{Kern2020b}
{Kern}, N.~S., {Dillon}, J.~S., {Parsons}, A.~R., {et~al.} 2020{\natexlab{a}},
  \apj, 890, 122, \dodoi{10.3847/1538-4357/ab67bc}

\bibitem[{{Kern} {et~al.}(2020{\natexlab{b}}){Kern}, {Parsons}, {Dillon},
  {Lanman}, {Liu}, {Bull}, {Ewall-Wice}, {Abdurashidova}, {Aguirre},
  {Alexander}, {Ali}, {Balfour}, {Beardsley}, {Bernardi}, {Bowman}, {Bradley},
  {Burba}, {Carilli}, {Cheng}, {DeBoer}, {Dexter}, {de Lera Acedo}, {Fagnoni},
  {Fritz}, {Furlanetto}, {Glendenning}, {Gorthi}, {Greig}, {Grobbelaar},
  {Halday}, {Hazelton}, {Hewitt}, {Hickish}, {Jacobs}, {Julius}, {Kerrigan},
  {Kittiwisit}, {Kohn}, {Kolopanis}, {La Plante}, {Lekalake}, {MacMahon},
  {Malan}, {Malgas}, {Maree}, {Martinot}, {Matsetela}, {Mesinger}, {Molewa},
  {Morales}, {Mosiane}, {Murray}, {Neben}, {Parsons}, {Patra}, {Pieterse},
  {Pober}, {Razavi-Ghods}, {Ringuette}, {Robnett}, {Rosie}, {Sims}, {Smith},
  {Syce}, {Thyagarajan}, {Williams}, \& {Zheng}}]{Kern2020a}
{Kern}, N.~S., {Parsons}, A.~R., {Dillon}, J.~S., {et~al.} 2020{\natexlab{b}},
  \apj, 888, 70, \dodoi{10.3847/1538-4357/ab5e8a}

\bibitem[{{Kerrigan} {et~al.}(2018){Kerrigan}, {Pober}, {Ali}, {Cheng},
  {Beardsley}, {Parsons}, {Aguirre}, {Barry}, {Bradley}, {Bernardi}, {Carilli},
  {DeBoer}, {Dillon}, {Jacobs}, {Kohn}, {Kolopanis}, {Lanman}, {Li}, {Liu}, \&
  {Sullivan}}]{Kerrigan2018}
{Kerrigan}, J.~R., {Pober}, J.~C., {Ali}, Z.~S., {et~al.} 2018, \apj, 864, 131,
  \dodoi{10.3847/1538-4357/aad8bb}

\bibitem[{{Kim} {et~al.}(2016){Kim}, {Lilly}, {Miniati}, {Bernet}, {Beck},
  {O'Sullivan}, \& {Gaensler}}]{Kim2016}
{Kim}, K.~S., {Lilly}, S.~J., {Miniati}, F., {et~al.} 2016, \apj, 829, 133,
  \dodoi{10.3847/0004-637X/829/2/133}

\bibitem[{{Kolopanis} {et~al.}(2019){Kolopanis}, {Jacobs}, {Cheng}, {Parsons},
  {Kohn}, {Pober}, {Aguirre}, {Ali}, {Bernardi}, {Bradley}, {Carilli},
  {DeBoer}, {Dexter}, {Dillon}, {Kerrigan}, {Klima}, {Liu}, {MacMahon},
  {Moore}, {Thyagarajan}, {Nunhokee}, {Walbrugh}, \& {Walker}}]{Kolopanis2019}
{Kolopanis}, M., {Jacobs}, D.~C., {Cheng}, C., {et~al.} 2019, \apj, 883, 133,
  \dodoi{10.3847/1538-4357/ab3e3a}

\bibitem[{Koopmans {et~al.}(2015)Koopmans, Pritchard, Mellema, Aguirre, Ahn,
  Barkana, van Bemmel, Bernardi, Bonaldi, Briggs, de~Bruyn, Chang, Chapman,
  Chen, Ciardi, Dayal, Ferrara, Fialkov, Fiore, Ichiki, Illiev, Inoue, Jelic,
  Jones, Lazio, Maio, Majumdar, Mack, Mesinger, Morales, Parsons, Pen, Santos,
  Schneider, Semelin, de~Souza, Subrahmanyan, Takeuchi, Vedantham, Wagg,
  Webster, Wyithe, Datta, \& Trott}]{Koopmans2015}
Koopmans, L., Pritchard, J., Mellema, G., {et~al.} 2015, Adv. Astrophys. with
  Sq. Km. Array, 1.
\newblock \doarXiv{1505.07568}

\bibitem[{{Kulkarni} {et~al.}(2019){Kulkarni}, {Keating}, {Haehnelt}, {Bosman},
  {Puchwein}, {Chardin}, \& {Aubert}}]{Kulkarni19}
{Kulkarni}, G., {Keating}, L.~C., {Haehnelt}, M.~G., {et~al.} 2019, \mnras,
  485, L24, \dodoi{10.1093/mnrasl/slz025}

\bibitem[{{Lenc} {et~al.}(2017){Lenc}, {Anderson}, {Barry}, {Bowman}, {Cairns},
  {Farnes}, {Gaensler}, {Heald}, {Johnston-Hollitt}, {Kaplan}, {Lynch},
  {McCauley}, {Mitchell}, {Morgan}, {Morales}, {Murphy}, {Offringa}, {Ord},
  {Pindor}, {Riseley}, {Sadler}, {Sobey}, {Sokolowski}, {Sullivan},
  {O'Sullivan}, {Sun}, {Tremblay}, {Trott}, \& {Wayth}}]{Lenc2017}
{Lenc}, E., {Anderson}, C.~S., {Barry}, N., {et~al.} 2017, \pasa, 34, e040,
  \dodoi{10.1017/pasa.2017.36}

\bibitem[{{Li} {et~al.}(2018){Li}, {Pober}, {Hazelton}, {Barry}, {Morales},
  {Sullivan}, {Parsons}, {Ali}, {Dillon}, {Beardsley}, {Bowman}, {Briggs},
  {Byrne}, {Carroll}, {Crosse}, {Emrich}, {Ewall-Wice}, {Feng}, {Franzen},
  {Hewitt}, {Horsley}, {Jacobs}, {Johnston-Hollitt}, {Jordan}, {Joseph},
  {Kaplan}, {Kenney}, {Kim}, {Kittiwisit}, {Lanman}, {Line}, {McKinley},
  {Mitchell}, {Murray}, {Neben}, {Offringa}, {Pallot}, {Paul}, {Pindor},
  {Procopio}, {Rahimi}, {Riding}, {Sethi}, {Udaya Shankar}, {Steele},
  {Subrahmanian}, {Tegmark}, {Thyagarajan}, {Tingay}, {Trott}, {Walker},
  {Wayth}, {Webster}, {Williams}, {Wu}, \& {Wyithe}}]{Li2018}
{Li}, W., {Pober}, J.~C., {Hazelton}, B.~J., {et~al.} 2018, \apj, 863, 170,
  \dodoi{10.3847/1538-4357/aad3c3}

\bibitem[{{Li} {et~al.}(2019){Li}, {Pober}, {Barry}, {Hazelton}, {Morales},
  {Trott}, {Lanman}, {Wilensky}, {Sullivan}, {Beardsley}, {Booler}, {Bowman},
  {Byrne}, {Crosse}, {Emrich}, {Franzen}, {Hasegawa}, {Horsley},
  {Johnston-Hollitt}, {Jacobs}, {Jordan}, {Joseph}, {Kaneuji}, {Kaplan},
  {Kenney}, {Kubota}, {Line}, {Lynch}, {McKinley}, {Mitchell}, {Murray},
  {Pallot}, {Pindor}, {Rahimi}, {Riding}, {Sleap}, {Steele}, {Takahashi},
  {Tingay}, {Walker}, {Wayth}, {Webster}, {Williams}, {Wu}, {Wyithe},
  {Yoshiura}, \& {Zheng}}]{Li2019}
{Li}, W., {Pober}, J.~C., {Barry}, N., {et~al.} 2019, arXiv e-prints,
  arXiv:1911.10216.
\newblock \doarXiv{1911.10216}

\bibitem[{Liu \& Parsons(2016)}]{Liu2016b}
Liu, A., \& Parsons, A. 2016, \mnras, 457, 1864, \dodoi{10.1093/mnras/stw071}

\bibitem[{Liu {et~al.}(2014{\natexlab{a}})Liu, Parsons, \& Trott}]{Liu2014a}
Liu, A., Parsons, A., \& Trott, C. 2014{\natexlab{a}}, \prd, 90, 23018,
  \dodoi{10.1103/PhysRevD.90.023018}

\bibitem[{Liu {et~al.}(2014{\natexlab{b}})Liu, Parsons, \& Trott}]{Liu2014b}
---. 2014{\natexlab{b}}, \prd, 90, 23019, \dodoi{10.1103/PhysRevD.90.023019}

\bibitem[{{Liu} \& {Shaw}(2020)}]{Liu2020}
{Liu}, A., \& {Shaw}, J.~R. 2020, \pasp, 132, 062001,
  \dodoi{10.1088/1538-3873/ab5bfd}

\bibitem[{Liu \& Tegmark(2011)}]{Liu2011}
Liu, A., \& Tegmark, M. 2011, \prd, 83, 103006,
  \dodoi{10.1103/PhysRevD.83.103006}

\bibitem[{Liu {et~al.}(2010)Liu, Tegmark, Morrison, Lutomirski, \&
  Zaldarriaga}]{Liu2010}
Liu, A., Tegmark, M., Morrison, S., Lutomirski, A., \& Zaldarriaga, M. 2010,
  \mnras, 408, 1029, \dodoi{10.1111/j.1365-2966.2010.17174.x}

\bibitem[{Madau {et~al.}(1997)Madau, Meiksin, \& Rees}]{Madau1997}
Madau, P., Meiksin, A., \& Rees, M. 1997, \apj, 475, 429,
  \dodoi{10.1086/303549}

\bibitem[{{Mahesh} {et~al.}(2021){Mahesh}, {Bowman}, {Mozdzen}, {Rogers},
  {Monsalve}, {Murray}, \& {Lewis}}]{Mahesh2021}
{Mahesh}, N., {Bowman}, J.~D., {Mozdzen}, T.~J., {et~al.} 2021, arXiv e-prints,
  arXiv:2103.00423.
\newblock \doarXiv{2103.00423}

\bibitem[{Mao {et~al.}(2008)Mao, Tegmark, McQuinn, Zaldarriaga, \&
  Zahn}]{Mao2008}
Mao, Y., Tegmark, M., McQuinn, M., Zaldarriaga, M., \& Zahn, O. 2008, \prd, 78,
  23529, \dodoi{10.1103/PhysRevD.78.023529}

\bibitem[{{Mason} {et~al.}(2018){Mason}, {Treu}, {Dijkstra}, {Mesinger},
  {Trenti}, {Pentericci}, {de Barros}, \& {Vanzella}}]{mason18}
{Mason}, C.~A., {Treu}, T., {Dijkstra}, M., {et~al.} 2018, \apj, 856, 2,
  \dodoi{10.3847/1538-4357/aab0a7}

\bibitem[{{Mason} {et~al.}(2019)}]{Mason19}
{Mason}, C.~A., {et~al.} 2019, \mnras, 485, 3947, \dodoi{10.1093/mnras/stz632}

\bibitem[{McGreer {et~al.}(2015)McGreer, Mesinger, \& D'Odorico}]{McGreer2015}
McGreer, I., Mesinger, A., \& D'Odorico, V. 2015, \mnras, 447, 499,
  \dodoi{10.1093/mnras/stu2449}

\bibitem[{{McMullin} {et~al.}(2007){McMullin}, {Waters}, {Schiebel}, {Young},
  \& {Golap}}]{CASA}
{McMullin}, J.~P., {Waters}, B., {Schiebel}, D., {Young}, W., \& {Golap}, K.
  2007, in Astronomical Society of the Pacific Conference Series, Vol. 376,
  Astronomical Data Analysis Software and Systems XVI, ed. R.~A. {Shaw},
  F.~{Hill}, \& D.~J. {Bell}, 127

\bibitem[{{Mertens} {et~al.}(2020){Mertens}, {Mevius}, {Koopmans}, {Offringa},
  {Mellema}, {Zaroubi}, {Brentjens}, {Gan}, {Gehlot}, {Pand ey}, {Sardarabadi},
  {Vedantham}, {Yatawatta}, {Asad}, {Ciardi}, {Chapman}, {Gazagnes}, {Ghara},
  {Ghosh}, {Giri}, {Iliev}, {Jeli{\'c}}, {Kooistra}, {Mondal}, {Schaye}, \&
  {Silva}}]{Mertens2020}
{Mertens}, F.~G., {Mevius}, M., {Koopmans}, L.~V.~E., {et~al.} 2020, \mnras,
  493, 1662, \dodoi{10.1093/mnras/staa327}

\bibitem[{Mesinger(2016)}]{Mesinger2016b}
Mesinger, A. 2016, Underst. Epoch Cosm. Reionization Challenges Prog., 423,
  \dodoi{10.1007/978-3-319-21957-8}

\bibitem[{{Mesinger} {et~al.}(2015){Mesinger}, {Aykutalp}, {Vanzella},
  {Pentericci}, {Ferrara}, \& {Dijkstra}}]{Mesinger15}
{Mesinger}, A., {Aykutalp}, A., {Vanzella}, E., {et~al.} 2015, \mnras, 446,
  566, \dodoi{10.1093/mnras/stu2089}

\bibitem[{{Mesinger} \& {Haiman}(2004)}]{MH04}
{Mesinger}, A., \& {Haiman}, Z. 2004, \apjl, 611, L69

\bibitem[{{Millea} \& {Bouchet}(2018)}]{Millea18}
{Millea}, M., \& {Bouchet}, F. 2018, \aap, 617, A96,
  \dodoi{10.1051/0004-6361/201833288}

\bibitem[{{Mitra} {et~al.}(2015){Mitra}, {Choudhury}, \& {Ferrara}}]{Mitra15}
{Mitra}, S., {Choudhury}, T.~R., \& {Ferrara}, A. 2015, \mnras, 454, L76,
  \dodoi{10.1093/mnrasl/slv134}

\bibitem[{{Monsalve} {et~al.}(2017){Monsalve}, {Rogers}, {Bowman}, \&
  {Mozdzen}}]{Monsalve2017}
{Monsalve}, R.~A., {Rogers}, A. E.~E., {Bowman}, J.~D., \& {Mozdzen}, T.~J.
  2017, \apj, 847, 64, \dodoi{10.3847/1538-4357/aa88d1}

\bibitem[{{Moore} {et~al.}(2013){Moore}, {Aguirre}, {Parsons}, {Jacobs}, \&
  {Pober}}]{Moore2013}
{Moore}, D.~F., {Aguirre}, J.~E., {Parsons}, A.~R., {Jacobs}, D.~C., \&
  {Pober}, J.~C. 2013, \apj, 769, 154, \dodoi{10.1088/0004-637X/769/2/154}

\bibitem[{{Moore} {et~al.}(2017){Moore}, {Aguirre}, {Kohn}, {Parsons}, {Ali},
  {Bradley}, {Carilli}, {DeBoer}, {Dexter}, {Gugliucci}, {Jacobs}, {Klima},
  {Liu}, {MacMahon}, {Manley}, {Pober}, {Stefan}, \& {Walbrugh}}]{Moore2017}
{Moore}, D.~F., {Aguirre}, J.~E., {Kohn}, S.~A., {et~al.} 2017, \apj, 836, 154,
  \dodoi{10.3847/1538-4357/836/2/154}

\bibitem[{Morales \& Wyithe(2010)}]{Morales2010}
Morales, M., \& Wyithe, J. 2010, \araa, 48, 127,
  \dodoi{10.1146/annurev-astro-081309-130936}

\bibitem[{{Mouri Sardarabadi} \& {Koopmans}(2019)}]{Mouri2019}
{Mouri Sardarabadi}, A., \& {Koopmans}, L.~V.~E. 2019, \mnras, 483, 5480,
  \dodoi{10.1093/mnras/sty3444}

\bibitem[{{Neben} {et~al.}(2016){Neben}, {Bradley}, {Hewitt}, {DeBoer},
  {Parsons}, {Aguirre}, {Ali}, {Cheng}, {Ewall-Wice}, {Patra}, {Thyagarajan},
  {Bowman}, {Dickenson}, {Dillon}, {Doolittle}, {Egan}, {Hedrick}, {Jacobs},
  {Kohn}, {Klima}, {Moodley}, {Saliwanchik}, {Schaffner}, {Shelton}, {Taylor},
  {Taylor}, {Tegmark}, {Wirt}, \& {Zheng}}]{Neben2016}
{Neben}, A.~R., {Bradley}, R.~F., {Hewitt}, J.~N., {et~al.} 2016, \apj, 826,
  199, \dodoi{10.3847/0004-637X/826/2/199}

\bibitem[{{Nunhokee} {et~al.}(2017){Nunhokee}, {Bernardi}, {Kohn}, {Aguirre},
  {Thyagarajan}, {Dillon}, {Foster}, {Grobler}, {Martinot}, \&
  {Parsons}}]{Nunhokee2017}
{Nunhokee}, C.~D., {Bernardi}, G., {Kohn}, S.~A., {et~al.} 2017, \apj, 848, 47,
  \dodoi{10.3847/1538-4357/aa8b73}

\bibitem[{{Offringa} {et~al.}(2019){Offringa}, {Mertens}, \&
  {Koopmans}}]{Offringa2019}
{Offringa}, A.~R., {Mertens}, F., \& {Koopmans}, L.~V.~E. 2019, \mnras, 484,
  2866, \dodoi{10.1093/mnras/stz175}

\bibitem[{{Orosz} {et~al.}(2019){Orosz}, {Dillon}, {Ewall-Wice}, {Parsons}, \&
  {Thyagarajan}}]{Orosz2019}
{Orosz}, N., {Dillon}, J.~S., {Ewall-Wice}, A., {Parsons}, A.~R., \&
  {Thyagarajan}, N. 2019, \mnras, 487, 537, \dodoi{10.1093/mnras/stz1287}

\bibitem[{Paciga {et~al.}(2013)Paciga, Albert, Bandura, Chang, Gupta, Hirata,
  Odegova, Pen, Peterson, Roy, Shaw, Sigurdson, \& Voytek}]{Paciga2013}
Paciga, G., Albert, J., Bandura, K., {et~al.} 2013, \mnras, 433, 639,
  \dodoi{10.1093/mnras/stt753}

\bibitem[{{Park} {et~al.}(2019){Park}, {Mesinger}, {Greig}, \&
  {Gillet}}]{Park19}
{Park}, J., {Mesinger}, A., {Greig}, B., \& {Gillet}, N. 2019, \mnras, 484,
  933, \dodoi{10.1093/mnras/stz032}

\bibitem[{Parsons {et~al.}(2012)Parsons, Pober, McQuinn, Jacobs, \&
  Aguirre}]{Parsons2012a}
Parsons, A., Pober, J., McQuinn, M., Jacobs, D., \& Aguirre, J. 2012, \apj,
  753, 81, \dodoi{10.1088/0004-637X/753/1/81}

\bibitem[{Parsons {et~al.}(2014)Parsons, Liu, Aguirre, Ali, Bradley, Carilli,
  DeBoer, Dexter, Gugliucci, Jacobs, Klima, MacMahon, Manley, Moore, Pober,
  Stefan, \& Walbrugh}]{Parsons2014}
Parsons, A., Liu, A., Aguirre, J., {et~al.} 2014, \apj, 788, 106,
  \dodoi{10.1088/0004-637X/788/2/106}

\bibitem[{{Parsons} \& {Backer}(2009)}]{Parsons2009}
{Parsons}, A.~R., \& {Backer}, D.~C. 2009, \aj, 138, 219,
  \dodoi{10.1088/0004-6256/138/1/219}

\bibitem[{{Parsons} {et~al.}(2016){Parsons}, {Liu}, {Ali}, \&
  {Cheng}}]{Parsons2016}
{Parsons}, A.~R., {Liu}, A., {Ali}, Z.~S., \& {Cheng}, C. 2016, \apj, 820, 51,
  \dodoi{10.3847/0004-637X/820/1/51}

\bibitem[{{Parsons} {et~al.}(2010){Parsons}, {Backer}, {Foster}, {Wright},
  {Bradley}, {Gugliucci}, {Parashare}, {Benoit}, {Aguirre}, {Jacobs},
  {Carilli}, {Herne}, {Lynch}, {Manley}, \& {Werthimer}}]{Parsons2010}
{Parsons}, A.~R., {Backer}, D.~C., {Foster}, G.~S., {et~al.} 2010, \aj, 139,
  1468, \dodoi{10.1088/0004-6256/139/4/1468}

\bibitem[{Patil {et~al.}(2017)Patil, Yatawatta, Koopmans, de~Bruyn, Brentjens,
  Zaroubi, Asad, Hatef, Jeli{\'{c}}, Mevius, Offringa, Pandey, Vedantham,
  Abdalla, Brouw, Chapman, Ciardi, Gehlot, Ghosh, Harker, Iliev, Kakiichi,
  Majumdar, Mellema, Silva, Schaye, Vrbanec, \& Wijnholds}]{Patil2017}
Patil, A., Yatawatta, S., Koopmans, L., {et~al.} 2017, \apj, 838, 65,
  \dodoi{10.3847/1538-4357/aa63e7}

\bibitem[{{Patil} {et~al.}(2014){Patil}, {Zaroubi}, {Chapman}, {Jeli{\'c}},
  {Harker}, {Abdalla}, {Asad}, {Bernardi}, {Brentjens}, {de Bruyn}, {Bus},
  {Ciardi}, {Daiboo}, {Fernandez}, {Ghosh}, {Jensen}, {Kazemi}, {Koopmans},
  {Labropoulos}, {Mevius}, {Martinez}, {Mellema}, {Offringa}, {Pandey},
  {Schaye}, {Thomas}, {Vedantham}, {Veligatla}, {Wijnholds}, \&
  {Yatawatta}}]{Patil2014}
{Patil}, A.~H., {Zaroubi}, S., {Chapman}, E., {et~al.} 2014, \mnras, 443, 1113,
  \dodoi{10.1093/mnras/stu1178}

\bibitem[{{Patra} {et~al.}(2018){Patra}, {Parsons}, {DeBoer}, {Thyagarajan},
  {Ewall-Wice}, {Hsyu}, {Leung}, {Day}, {de Lera Acedo}, {Aguirre},
  {Alexander}, {Ali}, {Beardsley}, {Bowman}, {Bradley}, {Carilli}, {Cheng},
  {Dillon}, {Fadana}, {Fagnoni}, {Fritz}, {Furlanetto}, {Glendenning}, {Greig},
  {Grobbelaar}, {Hazelton}, {Jacobs}, {Julius}, {Kariseb}, {Kohn}, {Lebedeva},
  {Lekalake}, {Liu}, {Loots}, {MacMahon}, {Malan}, {Malgas}, {Maree},
  {Martinot}, {Mathison}, {Matsetela}, {Mesinger}, {Morales}, {Neben},
  {Pieterse}, {Pober}, {Razavi-Ghods}, {Ringuette}, {Robnett}, {Rosie}, {Sell},
  {Smith}, {Syce}, {Tegmark}, {Williams}, \& {Zheng}}]{Patra2018}
{Patra}, N., {Parsons}, A.~R., {DeBoer}, D.~R., {et~al.} 2018, Experimental
  Astronomy, 45, 177, \dodoi{10.1007/s10686-017-9563-0}

\bibitem[{{Pentericci} {et~al.}(2014)}]{Pentericci14}
{Pentericci}, L., {et~al.} 2014, \apj, 793, 113,
  \dodoi{10.1088/0004-637X/793/2/113}

\bibitem[{{Planck Collaboration}(2018)}]{Planck18}
{Planck Collaboration}. 2018, arXiv e-prints, arXiv:1807.06209.
\newblock \doarXiv{1807.06209}

\bibitem[{{Planck Collaboration} {et~al.}(2016){Planck Collaboration}, Ade,
  Aghanim, Arnaud, Ashdown, Aumont, Baccigalupi, Banday, Barreiro, Bartlett, \&
  al.}]{Planck2016}
{Planck Collaboration}, Ade, P., Aghanim, N., {et~al.} 2016, \aap, 594, A13,
  \dodoi{10.1051/0004-6361/201525830}

\bibitem[{Pober {et~al.}(2013{\natexlab{a}})Pober, Parsons, DeBoer, McDonald,
  McQuinn, Aguirre, Ali, Bradley, Chang, \& Morales}]{Pober2013a}
Pober, J., Parsons, A., DeBoer, D., {et~al.} 2013{\natexlab{a}}, \aj, 145, 65,
  \dodoi{10.1088/0004-6256/145/3/65}

\bibitem[{Pober {et~al.}(2013{\natexlab{b}})Pober, Parsons, Aguirre, Ali,
  Bradley, Carilli, DeBoer, Dexter, Gugliucci, Jacobs, Klima, MacMahon, Manley,
  Moore, Stefan, \& Walbrugh}]{Pober2013b}
Pober, J., Parsons, A., Aguirre, J., {et~al.} 2013{\natexlab{b}}, \apjl, 768,
  L36, \dodoi{10.1088/2041-8205/768/2/L36}

\bibitem[{Pober {et~al.}(2014)Pober, Liu, Dillon, Aguirre, Bowman, Bradley,
  Carilli, DeBoer, Hewitt, Jacobs, McQuinn, Morales, Parsons, Tegmark, \&
  Werthimer}]{Pober2014}
Pober, J., Liu, A., Dillon, J., {et~al.} 2014, \apj, 782, 66,
  \dodoi{10.1088/0004-637X/782/2/66}

\bibitem[{Pober {et~al.}(2016)Pober, Hazelton, Beardsley, Barry, Martinot,
  Sullivan, Morales, Bell, Bernardi, Bhat, Bowman, Briggs, Cappallo, Carroll,
  Corey, de~Oliveira-Costa, Deshpande, Dillon, Emrich, Ewall-Wice, Feng, Goeke,
  Greenhill, Hewitt, Hindson, Hurley-Walker, Jacobs, Johnston-Hollitt, Kaplan,
  Kasper, Kim, Kittiwisit, Kratzenberg, Kudryavtseva, Lenc, Line, Loeb,
  Lonsdale, Lynch, McKinley, McWhirter, Mitchell, Morgan, Neben, Oberoi,
  Offringa, Ord, Paul, Pindor, Prabu, Procopio, Riding, Rogers, Roshi, Sethi,
  {Udaya Shankar}, Srivani, Subrahmanyan, Tegmark, Thyagarajan, Tingay, Trott,
  Waterson, Wayth, Webster, Whitney, Williams, Williams, \& Wyithe}]{Pober2016}
Pober, J., Hazelton, B., Beardsley, A., {et~al.} 2016, \apj, 819, 8,
  \dodoi{10.3847/0004-637X/819/1/8}

\bibitem[{Price {et~al.}(2016)Price, Trac, \& Cen}]{Price16}
Price, L.~C., Trac, H., \& Cen, R. 2016, arXiv, 6, 1605.03970.
\newblock \doarXiv{1605.03970}

\bibitem[{Pritchard \& Loeb(2012)}]{Pritchard2012}
Pritchard, J., \& Loeb, A. 2012, Reports Prog. Phys., 75, 86901,
  \dodoi{10.1088/0034-4885/75/8/086901}

\bibitem[{{Qin} {et~al.}(2021){Qin}, {Mesinger}, {Bosman}, \& {Viel}}]{Qin21}
{Qin}, Y., {Mesinger}, A., {Bosman}, S. E.~I., \& {Viel}, M. 2021, arXiv
  e-prints, arXiv:2101.09033.
\newblock \doarXiv{2101.09033}

\bibitem[{{Qin} {et~al.}(2020{\natexlab{a}}){Qin}, {Mesinger}, {Park}, {Greig},
  \& {Mu{\~n}oz}}]{Qin20b}
{Qin}, Y., {Mesinger}, A., {Park}, J., {Greig}, B., \& {Mu{\~n}oz}, J.~B.
  2020{\natexlab{a}}, \mnras, 495, 123, \dodoi{10.1093/mnras/staa1131}

\bibitem[{{Qin} {et~al.}(2020{\natexlab{b}}){Qin}, {Poulin}, {Mesinger},
  {Greig}, {Murray}, \& {Park}}]{Qin20}
{Qin}, Y., {Poulin}, V., {Mesinger}, A., {et~al.} 2020{\natexlab{b}}, \mnras,
  499, 550, \dodoi{10.1093/mnras/staa2797}

\bibitem[{Robertson {et~al.}(2015)Robertson, Ellis, Furlanetto, \&
  Dunlop}]{Robertson2015}
Robertson, B., Ellis, R., Furlanetto, S., \& Dunlop, J. 2015, \apjl, 802, L19,
  \dodoi{10.1088/2041-8205/802/2/L19}

\bibitem[{{Schenker} {et~al.}(2012){Schenker}, {Stark}, {Ellis}, {Robertson},
  {Dunlop}, {McLure}, {Kneib}, \& {Richard}}]{Schenker12}
{Schenker}, M.~A., {Stark}, D.~P., {Ellis}, R.~S., {et~al.} 2012, \apj, 744,
  179, \dodoi{10.1088/0004-637X/744/2/179}

\bibitem[{{Sims} \& {Pober}(2020)}]{Sims2020}
{Sims}, P.~H., \& {Pober}, J.~C. 2020, \mnras, 492, 22,
  \dodoi{10.1093/mnras/stz3388}

\bibitem[{{Singh} \& {Subrahmanyan}(2019)}]{Singh2019}
{Singh}, S., \& {Subrahmanyan}, R. 2019, \apj, 880, 26,
  \dodoi{10.3847/1538-4357/ab2879}

\bibitem[{{Singh} {et~al.}(2017){Singh}, {Subrahmanyan}, {Udaya Shankar},
  {Sathyanarayana Rao}, {Fialkov}, {Cohen}, {Barkana}, {Girish}, {Raghunathan},
  {Somashekar}, \& {Srivani}}]{Singh2017}
{Singh}, S., {Subrahmanyan}, R., {Udaya Shankar}, N., {et~al.} 2017, \apjl,
  845, L12, \dodoi{10.3847/2041-8213/aa831b}

\bibitem[{{Smirnov}(2011)}]{Smirnov2011}
{Smirnov}, O.~M. 2011, \aap, 527, A106, \dodoi{10.1051/0004-6361/201016082}

\bibitem[{{Stark} {et~al.}(2010){Stark}, {Ellis}, {Chiu}, {Ouchi}, \&
  {Bunker}}]{Stark10}
{Stark}, D.~P., {Ellis}, R.~S., {Chiu}, K., {Ouchi}, M., \& {Bunker}, A. 2010,
  \mnras, 408, 1628, \dodoi{10.1111/j.1365-2966.2010.17227.x}

\bibitem[{{Tan} {et~al.}(in prep.){Tan}, {Ali}, \& the
  HERA~Collaboration}]{Tan2021}
{Tan}, J., {Ali}, Z.~S., \& the HERA~Collaboration. in prep.

\bibitem[{{Tegmark}(1997)}]{Tegmark1997}
{Tegmark}, M. 1997, \prd, 55, 5895, \dodoi{10.1103/PhysRevD.55.5895}

\bibitem[{{Thyagarajan} {et~al.}(2016){Thyagarajan}, {Parsons}, {DeBoer},
  {Bowman}, {Ewall-Wice}, {Neben}, \& {Patra}}]{Thyagarajan2016}
{Thyagarajan}, N., {Parsons}, A.~R., {DeBoer}, D.~R., {et~al.} 2016, \apj, 825,
  9, \dodoi{10.3847/0004-637X/825/1/9}

\bibitem[{{Thyagarajan} {et~al.}(2020){Thyagarajan}, {Carilli}, {Nikolic},
  {Kent}, {Mesinger}, {Kern}, {Bernardi}, {Matika}, {Abdurashidova}, {Aguirre},
  {Alexander}, {Ali}, {Balfour}, {Beardsley}, {Billings}, {Bowman}, {Bradley},
  {Burba}, {Carey}, {Cheng}, {DeBoer}, {Dexter}, {Acedo}, {Dillon}, {Ely},
  {Ewall-Wice}, {Fagnoni}, {Fritz}, {Furlanetto}, {Gale-Sides}, {Glendenning},
  {Gorthi}, {Greig}, {Grobbelaar}, {Halday}, {Hazelton}, {Hewitt}, {Hickish},
  {Jacobs}, {Julius}, {Kerrigan}, {Kittiwisit}, {Kohn}, {Kolopanis}, {Lanman},
  {La Plante}, {Lekalake}, {Lewis}, {Liu}, {MacMahon}, {Malan}, {Malgas},
  {Maree}, {Martinot}, {Matsetela}, {Molewa}, {Morales}, {Mosiane}, {Neben},
  {Parsons}, {Patra}, {Pieterse}, {Pober}, {Razavi-Ghods}, {Ringuette},
  {Robnett}, {Rosie}, {Sims}, {Smith}, {Syce}, {Williams}, \&
  {Zheng}}]{Thyagarajan2020}
{Thyagarajan}, N., {Carilli}, C.~L., {Nikolic}, B., {et~al.} 2020, \prd, 102,
  022002, \dodoi{10.1103/PhysRevD.102.022002}

\bibitem[{{Tingay} {et~al.}(2013){Tingay}, {Goeke}, {Bowman}, {Emrich}, {Ord},
  {Mitchell}, {Morales}, {Booler}, {Crosse}, {Wayth}, {Lonsdale}, {Tremblay},
  {Pallot}, {Colegate}, {Wicenec}, {Kudryavtseva}, {Arcus}, {Barnes},
  {Bernardi}, {Briggs}, {Burns}, {Bunton}, {Cappallo}, {Corey}, {Deshpande},
  {Desouza}, {Gaensler}, {Greenhill}, {Hall}, {Hazelton}, {Herne}, {Hewitt},
  {Johnston-Hollitt}, {Kaplan}, {Kasper}, {Kincaid}, {Koenig}, {Kratzenberg},
  {Lynch}, {Mckinley}, {Mcwhirter}, {Morgan}, {Oberoi}, {Pathikulangara},
  {Prabu}, {Remillard}, {Rogers}, {Roshi}, {Salah}, {Sault}, {Udaya-Shankar},
  {Schlagenhaufer}, {Srivani}, {Stevens}, {Subrahmanyan}, {Waterson},
  {Webster}, {Whitney}, {Williams}, {Williams}, \& {Wyithe}}]{Tingay2013}
{Tingay}, S.~J., {Goeke}, R., {Bowman}, J.~D., {et~al.} 2013, \pasa, 30, e007,
  \dodoi{10.1017/pasa.2012.007}

\bibitem[{{Trott} {et~al.}(2020){Trott}, {Jordan}, {Midgley}, {Barry}, {Greig},
  {Pindor}, {Cook}, {Sleap}, {Tingay}, {Ung}, {Hancock}, {Williams}, {Bowman},
  {Byrne}, {Chokshi}, {Hazelton}, {Hasegawa}, {Jacobs}, {Joseph}, {Li}, {Line},
  {Lynch}, {McKinley}, {Mitchell}, {Morales}, {Ouchi}, {Pober}, {Rahimi},
  {Takahashi}, {Wayth}, {Webster}, {Wilensky}, {Wyithe}, {Yoshiura}, {Zhang},
  \& {Zheng}}]{Trott2020}
{Trott}, C.~M., {Jordan}, C.~H., {Midgley}, S., {et~al.} 2020, \mnras,
  \dodoi{10.1093/mnras/staa414}

\bibitem[{{van Haarlem} {et~al.}(2013){van Haarlem}, {Wise}, {Gunst}, {Heald},
  {McKean}, {Hessels}, {de Bruyn}, {Nijboer}, {Swinbank}, {Fallows},
  {Brentjens}, {Nelles}, {Beck}, {Falcke}, {Fender}, {H{\"o}randel},
  {Koopmans}, {Mann}, {Miley}, {R{\"o}ttgering}, {Stappers}, {Wijers},
  {Zaroubi}, {van den Akker}, {Alexov}, {Anderson}, {Anderson}, {van Ardenne},
  {Arts}, {Asgekar}, {Avruch}, {Batejat}, {B{\"a}hren}, {Bell}, {Bell}, {van
  Bemmel}, {Bennema}, {Bentum}, {Bernardi}, {Best}, {B{\^\i}rzan}, {Bonafede},
  {Boonstra}, {Braun}, {Bregman}, {Breitling}, {van de Brink}, {Broderick},
  {Broekema}, {Brouw}, {Br{\"u}ggen}, {Butcher}, {van Cappellen}, {Ciardi},
  {Coenen}, {Conway}, {Coolen}, {Corstanje}, {Damstra}, {Davies}, {Deller},
  {Dettmar}, {van Diepen}, {Dijkstra}, {Donker}, {Doorduin}, {Dromer}, {Drost},
  {van Duin}, {Eisl{\"o}ffel}, {van Enst}, {Ferrari}, {Frieswijk}, {Gankema},
  {Garrett}, {de Gasperin}, {Gerbers}, {de Geus}, {Grie{\ss}meier}, {Grit},
  {Gruppen}, {Hamaker}, {Hassall}, {Hoeft}, {Holties}, {Horneffer}, {van der
  Horst}, {van Houwelingen}, {Huijgen}, {Iacobelli}, {Intema}, {Jackson},
  {Jelic}, {de Jong}, {Juette}, {Kant}, {Karastergiou}, {Koers}, {Kollen},
  {Kondratiev}, {Kooistra}, {Koopman}, {Koster}, {Kuniyoshi}, {Kramer},
  {Kuper}, {Lambropoulos}, {Law}, {van Leeuwen}, {Lemaitre}, {Loose}, {Maat},
  {Macario}, {Markoff}, {Masters}, {McFadden}, {McKay-Bukowski}, {Meijering},
  {Meulman}, {Mevius}, {Middelberg}, {Millenaar}, {Miller-Jones}, {Mohan},
  {Mol}, {Morawietz}, {Morganti}, {Mulcahy}, {Mulder}, {Munk}, {Nieuwenhuis},
  {van Nieuwpoort}, {Noordam}, {Norden}, {Noutsos}, {Offringa}, {Olofsson},
  {Omar}, {Orr{\'u}}, {Overeem}, {Paas}, {Pand ey-Pommier}, {Pandey}, {Pizzo},
  {Polatidis}, {Rafferty}, {Rawlings}, {Reich}, {de Reijer}, {Reitsma},
  {Renting}, {Riemers}, {Rol}, {Romein}, {Roosjen}, {Ruiter}, {Scaife}, {van
  der Schaaf}, {Scheers}, {Schellart}, {Schoenmakers}, {Schoonderbeek},
  {Serylak}, {Shulevski}, {Sluman}, {Smirnov}, {Sobey}, {Spreeuw}, {Steinmetz},
  {Sterks}, {Stiepel}, {Stuurwold}, {Tagger}, {Tang}, {Tasse}, {Thomas},
  {Thoudam}, {Toribio}, {van der Tol}, {Usov}, {van Veelen}, {van der Veen},
  {ter Veen}, {Verbiest}, {Vermeulen}, {Vermaas}, {Vocks}, {Vogt}, {de Vos},
  {van der Wal}, {van Weeren}, {Weggemans}, {Weltevrede}, {White}, {Wijnholds},
  {Wilhelmsson}, {Wucknitz}, {Yatawatta}, {Zarka}, {Zensus}, \& {van
  Zwieten}}]{vanHaarlem2013}
{van Haarlem}, M.~P., {Wise}, M.~W., {Gunst}, A.~W., {et~al.} 2013, \aap, 556,
  A2, \dodoi{10.1051/0004-6361/201220873}

\bibitem[{Virtanen {et~al.}(2020)Virtanen, Gommers, Oliphant, Haberland, Reddy,
  Cournapeau, Burovski, Peterson, Weckesser, Bright, {van der Walt}, Brett,
  Wilson, Millman, Mayorov, Nelson, Jones, Kern, Larson, Carey, Polat, Feng,
  Moore, {VanderPlas}, Laxalde, Perktold, Cimrman, Henriksen, Quintero, Harris,
  Archibald, Ribeiro, Pedregosa, {van Mulbregt}, \& {SciPy 1.0
  Contributors}}]{scipy2020}
Virtanen, P., Gommers, R., Oliphant, T.~E., {et~al.} 2020, Nature Methods, 17,
  261, \dodoi{10.1038/s41592-019-0686-2}

\bibitem[{{Wieringa}(1992)}]{Wieringa1992}
{Wieringa}, M.~H. 1992, Experimental Astronomy, 2, 203,
  \dodoi{10.1007/BF00420576}

\bibitem[{{Wilensky}(2020)}]{Wilensky2020Memo}
{Wilensky}, M. 2020, {Useful Information About TV and Digital Audio RFI As
  Observed By HERA}, Tech. rep.

\bibitem[{{Wilensky} {et~al.}(2019){Wilensky}, {Morales}, {Hazelton}, {Barry},
  {Byrne}, \& {Roy}}]{Wilensky2019}
{Wilensky}, M.~J., {Morales}, M.~F., {Hazelton}, B.~J., {et~al.} 2019, arXiv
  e-prints.
\newblock \doarXiv{1906.01093}

\bibitem[{{Yatawatta}(2015)}]{Yatawatta2015}
{Yatawatta}, S. 2015, \mnras, 449, 4506, \dodoi{10.1093/mnras/stv596}

\bibitem[{{Yatawatta}(2016)}]{Yatawatta2016}
---. 2016, arXiv e-prints, arXiv:1605.09219.
\newblock \doarXiv{1605.09219}

\bibitem[{Zheng {et~al.}(2014)Zheng, Tegmark, Buza, Dillon, Gharibyan, Hickish,
  Kunz, Liu, Losh, Lutomirski, Morrison, Narayanan, Perko, Rosner, Sanchez,
  Schutz, Tribiano, Valdez, Yang, Adami, Zelko, Zheng, Armstrong, Bradley,
  Dexter, Ewall-Wice, Magro, Matejek, Morgan, Neben, Pan, Penna, Peterson, Su,
  Villasenor, Williams, \& Zhu}]{Zheng2014}
Zheng, H., Tegmark, M., Buza, V., {et~al.} 2014, \mnras, 445, 1084,
  \dodoi{10.1093/mnras/stu1773}

\end{thebibliography}
